\documentclass[preprint,12pt]{elsarticle}
\usepackage{subfigure}
\usepackage{lineno,hyperref}
\usepackage{indentfirst}
\usepackage{bm}
\usepackage{threeparttable}
\usepackage{multirow}
\usepackage{mathrsfs}
\usepackage{amsmath}
\usepackage{array}
\usepackage{amssymb}
\usepackage{graphicx}
\usepackage{url}
\usepackage{float}
\usepackage{epstopdf}
\usepackage{booktabs}
\usepackage{color}
\usepackage{lscape}
\usepackage{fancyhdr}
\usepackage{dsfont}
\usepackage[top=2.5cm, bottom=2.5cm, left=2cm, right=2cm]{geometry}
\modulolinenumbers[5]

\journal{?}

\biboptions{numbers,sort&compress}

\begin{document}

\begin{frontmatter}

\title{A unified gas-kinetic wave-particle method for multiscale binary-species gas mixtures}

\author{Junzhe Cao$^{a}$}
\ead{jcaobb@connect.ust.hk}

\author{Yufeng Wei$^a$}
\ead{yweibe@connect.ust.hk}

\author{Wenpei Long$^a$}
\ead{wlongab@connect.ust.hk}

\author{Chengwen Zhong$^b$}
\ead{zhongcw@nwpu.edu.cn}

\author[]{Kun Xu$^{a,}$$^{c,}$$^{d,}$$^*$\corref{mycorrespondingauthor}}
\ead{makxu@ust.hk}

\address{$^a$Department of Mathematics, Hong Kong University of Science and Technology, Hong Kong, China\\
$^b$School of Aeronautics, Northwestern Polytechnical University, Xi'an, Shaanxi 710072, China\\
$^c$Department of Mechanical and Aerospace Engineering, Hong Kong University of Science and Technology, Hong Kong, China\\
$^d$HKUST Shenzhen Research Institute, Shenzhen, 518057, China}

\begin{abstract}

This paper presents a unified gas-kinetic wave-particle (UGKWP) method for simulating multiscale binary-species gas mixtures. Benefiting from direct modeling in a discretized space, the UGKWP method enables the automatic decomposition of the gas distribution function into analytical hydrodynamic waves and discrete particles, which respectively describe its near-equilibrium and non-equilibrium parts. This approach offers significant advantages for simulating various multiscale physical phenomena, such as hypersonic flows, plasma transport, and radiation transport. In this study, we employ the model proposed by Groppi et al. [EPL, 96 (2011) 64002] to calculate the macroscopic velocity and temperature of the local target equilibrium distribution function, thereby recovering the correct viscosity and diffusion coefficients in the continuum flow regime. To address the heat conduction coefficient, the Shakhov model is incorporated to correct the Prandtl number. Diffusion effects are accounted for not only in the source term via an operator-splitting method, but also in the flux evolution through the characteristic integral solution, while strictly maintaining consistency between the wave and particle descriptions. Furthermore, the microscopic model for high-speed particles is improved by utilizing a physically corrected collision time to determine their free-transport time. Through a series of numerical tests spanning the continuum to rarefied regimes, the proposed UGKWP method is shown to accurately capture the differences in velocity and temperature between different species. Notably, for hypersonic flows, the predicted wall pressure, shear stress, and heat flux coefficients agree well with direct simulation Monte Carlo (DSMC) results. Consequently, this method establishes a solid foundation for future extensions to more complex flows around near-space vehicles, including thermal non-equilibrium, chemical reactions, and plasma transport.
\end{abstract}


\end{frontmatter}

\section{Introduction}\label{Sec:introduction}
The unified gas-kinetic wave-particle (UGKWP) method~\cite{ugkwp1,ugkwp2} is a multiscale numerical approach that offers significant advantages across diverse multiscale physical problems, including hypersonic flows~\cite{wp-hyper1,wp-hyper2,wp-hyper3}, gas-solid two-phase flows~\cite{wp-phase1,wp-phase2}, radiation transport~\cite{wp-rad1,wp-rad2,wp-rad3}, plasma transport~\cite{wp-plasma1,wp-plasma2,wp-plasma3}, and turbulent flows~\cite{wp-turb1,wp-turb2}. In the UGKWP method, the near-equilibrium gain term is accumulated into hydrodynamic waves, while discrete particles are employed to capture the non-equilibrium part of the microscopic gas distribution function. By utilizing the integral solution of the kinetic model along the characteristic line, the time step is incorporated into the scheme as an observation scale. This achieves direct modeling~\cite{dm1,dm2} of physical laws in discretized space and enables an adaptive transformation between waves and particles. In the continuum flow regime, where no particles remain in the computational domain, the UGKWP method recovers the gas-kinetic scheme (GKS)~\cite{gks2001} as a Navier-Stokes solver. Conversely, in the rarefied regime, particles dominate the flow evolution, and UGKWP functions as a Boltzmann solver. Simplified variants of this method can be found in Refs.\cite{suwp,dugkwp,shpm}. In Ref.\cite{wp-6}, the underlying six-population kinetic system within the UGKWP framework is categorized into collisionless, collisional, and collided waves and particles, accompanied by a detailed analysis of their asymptotic behaviors. Furthermore, a unified gas-kinetic framework (UGKF)~\cite{ugkf} has been derived to classify molecules by their collision histories over an observation timescale, thereby bridging different flow regimes within a single framework. Alternatively, the UGKWP method can be viewed as a particle-based variant of the unified gas-kinetic scheme (UGKS)~\cite{ugks}, where particles represent a highly adaptive velocity space that offers exceptional computational efficiency for high-speed flows. The UGKWP method has shown advantages in simulating hypersonic flows around near-space vehicles--such as the X-38-like vehicle, the Apollo reentry capsule, and a $70^{\circ}$ blunted cone identical to the forebody of the Mars Pathfinder probe~\cite{wp-hyper4,awp1,awp2}--addressing a rapidly growing demand for advanced numerical simulations in aerospace engineering. However, due to the extreme temperatures behind shock waves, these multiscale phenomena are tightly coupled with thermal non-equilibrium and chemical reaction effects~\cite{rarefied-chemical}, making the development of corresponding physical models and numerical methods highly challenging. To address this, the present work proposes a UGKWP method for multiscale binary-species gas mixtures, establishing a robust foundation for the further development of more complex physical models.

Regarding multiscale numerical methods for gas mixtures, either a single relaxation term~\cite{aap,groppi1,brull1,brull2,todorova1,pfeiffer,usp-mixture} or multiple relaxation terms~\cite{mcc,haack,groppi2,liqi,dugks-mcc,gsis-liqi} can be employed in the kinetic model to represent the complex collisions of multi-species molecules. To simplify the construction of the numerical flux and facilitate extensions to more components, the former approach is adopted in this study. The single-relaxation kinetic model proposed by Andries, Aoki, and Perthame (the AAP model)~\cite{aap} successfully complies with the indifferentiability principle and has been widely applied in multiscale methods~\cite{wp-plasma1,ugks-aap,dugks-aap1}. This includes operator-splitting and implicit methods developed to resolve the stiff source terms governing momentum and energy exchange between different species~\cite{dugks-aap2}, as well as further extensions to multi-physics systems~\cite{plasma-aap,radiation-aap}. Within the framework of the AAP model, this study employs the target near-equilibrium gas distribution function proposed by Groppi et al.~\cite{groppi1}. This distribution is modeled by introducing a second relaxation parameter, allowing it to recover both the shear viscosity and diffusion coefficients in the continuum flow regime. To address the heat conduction coefficient, the Prandtl number ($\rm{Pr}$) is corrected following Refs.\cite{todorova1,todorova2}, which extends the Shakhov model~\cite{shakhov} to gas mixtures and derives the corresponding Chapman-Enskog expansion. Additionally, the modification of the relaxation time $\tau$ for high-speed particles in the UGKWP method~\cite{xu-tau} is extended here to gas mixtures. This correction yields more accurate results in high-$\rm{Ma}$ cases, particularly for the temperature profile in the pre-shock region and the heat flux at the stagnation point. Finally, other recent multiscale approaches for gas mixtures originating from the direct simulation Monte Carlo (DSMC) method~\cite{dsmc} include the asymptotic-preserving Monte Carlo (AAPMC) method~\cite{aapdsmc} and the direct intermittent GSIS-DSMC coupling (DIG) method~\cite{dig}.

The remainder of this paper is organized as follows. Section~\ref{sec:ugkwp} introduces the formulation of the UGKWP method for multiscale binary-species gas mixtures. Numerical test cases are presented in Sec.\ref{sec:cases}, and concluding remarks are provided in Sec.\ref{sec:conclusion}.

\section{UGKWP method for multiscale binary-species gas mixture}\label{sec:ugkwp}
\subsection{Discretized governing equations in the finite volume method framework}\label{sec:discretization}
In the general finite volume method (FVM) framework, the discretized governing equation of microscopic gas distribution function is as follows,
\begin{equation}\label{eq:conservation-micro}
f^{n+1}_{\alpha,i} = f^{n}_{\alpha,i} - \frac{1}{\Omega_i}\sum\limits_{j\in \mathcal{M}\left(i\right)}\int^{\Delta t}_0\boldsymbol{u}\cdot\boldsymbol{n}_jf_{\alpha,j}S_j{\rm d}t + \int^{\Delta t}_0\mathcal{J}_{\alpha,i}{\rm d}t,
\end{equation}
where $\alpha\in\left\{1,2\right\}$ is the label of species, ``$i$'' is the label of cell, $\mathcal{M}\left(i\right)$ is a collection of interfaces surrounding cell ``$i$'', $\Omega_i$ is the area or volume of cell ``$i$'', $S_j$ is the length or area of interface ``$j$'', $\boldsymbol{n}_j$ is the outer unit normal vector, $\Delta t$ is the time step. $f_{\alpha}\left(\rho,\boldsymbol{x},\boldsymbol{u},\boldsymbol{\xi},t\right)$ is the microscopic gas distribution function of species $\alpha$, where $\boldsymbol{x}$ denotes the position, $t$ denotes the time, $\boldsymbol{u}$ denotes the molecular velocity and internal energy is represented by the equivalent velocity $\boldsymbol{\xi}$ in length $D$. $\mathcal{J}_{\alpha}$ is the microscopic collision term considering both self-collisions and cross-collisions. The macroscopic conserved variable vector $\boldsymbol{W}_{\alpha}=\left(\rho_{\alpha}, \rho_{\alpha} \boldsymbol{U}_{\alpha}, \rho_{\alpha} E_{\alpha}\right)^T$ can be directly obtained from the distribution function $f_{\alpha}$ by calculating moments as follows, where $\rho$ is density, $\boldsymbol{U}$ is macroscopic velocity, $E$ is specific total energy.
\begin{equation}
\begin{aligned}
&\boldsymbol{W}_{\alpha} = \int_{\mathbb{R}^D}\int_{\mathbb{R}^3} \boldsymbol{\Psi}f_{\alpha} {\rm d}\boldsymbol{u} {\rm d}\boldsymbol{\xi},\\
&\boldsymbol{\Psi}=\left({ 1,\boldsymbol{u}, {1\over 2}\left|\boldsymbol{u}\right|^2+{1\over 2}\left|\boldsymbol{\xi}\right|^2 }\right).
\end{aligned}
\nonumber
\end{equation}
The heat flux vector $\boldsymbol{Q}_{\alpha}$ can also be obtained by the following moments,
\begin{equation}
{\boldsymbol{Q}_{\alpha}} = \int_{\mathbb{R}^D}\int_{\mathbb{R}^3} \left(\boldsymbol{u}-\boldsymbol{U}_{\alpha}\right)\left[{ {1\over 2}\left({ \left|\boldsymbol{u}-\boldsymbol{U}_{\alpha}\right|^2 + \left|\boldsymbol{\xi}\right|^2 }\right)f }\right] {\rm d}\boldsymbol{u} {\rm d}\boldsymbol{\xi},
\nonumber
\end{equation}
On the other hand, regarding the macroscopic governing equation, an operator splitting is implemented before the discretization as follows, for a better consistency of microscopic particles with macroscopic variables, which will be introduced in detail in Sec.~\ref{sec:source}.
\begin{equation}
\begin{aligned}
&\frac{\partial\boldsymbol{W}_{\alpha}}{\partial t}+\nabla\cdot\boldsymbol{F}_{\alpha}= \mathds{S}_{\alpha}\\
\Rightarrow&\left\{ \renewcommand{\arraystretch}{2}\begin{array}{l}
\dfrac{\partial\boldsymbol{W}_{\alpha}}{\partial t}+\nabla\cdot\boldsymbol{F}_{\alpha}= 0,\\
\dfrac{\partial\boldsymbol{W}_{\alpha}}{\partial t}= \mathds{S}_{\alpha},
\end{array}\right.\renewcommand{\arraystretch}{1}
\end{aligned}
\nonumber
\end{equation}
where source term vector $\mathds{S}_{\alpha}$ is for the exchanged momentum and energy between components. Their discretized governing equations are,
\begin{equation}\label{eq:conservation-macro}
\boldsymbol{W}^{\ast}_{\alpha,i} = \boldsymbol{W}^{n}_{\alpha,i} - \frac{1}{\Omega_i}\sum\limits_{j\in \mathcal{M}\left(i\right)}\boldsymbol{F}_{\alpha,j}S_j,
\end{equation}
and,
\begin{equation}\label{eq:conservation-macrosource}
\frac{\partial\boldsymbol{W}_{\alpha,i}}{\partial t}= \mathds{S}_{\alpha,i},
\end{equation}
where macroscopic flux $\boldsymbol{F}_{\alpha,j}$ is calculated by moments of $f_{\alpha,j}$ at the interface,
\begin{equation}\label{eq:macrof}
\boldsymbol{F}_{\alpha,j} = \int_{\mathbb{R}^D} \int_{\mathbb{R}^3}\int^{\Delta t}_0\boldsymbol{u}\cdot\boldsymbol{n}_jf_{\alpha,j}\boldsymbol{\Psi} {\rm d}t{\rm d}\boldsymbol{u}{\rm d}\boldsymbol{\xi}.
\end{equation}
An integral solution of Eq.~\eqref{eq:conservation-macrosource} will be derived in Sec.~\ref{sec:source}, whose initial condition is the result of $\boldsymbol{W}^{\ast}_{\alpha,i}$ in Eq.~\eqref{eq:conservation-macro}.

\subsection{Kinetic model}
For further calculating the source term and the fluxes of gas distribution function $f_{\alpha,j}$ at the interface in Eq.~\eqref{eq:conservation-micro} and Eq.~\eqref{eq:conservation-macro}, the kinetic model is introduced. In this study, the single collision operator in the type of Bhatnagar--Gross--Krook (BGK) model~\cite{bgk} is employed as,
\begin{equation}\label{eq:bgk}
\begin{aligned}
& \frac{\partial f_{\alpha}}{\partial t}+\boldsymbol{u}\cdot\frac{\partial f_{\alpha}}{\partial \boldsymbol{x}}=\frac{g_{\alpha}-f_{\alpha}}{\tau_0},\\
& g_{\alpha}=g^M_{\alpha}C^S_{\alpha},\\
& g^{M}_{\alpha}=\rho_{\alpha}\left( {\frac{1}{{2\pi R_{\alpha}\tilde{T}_{\alpha}}}} \right)^{\frac{3}{2}}\exp \left(  - \frac{\left|\boldsymbol{u}-\tilde{\boldsymbol{U}}_{\alpha}\right|^2}{2R_{\alpha}\tilde{T}_{\alpha}} \right),\\
& C^S_{\alpha}={1+(1-{\rm{Pr_{0}}})\frac{\left(\boldsymbol{u}-\tilde{\boldsymbol{U}}_{\alpha}\right)\cdot\boldsymbol{Q}_{\alpha}}{5\tilde{p}_{\alpha}R_{\alpha}\tilde{T}_{\alpha}}\left( {\frac{\left|\boldsymbol{u}-\tilde{\boldsymbol{U}}_{\alpha}\right|^2}{R_{\alpha}\tilde{T}_{\alpha}}-5} \right)},
\end{aligned}
\end{equation}
where $R_{\alpha}=k_B/m_{\alpha}$ is the specific gas constant of species $\alpha$, $k_B=1.380649\times10^{-23}J/K$ is Boltzmann constant, $m_{\alpha}$ is the molecular mass, $T$ is the macroscopic temperature, $p_{\alpha}=\rho_{\alpha} R_{\alpha}\tilde{T}_{\alpha}$ is the species pressure, $\tau$ is the relaxation time and $\mu$ is the viscosity coefficient, which will be introduced together with the heat conduction coefficient and ${\rm{Pr}}_0$ in detail from Eq.~\eqref{eq:coeffs}. The target near-equilibrium gas distribution function $g_{\alpha}$ contains two parts. The first part $g^{M}_{\alpha}$ is in the form of Maxwellian distribution. In this work, we consider only monatomic gas in three dimensional physical space so internal energy is not introduced. Different from variables at the current state, we use tilde to denote the target macroscopic variables $\tilde{\boldsymbol{U}}_{\alpha}$ and $\tilde{T}_{\alpha}$, which recover the diffusion effect. The model proposed by Groppi et al.~\cite{groppi1} is employed as follows,
\begin{equation}\label{eq:groppi}
\begin{aligned}
& \tilde{\boldsymbol{U}}_{\alpha}=\left(1-\vartheta\right)\boldsymbol{U}_{\alpha}+\vartheta\boldsymbol{U}_{0},\\
& \tilde{T}_{\alpha}=T_{0}-\frac{1}{3n_0k_B}\left(1-\vartheta\right)^2\sum\limits_{\beta=1}^2\rho_{\beta}\left|\boldsymbol{U}_{\beta}-\boldsymbol{U}_{0}\right|^2,
\end{aligned}
\end{equation}
where $\beta$ is also a label of species, to differentiate from $\alpha$, $n_{\alpha}=\rho_{\alpha}/m_{\alpha}$ is the number density. The subscript ``$0$'' denotes the macroscopic variable of gas mixture, such as,
\begin{equation}\label{eq:gasmixture}
\begin{aligned}
& n_{0}=\sum\limits_{\alpha=1}^2n_{\alpha},\quad \rho_{0}=\sum\limits_{\alpha=1}^2\rho_{\alpha}, \quad m_0=\frac{\rho_0}{n_0}=\sum\limits_{\alpha=1}^2\chi_{\alpha}m_{\alpha},\\
& \rho_{0}\boldsymbol{U}_{0}=\sum\limits_{\alpha=1}^2\rho_{\alpha}\boldsymbol{U}_{\alpha},\quad \rho_{0}E_{0}=\sum\limits_{\alpha=1}^2\rho_{\alpha}E_{\alpha},\\
& \frac{3}{2}n_0k_BT_{0}=\sum\limits_{\alpha=1}^2{\frac{3}{2}n_{\alpha}k_BT_{\alpha}}+\frac{1}{2}\sum\limits_{\alpha=1}^2{\rho_{\alpha}\left|\boldsymbol{U}_{\alpha}-\boldsymbol{U}_{0}\right|^2},
\end{aligned}
\end{equation}
where $\chi_{\alpha}=\frac{n_{\alpha}}{n_0}$ is the mole fraction and $T_{\alpha}$ is the species temperature calculated as,
\begin{equation}
T_{\alpha}=\frac{1}{3\rho_{\alpha}R_{\alpha}}\int_{\mathbb{R}^3} \left|\boldsymbol{u}-\boldsymbol{U}_{\alpha}\right|^2f_{\alpha}{\rm d}\boldsymbol{u}=\frac{2E_{\alpha}-\left|\boldsymbol{U}_{\alpha}\right|^2}{3R_{\alpha}}.
\nonumber
\end{equation}
The model on $\tilde{\boldsymbol{U}}_{\alpha}$ and $\tilde{T}_{\alpha}$ as Eq.~\eqref{eq:groppi} is derived by setting a constraint on the species drift velocity equalization. More details can be referred to Refs.\cite{groppi1,todorova1}, and the diffusion coefficient is recovered by,
\begin{equation}
\vartheta=\frac{5}{3}\frac{m_0}{A^{\ast}\sum\limits_{\alpha=1}^2m_{\alpha}},
\nonumber
\end{equation}
where the values of $A^{\ast}$ between different gases are given in Ref.\cite{weia} and a suggested value is $1.11$~\cite{todorova1}. The second part of $g_{\alpha}$ in Eq.~\eqref{eq:bgk}, $C^S_{\alpha}$, is for the correction of ${\rm{Pr}}_0$ in the approach of Shakhov model~\cite{shakhov}. In Ref.\cite{todorova1}, the Chapman-Enskog expansion in the gas mixture case is derived and the heat conduction coefficient in the continuum flow regime is recovered. In this kinetic model, each species shares the same $\tau$ and ${\rm{Pr}}$, which is calculated by,
\begin{equation}\label{eq:coeffs}
\tau_0=\frac{\mu_0}{n_0k_BT_0},\quad {\rm{Pr}}_0=\frac{C_{p,0}\mu_0}{k_0},
\end{equation}
where $C_{p,0}=\frac{5}{2}\frac{k_B}{m_0}$ is the specific heat capacity of gas mixture at constant pressure~\cite{todorova1} and $k$ is the heat conduction coefficient. The viscosity and heat conduction coefficients of gas mixture are calculated by Wilke's model and Wassiljewa's model respectively as follows~\cite{wilke,was1,was2,coeff},
\begin{equation}
\begin{aligned}
\mu_0=\sum\limits_{\alpha=1}^2{\frac{\chi_{\alpha}\mu_{\alpha}}{\sum\limits_{\beta=1}^2{\chi_{\beta}\frac{\left(1+\sqrt{\frac{\mu_{\alpha}}{\mu_{\beta}}}\sqrt[4]{\frac{m_{\beta}}{m_{\alpha}}}\right)^2}{\sqrt{8\left(1+\frac{m_{\alpha}}{m_{\beta}}\right)}}}}},\quad k_0=\sum\limits_{\alpha=1}^2{\frac{\chi_{\alpha}k_{\alpha}}{\sum\limits_{\beta=1}^2{\chi_{\beta}\frac{\left(1+\sqrt{\frac{k_{\alpha}}{k_{\beta}}}\sqrt[4]{\frac{m_{\beta}}{m_{\alpha}}}\right)^2}{\sqrt{8\left(1+\frac{m_{\alpha}}{m_{\beta}}\right)}}}}},
\nonumber
\end{aligned}
\end{equation}
where $\beta$ is also a label of species, to differentiate from $\alpha$. For each species the viscosity and heat conduction coefficients are calculated as,
\begin{equation}
\begin{aligned}
\mu_{\alpha}=\mu_{\alpha,{\rm{ref}}}\left(\frac{T_{\alpha}}{T_{\alpha,{\rm{ref}}}}\right)^{\omega},\quad k_{\alpha}=\frac{C_{p,\alpha}\mu_{\alpha}}{{\rm{Pr}_{\alpha}}},
\nonumber
\end{aligned}
\end{equation}
where subscript ``$\rm{ref}$'' denotes the reference value, $\omega$ is the viscosity index, $C_{p,\alpha}=\frac{5}{2}\frac{k_B}{m_{\alpha}}$ and for the monatomic gas in this work ${\rm{Pr}_{\alpha}}$ is set to be $2/3$.

\subsection{Integral solution along the characteristic line}
Setting the coordinate origin at the center of an interface, the integral solution along the characteristic line can be derived as,
\begin{equation}\label{eq:integral}
f_{\alpha}\left( {\boldsymbol{0},t} \right) = \frac{1}{\tau_0}\int^t_0 g_{\alpha}\left[{ -\boldsymbol{u}\left({ t-\hat{t} }\right),\hat{t} }\right]e^{\frac{\hat{t}-t}{\tau_0}} d\hat{t} + e^{-t/\tau_0}f_{\alpha}\left( {-\boldsymbol{u}t,0} \right),
\end{equation}
where the free transport is considered together with the collision. The first term denotes the accumulation of near-equilibrium distribution. The second term denotes the transport of initial non-equilibrium state. Expanding $g_{\alpha}$ and $f_{\alpha}$ as,
\begin{equation}\label{eq:taylor}
\begin{aligned}
g_{\alpha}\left({ \boldsymbol{x},t }\right)&=g_{\alpha}\left({ \boldsymbol{0},0 }\right)+\frac{\partial g_{\alpha}}{\partial \boldsymbol{x}}\cdot\boldsymbol{x}+\frac{\partial g_{\alpha}}{\partial t}t,\\
f_{\alpha}\left({ \boldsymbol{x},0 }\right)&=f_{\alpha}\left({ \boldsymbol{0},0 }\right)+\frac{\partial f_{\alpha}}{\partial \boldsymbol{x}}\cdot\boldsymbol{x},
\nonumber
\end{aligned}
\end{equation}
it can be derived from Eq.~\eqref{eq:integral} that,
\begin{equation}\label{eq:integrala1}
f_{\alpha}\left( {\boldsymbol{0},t} \right) = \varepsilon_ag_{\alpha}\left({ \boldsymbol{0},0 }\right) + \varepsilon_b\frac{\partial g_{\alpha}}{\partial \boldsymbol{x}}\cdot\boldsymbol{u}+\varepsilon_c\frac{\partial g_{\alpha}}{\partial t}+\varepsilon_df_{\alpha}\left({ \boldsymbol{0},0 }\right) + \varepsilon_e\frac{\partial f_{\alpha}}{\partial \boldsymbol{x}}\cdot\boldsymbol{u},
\end{equation}
where,
\begin{equation}\label{eq:integrala2}
\begin{aligned}
\varepsilon_a &= 1-e^{-t/\tau_0},\\
\varepsilon_b &= te^{-t/\tau_0}-\tau_0\left(1-e^{-t/\tau_0}\right),\\
\varepsilon_c &= t-\tau_0\left(1-e^{-t/\tau_0}\right),\\
\varepsilon_d &= e^{-t/\tau_0},\\
\varepsilon_e &= -te^{-t/\tau_0}.
\nonumber
\end{aligned}
\end{equation}
Then the integrated macroscopic flux can be derived by substituting Eq.~\eqref{eq:integrala1} into Eq.~\eqref{eq:macrof},
\begin{equation}\label{eq:integralb1}
\boldsymbol{F}_{\alpha} = \boldsymbol{F}_{\alpha}^{eq}+\boldsymbol{F}_{\alpha}^{fr},
\end{equation}
where,
\begin{equation}\label{eq:integralb2}
\begin{aligned}
\boldsymbol{F}_{\alpha}^{eq} =& \int_{\mathbb{R}^3}{\boldsymbol{\Psi}\left(\delta_ag_{\alpha}\left({ \boldsymbol{0},0 }\right)+\delta_b\frac{\partial g_{\alpha}}{\partial \boldsymbol{x}}\cdot\boldsymbol{u}+ \delta_c\frac{\partial g_{\alpha}}{\partial t} \right)\left(\boldsymbol{u}\cdot\boldsymbol{n}\right) {\rm d}\boldsymbol{u}},\\
\boldsymbol{F}_{\alpha}^{fr} =& \int_{\mathbb{R}^3}{ \boldsymbol{\Psi}\left(\delta_df_{\alpha}\left({ \boldsymbol{0},0 }\right)+\delta_e\frac{\partial f_{\alpha}}{\partial \boldsymbol{x}}\cdot\boldsymbol{u} \right)} \left(\boldsymbol{u}\cdot\boldsymbol{n}\right) {\rm d}\boldsymbol{u},
\end{aligned}
\end{equation}
and,
\begin{equation}\label{eq:integralb3}
\begin{aligned}
\delta_a &= \Delta t-\tau_0\left( {1-e^{-\Delta t/\tau_0}} \right),\\
\delta_b &= 2\tau_0^2\left( {1-e^{-\Delta t/\tau_0}} \right)-\tau_0\Delta t-\tau_0\Delta te^{-\Delta t/\tau_0},\\
\delta_c &= \Delta t^2/2-\tau_0\Delta t+\tau_0^2\left( {1-e^{-\Delta t/\tau_0}} \right),\\
\delta_d &= \tau_0\left( {1-e^{-\Delta t/\tau_0}} \right),\\
\delta_e &= \tau_0\Delta te^{-\Delta t/\tau_0}-\tau_0^2\left( {1-e^{-\Delta t/\tau_0}} \right).
\end{aligned}
\end{equation}
The $\boldsymbol{F}^{eq}_{\alpha}$ term in Eq.~\eqref{eq:integralb1} plays a major role in the continuum flow regime, where the gas evolution is constructed by deterministic hydrodynamic waves. The $\boldsymbol{F}^{fr}_{\alpha}$ term gradually dominates the gas evolution when the flow gets rarefied. Particles are employed for this part of simulation.

\subsection{Wave evolution}\label{sec:macro}
For the first term in Eq.~\eqref{eq:integralb1}, $\boldsymbol{F}^{eq}_{\alpha}$ is calculated as the GKS~\cite{gks2001}, and the first step is to calculate $g_{\alpha}\left({ \boldsymbol{0},0 }\right)$ and its derivatives $\frac{\partial g_{\alpha}}{\partial \boldsymbol{x}}$, $\frac{\partial g_{\alpha}}{\partial t}$. It is proved in Ref.\cite{wp-hyper1} that $\frac{\partial g_{\alpha}}{\partial \boldsymbol{x}}$ and $\frac{\partial g_{\alpha}}{\partial t}$ can be simplified by $\frac{\partial g^M_{\alpha}}{\partial \boldsymbol{x}}$ and $\frac{\partial g^M_{\alpha}}{\partial t}$. And for simplicity, instead of $g_{\alpha}\left({ \boldsymbol{0},0 }\right)$, $g^M_{\alpha}\left({ \boldsymbol{0},0 }\right)$ along with the ${\rm{Pr}}$ correction method in the GKS~\cite{may} is used. Calculated by the sufficient collision as follows, $g^M_{\alpha}\left({ \boldsymbol{0},0 }\right)$ denotes the equilibrium state at the interface,
\begin{equation}\label{eq:mac1}
\int_{\mathbb{R}^3} \boldsymbol{\Psi} g^M_{\alpha}\left({ \boldsymbol{0},0 }\right) {\rm d}\boldsymbol{u} = \boldsymbol{W}_{\alpha}\left({ \boldsymbol{0},0 }\right) = \int_{\boldsymbol{u}\cdot\boldsymbol{n}>0} \boldsymbol{\Psi} g^M_{\alpha,L}\left({ \boldsymbol{0},0 }\right) {\rm d}\boldsymbol{u} + \int_{\boldsymbol{u}\cdot\boldsymbol{n}<0} \boldsymbol{\Psi} g^M_{\alpha,R}\left({ \boldsymbol{0},0 }\right) {\rm d}\boldsymbol{u},
\nonumber
\end{equation}
where $g^M_{\alpha,L}\left({ \boldsymbol{0},0 }\right)$, $g^M_{\alpha,R}\left({ \boldsymbol{0},0 }\right)$, and $g^M_{\alpha}\left({ \boldsymbol{0},0 }\right)$ are calculated from macroscopic variables $\boldsymbol{W}_{\alpha,L}\left({ \boldsymbol{0},0 }\right)$, $\boldsymbol{W}_{\alpha,R}\left({ \boldsymbol{0},0 }\right)$ and $\boldsymbol{W}_{\alpha}\left({ \boldsymbol{0},0 }\right)$ through Eq.~\eqref{eq:bgk}. $\boldsymbol{W}_{\alpha,L}\left({ \boldsymbol{0},0 }\right)$ and $\boldsymbol{W}_{\alpha,R}\left({ \boldsymbol{0},0 }\right)$ are left-hand and right-hand limits at the interface by reconstruction. Meanwhile, spatial derivation $\frac{\partial g^M_{\alpha}}{\partial \boldsymbol{x}}$ and temporal derivation $\frac{\partial g^M_{\alpha}}{\partial t}$ are derived as:
\begin{equation}\label{eq:mac2}
\begin{aligned}
\frac{\partial g^M_{\alpha}}{\partial \boldsymbol{x}} &= \boldsymbol{a}_{\alpha}g^M_{\alpha},\\
\frac{\partial g^M_{\alpha}}{\partial t} &= A_{\alpha}g^M_{\alpha},
\nonumber
\end{aligned}
\end{equation}
where,
\begin{equation}
\begin{aligned}
\boldsymbol{a}_{\alpha}&=\frac{1}{g^M_{\alpha}}\frac{\partial g^M_{\alpha}}{\partial \boldsymbol{x}}=\frac{\partial \left[ {\rm{ln}}\left(g^M_{\alpha}\right) \right]}{\partial \boldsymbol{x}},\\
A_{\alpha}&=\frac{1}{g^M_{\alpha}}\frac{\partial g^M_{\alpha}}{\partial t}=\frac{\partial \left[ {\rm{ln}}\left(g^M_{\alpha}\right) \right]}{\partial t}.
\nonumber
\end{aligned}
\end{equation}
For the spatial derivative, $\boldsymbol{a}_{\alpha}$ is written in the form of:
\begin{equation}
\boldsymbol{a}_{\alpha}=\boldsymbol{a}_{\alpha,0}+\boldsymbol{a}_{\alpha,1}u_1+\boldsymbol{a}_{\alpha,2}u_2+\boldsymbol{a}_{\alpha,3}u_3+\boldsymbol{a}_{\alpha,4}\frac{|\boldsymbol{u}|^2}{2},
\nonumber
\end{equation}
and,
\begin{equation}
\begin{aligned}
\boldsymbol{a}_{\alpha,0} &= \frac{ 2\frac{\partial\rho_{\alpha}}{\partial\boldsymbol{x}}\tilde{\Lambda}_{\alpha} + \rho_{\alpha}\left[{ -4\tilde{\Lambda}_{\alpha}^2\tilde{\boldsymbol{U}}_{\alpha}\cdot\frac{\partial\tilde{\boldsymbol{U}}_{\alpha}}{\partial\boldsymbol{x}} + \left({3-2|\tilde{\boldsymbol{U}}_{\alpha}|^2\tilde{\Lambda}_{\alpha}}\right)\frac{\partial\tilde{\Lambda}_{\alpha}}{\partial\boldsymbol{x}} }\right] }{2\rho_{\alpha}\tilde{\Lambda}_{\alpha}},\\
\boldsymbol{a}_{\alpha,1} &= 2\left( {\frac{\partial \tilde{U}_{\alpha,1}}{\partial\boldsymbol{x}}\tilde{\Lambda}_{\alpha} + \tilde{U}_{\alpha,1}\frac{\partial\tilde{\Lambda}_{\alpha}}{\partial\boldsymbol{x}}} \right),\\
\boldsymbol{a}_{\alpha,2} &= 2\left( {\frac{\partial \tilde{U}_{\alpha,2}}{\partial\boldsymbol{x}}\tilde{\Lambda}_{\alpha} + \tilde{U}_{\alpha,2}\frac{\partial\tilde{\Lambda}_{\alpha}}{\partial\boldsymbol{x}}} \right),\\
\boldsymbol{a}_{\alpha,3} &= 2\left( {\frac{\partial \tilde{U}_{\alpha,3}}{\partial\boldsymbol{x}}\tilde{\Lambda}_{\alpha} + \tilde{U}_{\alpha,3}\frac{\partial\tilde{\Lambda}_{\alpha}}{\partial\boldsymbol{x}}} \right),\\
\boldsymbol{a}_{\alpha,4} &= -2\frac{\partial\tilde{\Lambda}_{\alpha}}{\partial\boldsymbol{x}},
\nonumber
\end{aligned}
\end{equation}
where $\tilde{\Lambda}_{\alpha}=\left({2R_{\alpha}\tilde{T}_{\alpha}}\right)^{-1}$. When using the ${\rm{Pr}}$ correction method, $\tilde{\Lambda}_{\alpha}$ derivation is modified as follows,
\begin{equation}\label{eq:mac3}
\begin{aligned}
&\frac{\partial\tilde{\Lambda}_{\alpha}}{\partial\boldsymbol{x}}\Rightarrow\frac{\partial\tilde{\Lambda}_{\alpha}}{\partial\boldsymbol{x}}/{\rm{Pr}}_0.
\nonumber
\end{aligned}
\end{equation}

On the other hand, the time derivative is calculated obeying the conservation law on the right side of BGK-type model:
\begin{equation}
\int_{\mathbb{R}^3} g^M_{\alpha}A_{\alpha} \boldsymbol{\Psi} {\rm d}\boldsymbol{u}= -\int_{\mathbb{R}^3} g^M_{\alpha} \boldsymbol{a}_{\alpha}\cdot\boldsymbol{u} \boldsymbol{\Psi} {\rm d}\boldsymbol{u} = \boldsymbol{b}_{\alpha}.
\nonumber
\end{equation}
Spreading $A_{\alpha}$ into $A_{\alpha}=A_{\alpha,0}+A_{\alpha,1}u_1+A_{\alpha,2}u_2+A_{\alpha,3}u_3+A_{\alpha,4}\frac{|\boldsymbol{u}|^2}{2}$, it is rewritten as:
\begin{equation}
A_{\alpha,q} M_{\alpha,pq} = b_{\alpha,p},
\nonumber
\end{equation}
where subscripts ``$p$'' and ``$q$'' here denote components, and Einstein summation convention is used. For example, as to the two-dimensional case,
\begin{equation}
\boldsymbol{M}_{\alpha}=
\left(
\begin{array}{cccc}
1                     &\tilde{U}_{\alpha,1}                                        &\tilde{U}_{\alpha,2}                                       &B_{\alpha,1} \\
\tilde{U}_{\alpha,1}  &\tilde{U}_{\alpha,1}^2+\frac{1}{2\tilde{\Lambda}_{\alpha}}  &\tilde{U}_{\alpha,1}\tilde{U}_{\alpha,2}                   &B_{\alpha,2} \\
\tilde{U}_{\alpha,2}  &\tilde{U}_{\alpha,1}\tilde{U}_{\alpha,2}                    &\tilde{U}_{\alpha,2}^2+\frac{1}{2\tilde{\Lambda}_{\alpha}} &B_{\alpha,3} \\
B_{\alpha,1}          &B_{\alpha,2}                                                &B_{\alpha,3}                                               &B_{\alpha,4}
\end{array} \right),
\nonumber
\end{equation}
where
\begin{equation}
\begin{aligned}
&B_{\alpha,1} = \frac{1}{2}\left( {\tilde{U}_{\alpha,1}^2+\tilde{U}_{\alpha,2}^2+\frac{3}{2\tilde{\Lambda}_{\alpha}}} \right),\\
&B_{\alpha,2} = \frac{1}{2}\left( {\tilde{U}_{\alpha,1}^3+\tilde{U}_{\alpha,1}\tilde{U}_{\alpha,2}^2+\frac{5}{2\tilde{\Lambda}_{\alpha}}\tilde{U}_{\alpha,1}} \right),\\
&B_{\alpha,3} = \frac{1}{2}\left( {\tilde{U}_{\alpha,2}^3+\tilde{U}_{\alpha,1}^2\tilde{U}_{\alpha,2}+\frac{5}{2\tilde{\Lambda}_{\alpha}}\tilde{U}_{\alpha,2}} \right),\\
&B_{\alpha,4} = \frac{1}{4}\left[ {\left({\tilde{U}_{\alpha,1}^2+\tilde{U}_{\alpha,2}^2}\right)^2 + \frac{5}{\tilde{\Lambda}_{\alpha}}\left({\tilde{U}_{\alpha,1}^2+\tilde{U}_{\alpha,2}^2}\right) + \frac{15}{4\tilde{\Lambda}_{\alpha}^2}} \right].
\nonumber
\end{aligned}
\end{equation}
And the result is:
\begin{equation}
\begin{aligned}
&A_{\alpha,4} = \frac{8\tilde{\Lambda}_{\alpha}^2}{3}\left[ {b_{\alpha,4}-\tilde{U}_{\alpha,1}b_{\alpha,2}-\tilde{U}_{\alpha,2}b_{\alpha,3}-(B_{\alpha,1}+\tilde{U}_{\alpha,1}^2+\tilde{U}_{\alpha,2}^2)b_{\alpha,1}} \right],\\
&A_{\alpha,3} = 2\tilde{\Lambda}_{\alpha}(b_{\alpha,3}-\tilde{U}_{\alpha,2}b_{\alpha,1})-\tilde{U}_{\alpha,2}A_{\alpha,4},\\
&A_{\alpha,2} = 2\tilde{\Lambda}_{\alpha}(b_{\alpha,2}-\tilde{U}_{\alpha,1}b_{\alpha,1})-\tilde{U}_{\alpha,1}A_{\alpha,4},\\
&A_{\alpha,1} = b_{\alpha,1}-\tilde{U}_{\alpha,1}A_{\alpha,2}-\tilde{U}_{\alpha,2}A_{\alpha,3}-B_{\alpha,1}A_{\alpha,4}.
\nonumber
\end{aligned}
\end{equation}

More details of calculating $g_{\alpha}\left({ \boldsymbol{0},0 }\right)$, $\frac{\partial g_{\alpha}}{\partial \boldsymbol{x}}$ and $\frac{\partial g_{\alpha}}{\partial t}$ can be referred to Refs.\cite{gks2001,gks1998}. By substituting them in to Eq.~\eqref{eq:integralb2}, the flux of wave evolution $\boldsymbol{F}^{eq}_{\alpha}$ can be obtained. To improve the accuracy and robustness in the flow field regions of drastic scale variation, the scale-related coefficient $\delta$ is separated into left and right sides of the interface ``$j$'', as follows. By this formula, $\delta_{a,L/R}$, $\delta_{b,L/R}$ and $\delta_{c,L/R}$ used by the wave flux can be calculated within the cell, instead of at the interface. When sampling particles, the scale-related coefficient $e^{-\Delta t/\tau}$ is calculated at the same location. This treatment enables the wave flux more consistent with particles, to prevent situations in which the scale changes significantly, the $\delta$ values within the cell and at the interface have significant discrepancies. Further illustrations can be referred to Ref.\cite{awp2}.
\begin{equation}\label{eq:feq}
\begin{aligned}
\boldsymbol{F}^{eq}_{\alpha,j} =& \int_{\boldsymbol{u}\cdot\boldsymbol{n}_j>0}{\boldsymbol{\Psi}\left(\delta_{a,L}g_{\alpha}\left({ \boldsymbol{0},0 }\right)+\delta_{b,L}\frac{\partial g^M_{\alpha}}{\partial \boldsymbol{x}}\cdot\boldsymbol{u}+ \delta_{c,L}\frac{\partial g^M_{\alpha}}{\partial t} \right)\left(\boldsymbol{u}\cdot\boldsymbol{n}_j\right) {\rm d}\boldsymbol{u}}\\
+& \int_{\boldsymbol{u}\cdot\boldsymbol{n}_j<0}{\boldsymbol{\Psi}\left(\delta_{a,R}g_{\alpha}\left({ \boldsymbol{0},0 }\right)+\delta_{b,R}\frac{\partial g^M_{\alpha}}{\partial \boldsymbol{x}}\cdot\boldsymbol{u}+ \delta_{c,R}\frac{\partial g^M_{\alpha}}{\partial t} \right)\left(\boldsymbol{u}\cdot\boldsymbol{n}_j\right) {\rm d}\boldsymbol{u}},
\end{aligned}
\end{equation}
where details of integrations are provided as follows,
\begin{equation}
\begin{aligned}
&\int_{\mathbb{R}} {\left( {\frac{1}{{2\pi RT}}} \right)^{\frac{{1}}{2}}}\exp \left( { - \frac{|u-U|^2}{{2RT}}} \right) du = 1,\\
&\int_{\mathbb{R}} u {\left( {\frac{1}{{2\pi RT}}} \right)^{\frac{{1}}{2}}}\exp \left( { - \frac{|u-U|^2}{{2RT}}} \right) du = U,\\
&\int_{u>0} {\left( {\frac{1}{{2\pi RT}}} \right)^{\frac{{1}}{2}}}\exp \left( { - \frac{|u-U|^2}{{2RT}}} \right) du = \frac{1}{2}\rm{erfc}\left( {-\sqrt{\Lambda}U} \right),\\
&\int_{u>0} u {\left( {\frac{1}{{2\pi RT}}} \right)^{\frac{{1}}{2}}}\exp \left( { - \frac{|u-U|^2}{{2RT}}} \right) du = \frac{U}{2}\rm{erfc}\left( {-\sqrt{\Lambda}U} \right) + \frac{1}{2}\frac{e^{-\Lambda U^2}}{\sqrt{\pi \Lambda}},\\
&\int_{u<0} {\left( {\frac{1}{{2\pi RT}}} \right)^{\frac{{1}}{2}}}\exp \left( { - \frac{|u-U|^2}{{2RT}}} \right) du = \frac{1}{2}\rm{erfc}\left( {\sqrt{\Lambda}U} \right),\\
&\int_{u<0} u {\left( {\frac{1}{{2\pi RT}}} \right)^{\frac{{1}}{2}}}\exp \left( { - \frac{|u-U|^2}{{2RT}}} \right) du = \frac{U}{2}\rm{erfc}\left( {\sqrt{\Lambda}U} \right) - \frac{1}{2}\frac{e^{-\Lambda U^2}}{\sqrt{\pi \Lambda}},
\nonumber
\end{aligned}
\end{equation}
and,
\begin{equation}
\begin{aligned}
&\int_{\mathbb{R}/u>0/u<0} u^{n+2} {\left( {\frac{1}{{2\pi RT}}} \right)^{\frac{{1}}{2}}}\exp \left( { - \frac{|u-U|^2}{{2RT}}} \right) du \\
= &U\int_{\mathbb{R}/u>0/u<0} u^{n+1} {\left( {\frac{1}{{2\pi RT}}} \right)^{\frac{{1}}{2}}}\exp \left( { - \frac{|u-U|^2}{{2RT}}} \right) du \\ + &\frac{n+1}{2\Lambda}\int_{\mathbb{R}/u>0/u<0} u^{n} {\left( {\frac{1}{{2\pi RT}}} \right)^{\frac{{1}}{2}}}\exp \left( { - \frac{|u-U|^2}{{2RT}}} \right) du,
\nonumber
\end{aligned}
\end{equation}
where $\rm{erfc}()=1-\rm{erf}()$ is the complementary error function.

\subsection{Particle evolution}\label{sec:micro}
In the UGKWP method, numerical particles are employed to describe the free-transport part of distribution function, whose parameters are: species label $\alpha$, particle mass $m$, location $\boldsymbol{x}$ and velocity $\boldsymbol{u}$. As Eq.~\eqref{eq:integral}, the distribution function after $\Delta t$ is a combination of the initial state $f_{\alpha}$ and target near-equilibrium state $g_{\alpha}$. Firstly, for a particle from the initial state, the probability for keeping free-transport is $e^{-\frac{\Delta t}{\tau_0}}$, otherwise it collides with other particles and is relaxed into $g_{\alpha}$. As a result, the cumulative distribution of a free-transport particle is $e^{-\frac{\Delta t}{\tau_0}}$. By taking a uniform random number $\epsilon\in\left(0,1\right)$, the free-transport time $t_{f,k}$ of a particle labeled by ``$k$'' can be calculated by,
\begin{equation}\label{eq:mic1}
t_{f,k} = {\rm{min}}\left(-\tau_{0,k}{\rm{ln}}\left(\epsilon\right),\Delta t\right).
\end{equation}
Then the location of the particle can be renewed by,
\begin{equation}\label{eq:mic2}
\boldsymbol{x}_k \Rightarrow \boldsymbol{x}_k + \boldsymbol{u}_kt_{f,k},
\end{equation}
and the contribution of all particles to the macroscopic conserved variables is,
\begin{equation}\label{eq:mic3}
\boldsymbol{W}_{\alpha,i}^{fr,p} = \boldsymbol{W}_{\alpha,i}^{p,+}-\boldsymbol{W}_{\alpha,i}^{p,-},
\end{equation}
and,
\begin{equation}\label{eq:mic4}
\boldsymbol{W}_{\alpha,i}^{p,+/-} = \frac{1}{\Omega_i}\sum\limits_{k\in \mathcal{N}^{+/-}_{\alpha}\left(i\right)}m_{p,k}\boldsymbol{\Psi}_k,
\nonumber
\end{equation}
where $\mathcal{N}_{\alpha}\left(i\right)$ is a collection of particles of species $\alpha$ within cell ``$i$''. Superscripts ``$-$'' and ``$+$'' denote the time before and after the free transport, respectively. Then, after calculating $\boldsymbol{W}_{\alpha,i}^{fr,p}$, if $t_{f,k}<\Delta t$, the particle is deleted, relaxing into $g_{\alpha}$, called collisional particles. Otherwise, if $t_{f,k}=\Delta t$, the particle will be kept, called collisionless particles.

Furthermore, when calculating $t_{f,k}$, there is more freedom to improve the microscopic model in the premise of conservation. In this study, the free-transport time is reduced for high-speed particles as follows, which is an extension to the gas mixture from Ref.\cite{xu-tau} and can recover the formula of single-species case in Ref.\cite{xu-tau}. This primarily breaks through the limitation that particles with different speeds have the same collision frequency. Particularly when the thermal speed of the particles is high, their collision frequency should significantly increase, which is especially reflected in the calculation of their free transport time. As shown in Ref.\cite{xu-tau}, it improves high-Ma case results such as the temperature profile in the pre-shock location and the heat flux at the stagnation point.
\begin{equation}\label{eq:taustar}
t_{f,k,\alpha}={\rm{min}}\left(-\tau^{\ast}_{\alpha,k}{\rm{ln}}\left(\epsilon\right),\Delta t\right),
\end{equation}
and,
\begin{equation}
\begin{aligned}
\tau^{\ast}_{\alpha,k}=\left\{ \begin{array}{ll}
\tau_{0,k}, & \left|\boldsymbol{u}_{k}-\boldsymbol{U}_{\alpha}\right|\leq b\sqrt{R_{\alpha}T_{\alpha}},\\
\frac{1}{1+\sum\limits_{\beta=1}^2{c_{\alpha\beta}}}\tau_{0,k}, & \left|\boldsymbol{u}_{k}-\boldsymbol{U}_{\alpha}\right|> b\sqrt{R_{\alpha}T_{\alpha}},
\end{array}\right.
\end{aligned}
\nonumber
\end{equation}
where,
\begin{equation}
\begin{aligned}
c_{\alpha\beta}=\left\{ \begin{array}{ll}
0, & \left|\boldsymbol{u}_{k}-\boldsymbol{U}_{\beta}\right|\leq b\sqrt{R_{\beta}T_{\beta}},\\
a\chi_{\beta}\frac{\left|\boldsymbol{u}_{k}-\boldsymbol{U}_{\beta}\right|}{\sqrt{R_{\beta}T_{\beta}}}, & \left|\boldsymbol{u}_{k}-\boldsymbol{U}_{\beta}\right|> b\sqrt{R_{\beta}T_{\beta}},
\end{array}\right.
\end{aligned}
\nonumber
\end{equation}
Here $a$ and $b$ are set to be $0.1$ and $5$ respectively as validated in Ref.\cite{xu-tau}. To illustrate the multispecies case, we set $m_1=m_2$ as a baseline for comparison. On one hand, when species ``$1$'' has smaller $m_{1}$, its distribution function is broader as $\sqrt{R_1T_1}=\sqrt{k_BT_1/m_1}$, so that the high-speed particle satisfying $\left|\boldsymbol{u}_{k}-\boldsymbol{U}_{\alpha}\right|> b\sqrt{R_{\alpha}T_{\alpha}}$ has a larger $\left|\boldsymbol{u}_{k}-\boldsymbol{U}_{\beta}\right|$. As a result, a larger $c_{12}$ gives a smaller $\tau^{\ast}_{1,k}$. It is compatible with the physics that a faster particle has a larger collision frequency, and in this case the speed is risen up by the mass ratio between the two species. On the other hand, when species ``$1$'' is heavier, its distribution is sharper. Though a high-speed particle satisfies $\left|\boldsymbol{u}_{k}-\boldsymbol{U}_{\alpha}\right|> b\sqrt{R_{\alpha}T_{\alpha}}$, it may be still within $b\sqrt{R_{\beta}T_{\beta}}$ because $\left|\boldsymbol{u}_{k}-\boldsymbol{U}_{\beta}\right|$ is not large enough. As a result, the contribution from $c_{12}$ decreases so that $\tau^{\ast}_{1,k}$ is not as small as the $m_1=m_2$ case.

Secondly, according to Eq.~\eqref{eq:integral}, particles should be sampled from $e^{-\frac{\Delta t}{\tau_{0,i}}}$ proportion of the hydrodynamic wave part, $\boldsymbol{W}_{\alpha,i}^{h}=\boldsymbol{W}_{\alpha,i}-\boldsymbol{W}_{\alpha,i}^{p}$ with $t_{f,k}=\Delta t$.  The density of sampled collisionless particles is,
\begin{equation}\label{eq:mic5}
\rho_{\alpha,i}^{hp} = e^{-\frac{\Delta t}{\tau_{0,i}}}\rho_{\alpha,i}^{h}.
\end{equation}
Unless $\rho_{\alpha,i}^{h}=0$ when no particle will be sampled, the number of sampled particles is calculated by,
\begin{equation}\label{eq:mic6}
N^{hp}_{\alpha,i} = \lceil{\frac{\rho^{hp}_{\alpha,i}}{\rho_{\alpha,i}}\chi_{\alpha,i}N^{hp,{\rm{ref}}}}\rceil,
\end{equation}
where $N^{hp,{\rm{ref}}}$ is a given reference number for the gas mixture. The corresponding mass of a particle in species $\alpha$ is,
\begin{equation}\label{eq:mic7}
m_{p,k} = \frac{\rho_{\alpha,i}^{hp}\Omega_i}{N^{hp}_{\alpha,i}}.
\nonumber
\end{equation}
The particle position is set according to a uniform distribution probability within cell ``$i$''. The velocity is got by acceptance-rejection sampling, as follows. Firstly, the velocity according to the Maxwellian distribution function is calculated by,
\begin{equation}\label{eq:maxw}
\begin{aligned}
u_{k,1} &= \tilde{U}_{\alpha,i,1} + \sqrt{2R_{\alpha}\tilde{T}_{\alpha,i}}{\rm{cos}}(2\pi \epsilon_{a1})\sqrt{-{\rm{ln}}(\epsilon_{a2})},\\
u_{k,2} &= \tilde{U}_{\alpha,i,2} + \sqrt{2R_{\alpha}\tilde{T}_{\alpha,i}}{\rm{cos}}(2\pi \epsilon_{b1})\sqrt{-{\rm{ln}}(\epsilon_{b2})},\\
u_{k,3} &= \tilde{U}_{\alpha,i,3} + \sqrt{2R_{\alpha}\tilde{T}_{\alpha,i}}{\rm{cos}}(2\pi \epsilon_{c1})\sqrt{-{\rm{ln}}(\epsilon_{c2})},
\nonumber
\end{aligned}
\end{equation}
where $\epsilon$ is a random number distributing uniformly in $\left(0,1\right)$. Then according to $C^S_{\alpha}$ in Eq.~\eqref{eq:bgk}, the criterion of the acceptance-rejection method is suggested to be:
\begin{equation}\label{eq:ajshak}
\frac{1+(1-{\rm{Pr}}_{0,i})\frac{\left(\boldsymbol{u}_k-\tilde{\boldsymbol{U}}_{\alpha,i}\right)\cdot\boldsymbol{Q}_{\alpha,i}}{5\tilde{p}_{\alpha,i}R_{\alpha}\tilde{T}_{\alpha,i}}\left( {\frac{|\boldsymbol{u}_k-\tilde{\boldsymbol{U}}_{\alpha,i}|^2}{R_{\alpha}\tilde{T}_{\alpha,i}}-5} \right)}{1+(1-{\rm{Pr}}_{0,i})\frac{20|\boldsymbol{Q}_{\alpha,i}|}{\tilde{p}_{\alpha,i}\sqrt{R_{\alpha}\tilde{T}_{\alpha,i}}}},
\end{equation}

Additionally, since the analytical macroscopic flux of newly sampled particles corresponds to the second-order discrete velocity method (DVM), the free transport fluxes contributed from the collisional particles of $\left(\boldsymbol{W}_{\alpha}^h-\boldsymbol{W}_{\alpha}^{hp}\right)$ can be calculated as,
\begin{equation}
\begin{aligned}
&\boldsymbol{F}_{\alpha}^{fr,wave}=\boldsymbol{F}^{fr}_{\rm{UGKS}}\left(\boldsymbol{W}_{\alpha}^h\right)-\boldsymbol{F}^{fr}_{\rm{DVM}}\left(\boldsymbol{W}_{\alpha}^{hp}\right)\\
=&\int_{\mathbb{R}^3}\boldsymbol{\Psi}\left(\delta_dg_{\alpha}^{h}\left({ \boldsymbol{0},0 }\right)+\delta_e\frac{\partial g_{\alpha}^h}{\partial \boldsymbol{x}}\cdot\boldsymbol{u}\right) \left(\boldsymbol{u}\cdot\boldsymbol{n}\right) {\rm d}\boldsymbol{u}\\
-&e^{-\frac{\Delta t}{\tau_0}}\int_{0}^{\Delta t}\int_{\mathbb{R}^3}\boldsymbol{\Psi}\left(g_{\alpha}^h\left({ \boldsymbol{0},0 }\right)+t\frac{\partial g_{\alpha}^h}{\partial \boldsymbol{x}}\cdot\boldsymbol{u}\right) \left(\boldsymbol{u}\cdot\boldsymbol{n}\right) {\rm d}\boldsymbol{u} {\rm d}t\\
=&\int_{\mathbb{R}^3}\boldsymbol{\Psi}\left[\left(\delta_d-\Delta te^{-\frac{\Delta t}{\tau_0}}\right)g_{\alpha}^h\left({ \boldsymbol{0},0 }\right)+\left(\delta_e+\frac{\Delta t^2}{2}e^{-\frac{\Delta t}{\tau_0}}\right)\frac{\partial g_{\alpha}^h}{\partial \boldsymbol{x}}\cdot\boldsymbol{u}\right] \left(\boldsymbol{u}\cdot\boldsymbol{n}\right) {\rm d}\boldsymbol{u}.
\nonumber
\end{aligned}
\end{equation}
The same as in Eq.~\eqref{eq:feq}, the scale-related coefficient $\delta$ is separated into left-side and right-side cells of the interface ``$j$'' for a better consistency with sampled particles, as follows,
\begin{equation}\label{eq:ffrwave}
\begin{aligned}
\boldsymbol{F}^{fr,wave}_{\alpha,j}=&\int_{\boldsymbol{u}\cdot\boldsymbol{n}_j>0}\boldsymbol{\Psi}\left[\left(\delta_{d,L}-\Delta te^{-\frac{\Delta t}{\tau_{0,L}}}\right)g_{\alpha}^h\left({ \boldsymbol{0},0 }\right)+\left(\delta_{e,L}+\frac{\Delta t^2}{2}e^{-\frac{\Delta t}{\tau_{0,L}}}\right)\frac{\partial g_{\alpha}^h}{\partial \boldsymbol{x}}\cdot\boldsymbol{u}\right] \left(\boldsymbol{u}\cdot\boldsymbol{n}_j\right) {\rm d}\boldsymbol{u}\\
+&\int_{\boldsymbol{u}\cdot\boldsymbol{n}_j<0}\boldsymbol{\Psi}\left[\left(\delta_{d,R}-\Delta te^{-\frac{\Delta t}{\tau_{0,R}}}\right)g_{\alpha}^h\left({ \boldsymbol{0},0 }\right)+\left(\delta_{e,R}+\frac{\Delta t^2}{2}e^{-\frac{\Delta t}{\tau_{0,R}}}\right)\frac{\partial g_{\alpha}^h}{\partial \boldsymbol{x}}\cdot\boldsymbol{u}\right] \left(\boldsymbol{u}\cdot\boldsymbol{n}_j\right) {\rm d}\boldsymbol{u}.\\
\end{aligned}
\end{equation}
As a result, the macroscopic update equation is renewed from Eq.~\eqref{eq:conservation-macro} as follows.
\begin{equation}\label{eq:update1}
\boldsymbol{W}^{\ast}_{\alpha,i} = \boldsymbol{W}^{n}_{\alpha,i} - \frac{1}{\Omega_i}\sum\limits_{j\in \mathcal{M}\left(i\right)}\boldsymbol{F}_{\alpha,j}^{eq}S_j - \frac{1}{\Omega_i}\sum\limits_{j\in \mathcal{M}\left(i\right)}\boldsymbol{F}_{\alpha,j}^{fr,wave}S_j+\boldsymbol{W}_{\alpha,i}^{fr,p}.
\end{equation}

\subsection{Macroscopic source term}\label{sec:source}
By calculating moments in Eq.~\eqref{eq:bgk}, the source term in Eq.~\eqref{eq:conservation-macrosource} can be derived as,
\begin{equation}\label{eq:source1}
\frac{\partial\boldsymbol{W}_{\alpha,i}}{\partial t}= \frac{\tilde{\boldsymbol{W}}^{\ast}_{\alpha,i}-\boldsymbol{W}_{\alpha,i}}{\tau_0},
\nonumber
\end{equation}
whose integral solution is,
\begin{equation}\label{eq:source2}
\boldsymbol{W}^{n+1}_{\alpha,i}= e^{\frac{-\Delta t}{\tau_0}}\boldsymbol{W}^{\ast}_{\alpha,i}+\left(1-e^{-\frac{\Delta t}{\tau_0}}\right)\tilde{\boldsymbol{W}}^{\ast}_{\alpha,i},
\end{equation}
where the initial condition $\boldsymbol{W}^{\ast}_{\alpha,i}$ is the result of Eq.~\eqref{eq:update1} and $\tilde{\boldsymbol{W}}^{\ast}_{\alpha,i}$ is the target conserved variables of $\boldsymbol{W}^{\ast}_{\alpha,i}$. Different from the AAP model, because the target velocity is obtained by a convex combination in Eq.~\eqref{eq:groppi}, an extra implicit module is not necessary. In this way, the consistency between wave and particles is achieved in three viewpoints. Firstly, for the flux, the contribution of particles (mainly about the newly sampled particles from $g_{\alpha,i}$) uses the same $\tilde{\boldsymbol{W}}^{\ast}_{\alpha,i}$ as the flux of deterministic wave. Secondly, for the source collision term, both of macroscopic evolution in Eq.~\eqref{eq:source2} and the particle sampling use the same $\tilde{\boldsymbol{W}}^{\ast}_{\alpha,i}$ as well. Finally, still for the source term, the coefficients in front of the target variables ($g_{\alpha}$ for particle as Eq.~\eqref{eq:mic5} and $\tilde{\boldsymbol{W}}^{\ast}_{\alpha,i}$ for macroscopic variables as Eq.~\eqref{eq:source2}) are the same, which is $1-e^{-\frac{\Delta t}{\tau_0}}$. If the operator splitting is not implemented before the discretization in Sec.~\ref{sec:discretization} but after that instead, the coefficient in front of $\tilde{\boldsymbol{W}}^{\ast}_{\alpha,i}$ will be $\frac{\Delta t}{\Delta t+\tau_0}$, different from $1-e^{-\frac{\Delta t}{\tau_0}}$. Additionally, Eq.~\eqref{eq:source2} ensures the conservation because,
\begin{equation}
\sum\limits_{\alpha=1}^2{\boldsymbol{W}^{\ast}_{\alpha,i}}=\sum\limits_{\alpha=1}^2{\tilde{\boldsymbol{W}}^{\ast}_{\alpha,i}}.
\nonumber
\end{equation}

\subsection{Algorithm}\label{sec:sum1}
\begin{figure}[H]
	\centering
	\subfigure[]{\label{fig1a}
			\includegraphics[width=0.22 \textwidth]{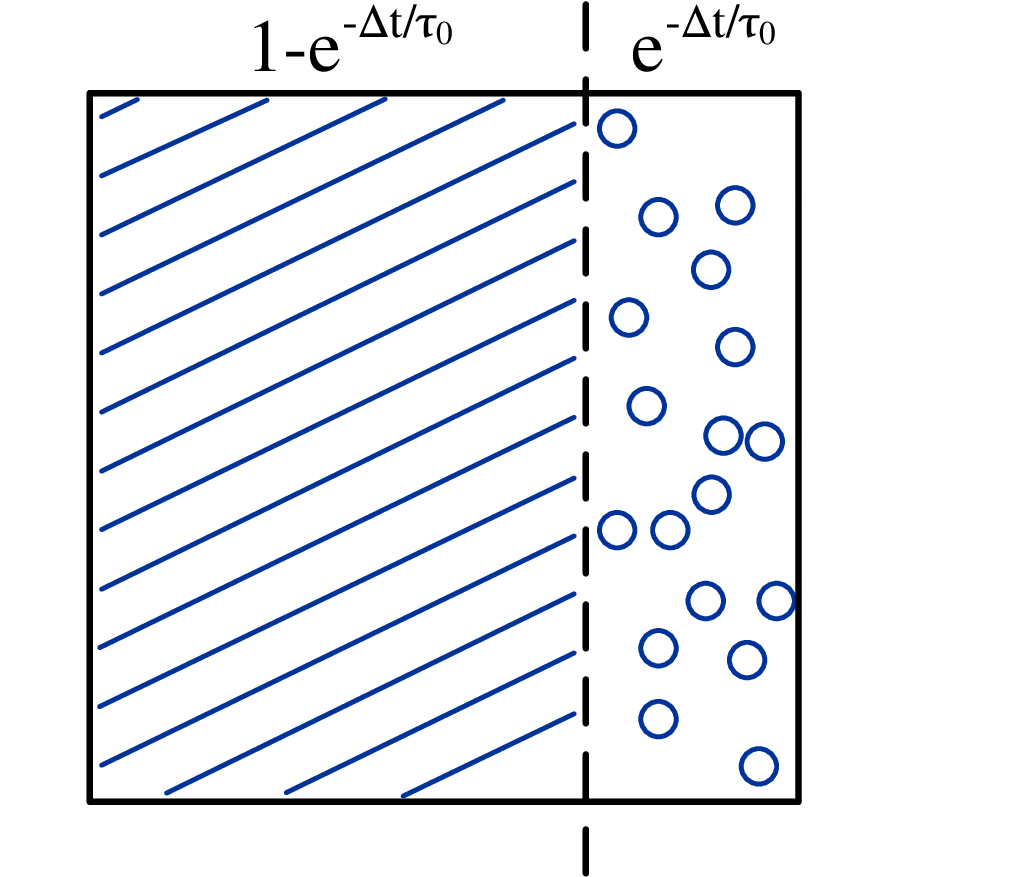}
		}
    \subfigure[]{\label{fig1b}
    		\includegraphics[width=0.22 \textwidth]{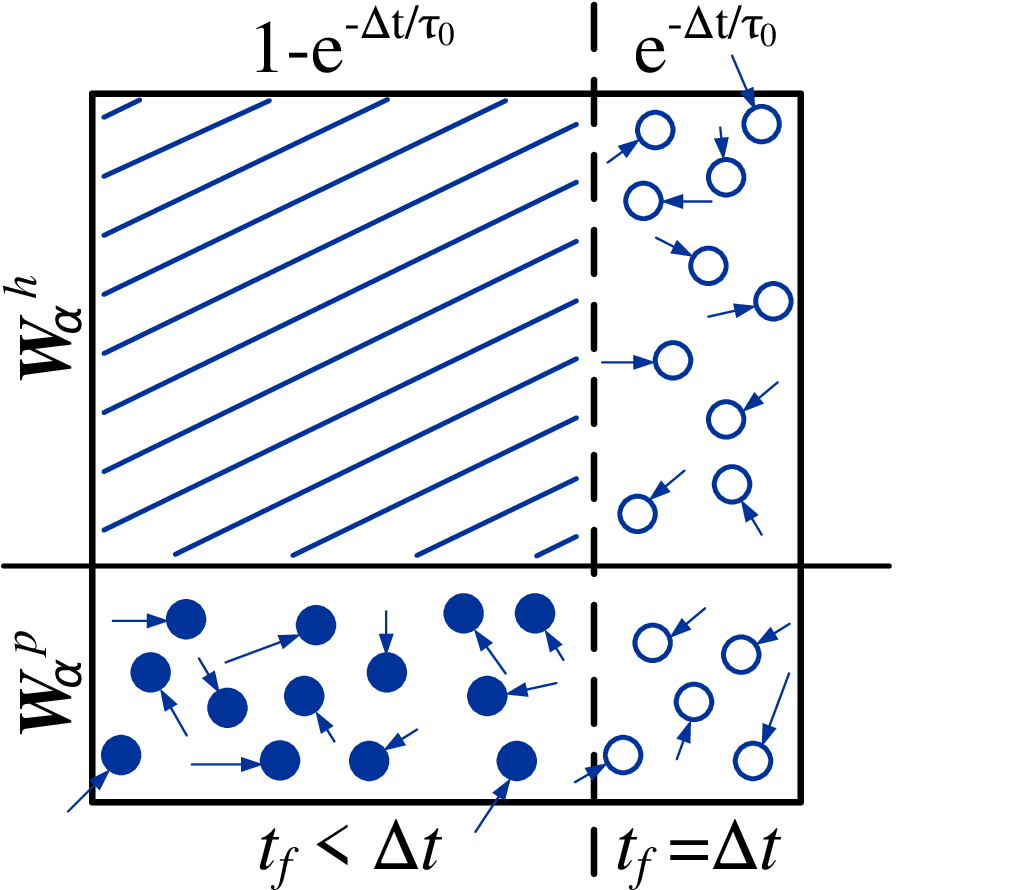}
    	}
    \subfigure[]{\label{fig1c}
    		\includegraphics[width=0.235 \textwidth]{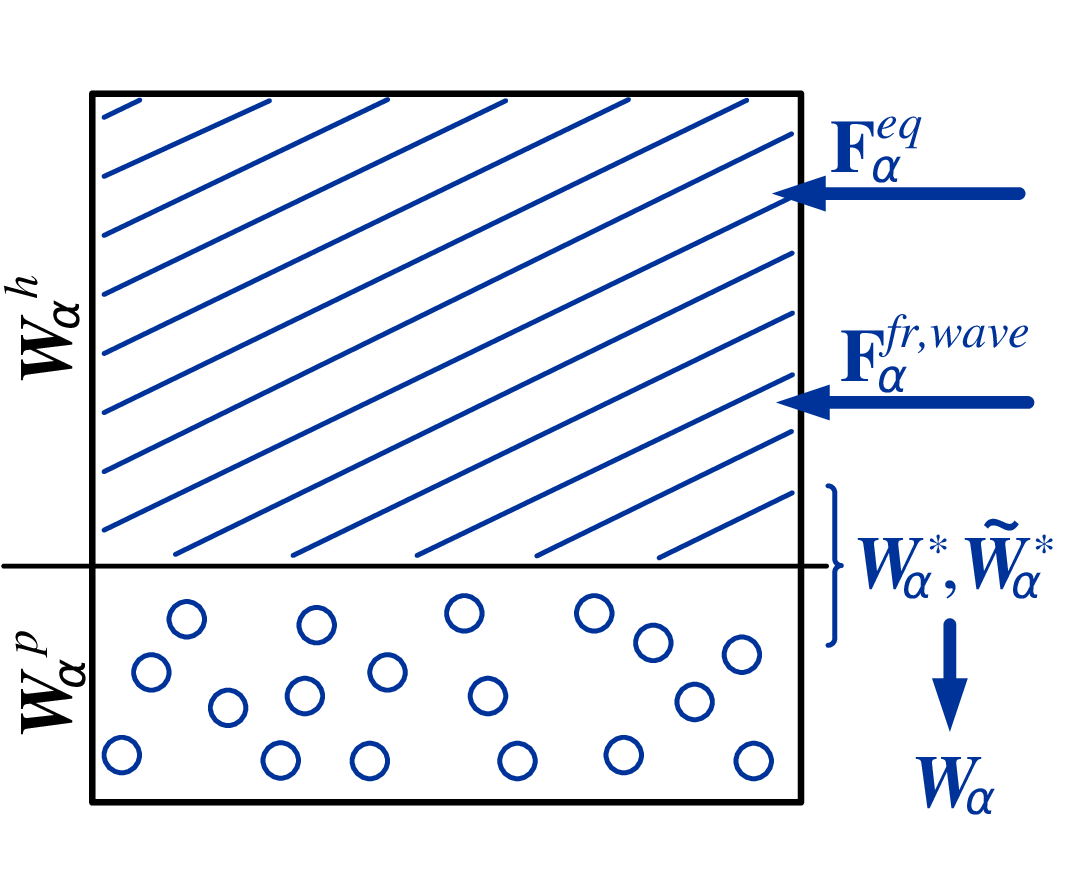}
    	}
    \subfigure[]{\label{fig1d}
    		\includegraphics[width=0.215 \textwidth]{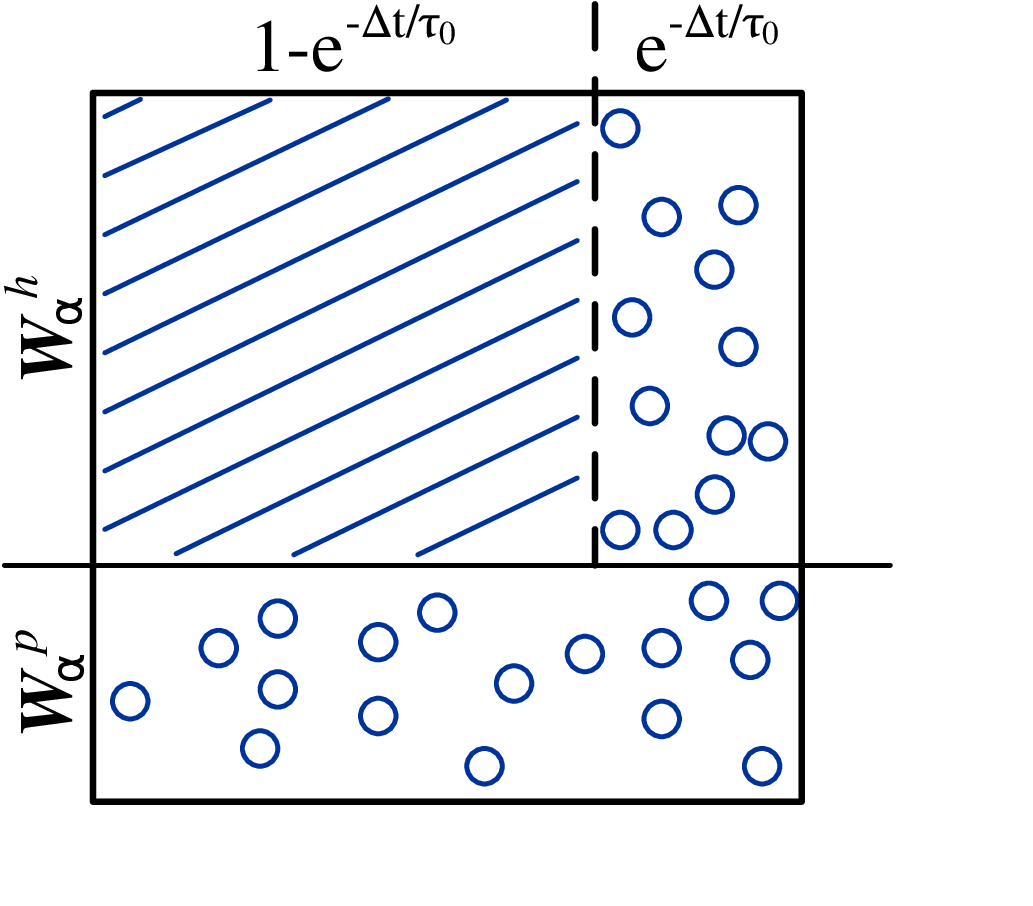}
    	}
	\caption{\label{fig1} Diagram to illustrate the algorithm of UGKWP method: (a) Initial state at $n=0$ step, (b) step $2$, (c) step $3$, (d) initial state at other steps.}
\end{figure}

Regarding Fig.~\ref{fig1}, the following part is a summary of the algorithm of the proposed UGKWP method for multiscale binary-species gas mixture.
\begin{description}
    \item[Step (1)] Initial state. As the result of the previous time step $n-1$ in Fig.~\ref{fig1d}, numerical particles $\boldsymbol{W}^p_{\alpha}$ coexist with hydrodynamic wave $\boldsymbol{W}^h_{\alpha}$. Then collisionless particles $\boldsymbol{W}^{hp}_{\alpha}$ are sampled as Eq.~\eqref{eq:mic5}, with detailed parameters from Eq.~\eqref{eq:mic6} to Eq.~\eqref{eq:ajshak}. For the first step, $\boldsymbol{W}^p_{\alpha}=\boldsymbol{0}$, as shown in Fig.~\ref{fig1a}.
    \item[Step (2)] Free transport. Firstly, divide particles $\boldsymbol{W}^p_{\alpha}$ into two parts as Eq.~\eqref{eq:mic1} or Eq.~\eqref{eq:taustar}. As in Fig.~\ref{fig1b}, collisionless particles are denoted by hollow circles, and collisional particles are denoted by solid circles. Then transport all particles as Eq.~\eqref{eq:mic2}, and sum their contribution to the macroscopic conserved variables as Eq.~\eqref{eq:mic3}. Meanwhile, the free transport fluxes $\boldsymbol{F}_{\alpha,j}^{fr,wave}$ contributed from the collisional particles of $\left(\boldsymbol{W}^h_{\alpha}-\boldsymbol{W}^{hp}_{\alpha}\right)$ are also calculated as Eq.~\eqref{eq:ffrwave}. Finally, delete the collisional particles which are denoted by solid circles.
    \item[Step (3)] Collision. Firstly, calculate the $\boldsymbol{F}^{eq}_{\alpha,j}$ term in Eq.~\eqref{eq:feq} by equations in Sec.~\ref{sec:macro}. Secondly, get $\boldsymbol{W}^{\ast}_{\alpha}$ as Eq.~\eqref{eq:update1}. Then get the target macroscopic variables $\tilde{\boldsymbol{W}}^{\ast}_{\alpha}$ from $\boldsymbol{W}^{\ast}_{\alpha}$ by Eq.~\eqref{eq:groppi}, along with other variables such as relaxation time $\tau_0$ and ${\rm{Pr}}_0$ in Eq.~\eqref{eq:coeffs}. Finally, update $\boldsymbol{W}^{n+1}_{\alpha}$ as Eq.~\eqref{eq:source2}.
    \item[Step (4)] If the simulation continues, go to Step $\left(1\right)$, where collisionless particles $\boldsymbol{W}^{hp}_{\alpha}$ will be sampled.
\end{description}

\section{Numerical cases}\label{sec:cases}
A series of cases are conducted to test the performance of the proposed UGKWP method. Firstly, the shock structures at ${\rm{Ma}}=1.5$ and ${\rm{Ma}}=3.0$ are simulated as a classical rarefied case. The proposed method is compared with the AAP model-based UGKS in detail. The performance of using a modified $\tau^{\ast}$ is also exhibited. Then the mass diffusion case is tested at different ${\rm{Kn}}$. The Couette flow is also simulated under four different conditions, ranging from continuum to rarefied regimes. The primary focus is on the velocity profile, which provides insights into the magnitude of velocity slip at the boundary and the separation of species. In these three cases, $N_{\rm{ref}}$ is set to be $5000$ to reduce the noise of particles. In the shock structure case, it is for avoiding the shock moving. In other two cases, the flow speed is very low, enlarging the influence of the particle noise. The last case is the hypersonic flow around a cylinder. In this case $N_{\rm{ref}}$ is set to be $800$ and a wide range of ${\rm{Ma}}$ and ${\rm{Kn}}$ is considered. The pressure, shear stress and heat flux coefficients at the wall are compared with the DSMC in detail. In all cases, the Courant-Friedrichs-Lewy (CFL) number is set to be $0.8$. The full accommodation is applied for the isothermal walls as the boundary condition.

\subsection{Shock structure}\label{sec:shock}
The first case is the shock structure for binary gas mixture containing species ``$A$'' and ``$B$''. It is a classical microscopic one-dimensional case to evaluate the model's and numerical method's capability to capture the highly non-equilibrium flow behaviors~\cite{egks,shock-model}, such as the profiles of density and temperature, the shock thickness, and the double-peak distribution function which is rather far away from the local equilibrium state~\cite{ugks-shock}. In the multi-species case, the non-equilibrium behaviors exist not only within the components but also between the components. Taking the Boltzmann solution as reference~\cite{shock-ref}, conditions for ${\rm{Ma}}^{-} = 1.5$ and ${\rm{Ma}}^{-} = 3.0$ are simulated with a large range of mole fraction $\chi$. Hereby superscripts ``$-$'' and ``$+$'' denote upstream and downstream respectively, and,
\begin{equation}
\begin{aligned}
&\chi^{-}_A=\chi^{+}_A,\chi^{-}_B=\chi^{+}_B,\\
&U^{-}_A=U^{-}_B=U^{-}_0,U^{+}_A=U^{+}_B=U^{+}_0,\\
&T^{-}_A=T^{-}_B=T^{-}_0,T^{+}_A=T^{+}_B=T^{+}_0,\\
&{\rm{Ma}}^{-}= \frac{U^{-}_0}{\sqrt{\frac{5}{3}k_BT^{-}_0/m^{-}_0}},{\rm{Ma}}^{+}= \frac{U^{+}_0}{\sqrt{\frac{5}{3}k_BT^{+}_0/m^{+}_0}}.
\nonumber
\end{aligned}
\end{equation}
The Rankine-Hugoniot condition is employed to achieve the upstream and downstream conditions for each species,
\begin{equation}
\begin{aligned}
&{\rm{Ma}}^{+}=\sqrt{\frac{2/3\left({\rm{Ma}}^{-}\right)^2+2}{10/3\left({\rm{Ma}}^{-}\right)^2-2/3}},\\
&\frac{\rho^{+}_A}{\rho^{-}_A}=\frac{\rho^{+}_B}{\rho^{-}_B}=\frac{8/3\left({\rm{Ma}}^{-}\right)^2}{2/3\left({\rm{Ma}}^{-}\right)^2+2},\\
&\frac{U^{+}_0}{U^{-}_0}=\frac{2/3\left({\rm{Ma}}^{-}\right)^2+2}{8/3\left({\rm{Ma}}^{-}\right)^2},\\
&\frac{T^{+}_0}{T^{-}_0}=\frac{\left[2/3\left({\rm{Ma}}^{-}\right)^2+2\right]\left[10/3\left({\rm{Ma}}^{-}\right)^2-2/3\right]}{64/9\left({\rm{Ma}}^{-}\right)^2}.
\nonumber
\end{aligned}
\end{equation}
The hard sphere (HS) model is employed for both species ``$A$'' and ``$B$'', and $\mu^{-}_A=\mu^{-}_B$. In the results, the vertical coordinate is normalized as,
\begin{equation}
\begin{aligned}
&{\hat{n}}_A=\frac{n_A-n^{-}_A}{n^{+}_A-n^{-}_A},\hat{n}_B=\frac{n_B-n^{-}_B}{n^{+}_B-n^{-}_B},\\
&{\hat{T}}_A=\frac{T_A-T^{-}_A}{T^{+}_A-T^{-}_A},\hat{T}_B=\frac{T_B-T^{-}_B}{T^{+}_B-T^{-}_B}.
\nonumber
\end{aligned}
\end{equation}
The horizontal coordinate is nondimensionalized by the mean free path defined as,
\begin{equation}
\begin{aligned}
\lambda_{\infty}=\frac{16}{5}\sqrt{\frac{m^{-}_0}{2\pi k_BT^{-}_0}}\frac{\mu^{-}_A}{\rho^{-}_0}.
\end{aligned}
\end{equation}
For the ${\rm{Ma}}=1.5$ case the mesh size is set to be $\Delta x=\lambda_{\infty}/2$, and for the ${\rm{Ma}}=3$ case the mesh size is set to be $\Delta x=\lambda_{\infty}/4$.

\begin{figure}[H]
	\centering
	\subfigure[]{
			\includegraphics[width=0.32 \textwidth]{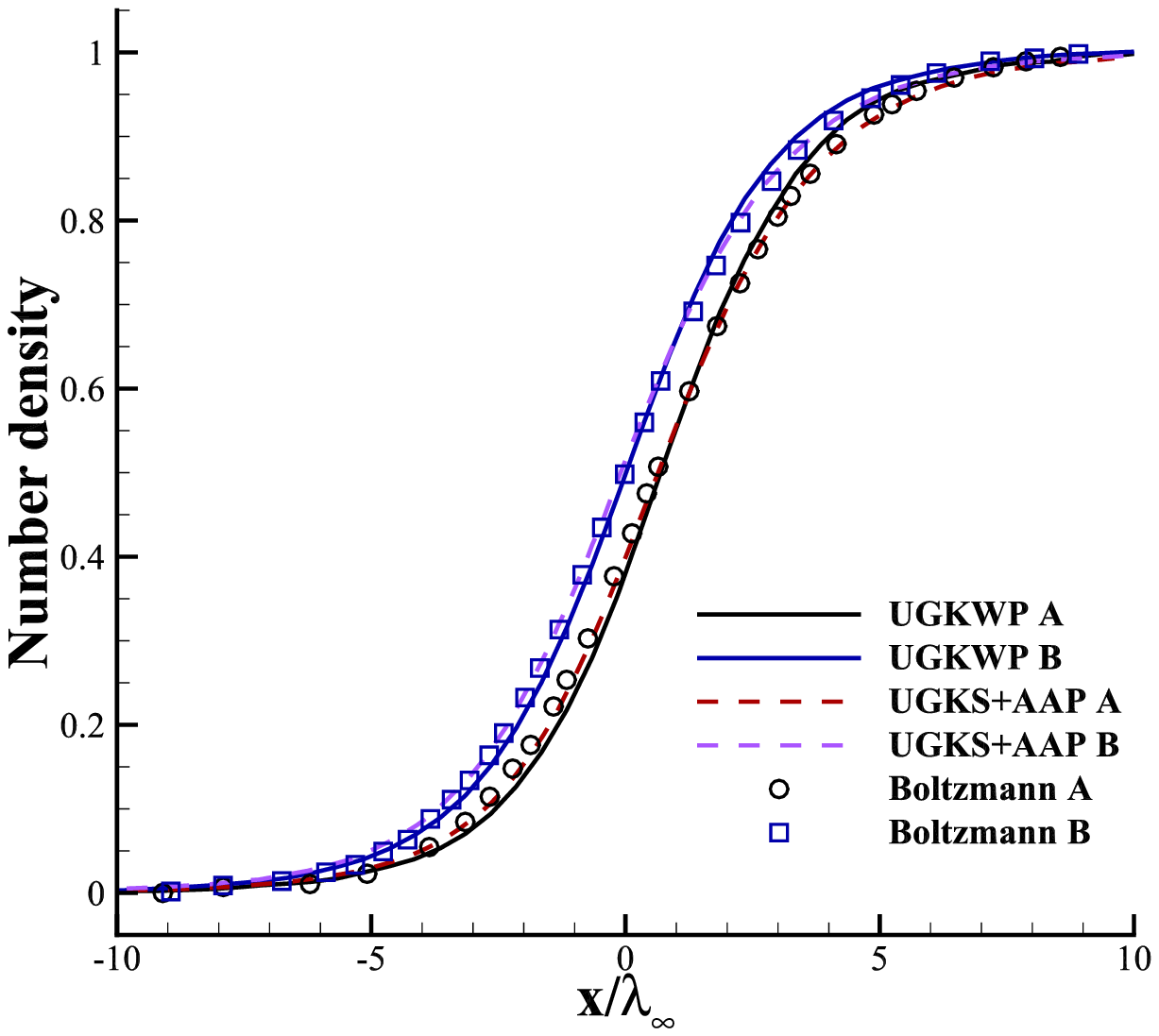}
		}
    \subfigure[]{
    		\includegraphics[width=0.32 \textwidth]{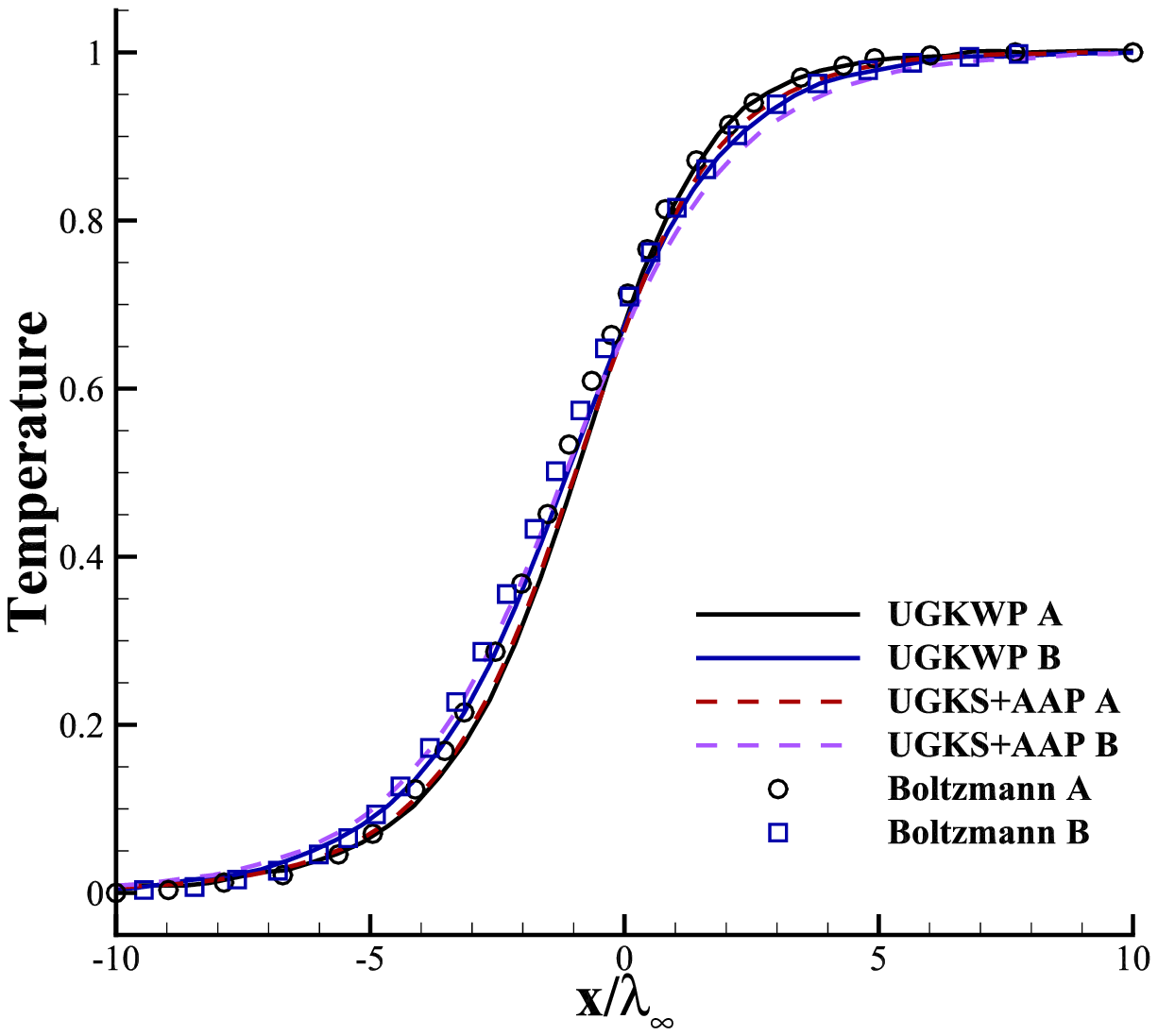}
    	}
	\caption{\label{ma1.5xb0.9ratio2_aap} Shock structure in binary gas mixture (${\rm{Ma}}^{-}=1.5$, $m_B/m_A=0.5$, $\chi^{-}_B=0.9$): (a) Number density and (b) temperature.}
\end{figure}

\begin{figure}[H]
	\centering
	\subfigure[]{
			\includegraphics[width=0.32 \textwidth]{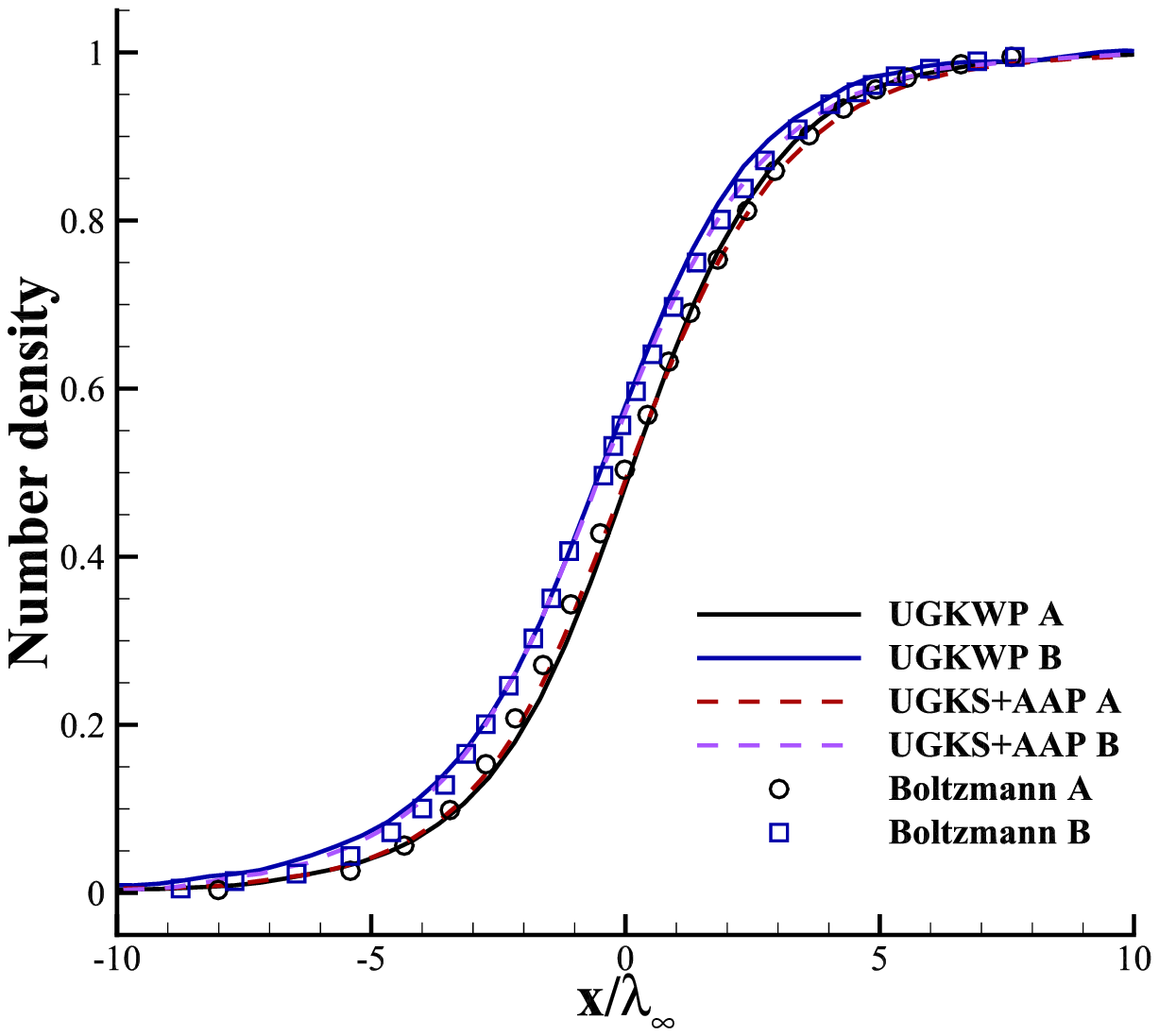}
		}
    \subfigure[]{
    		\includegraphics[width=0.32 \textwidth]{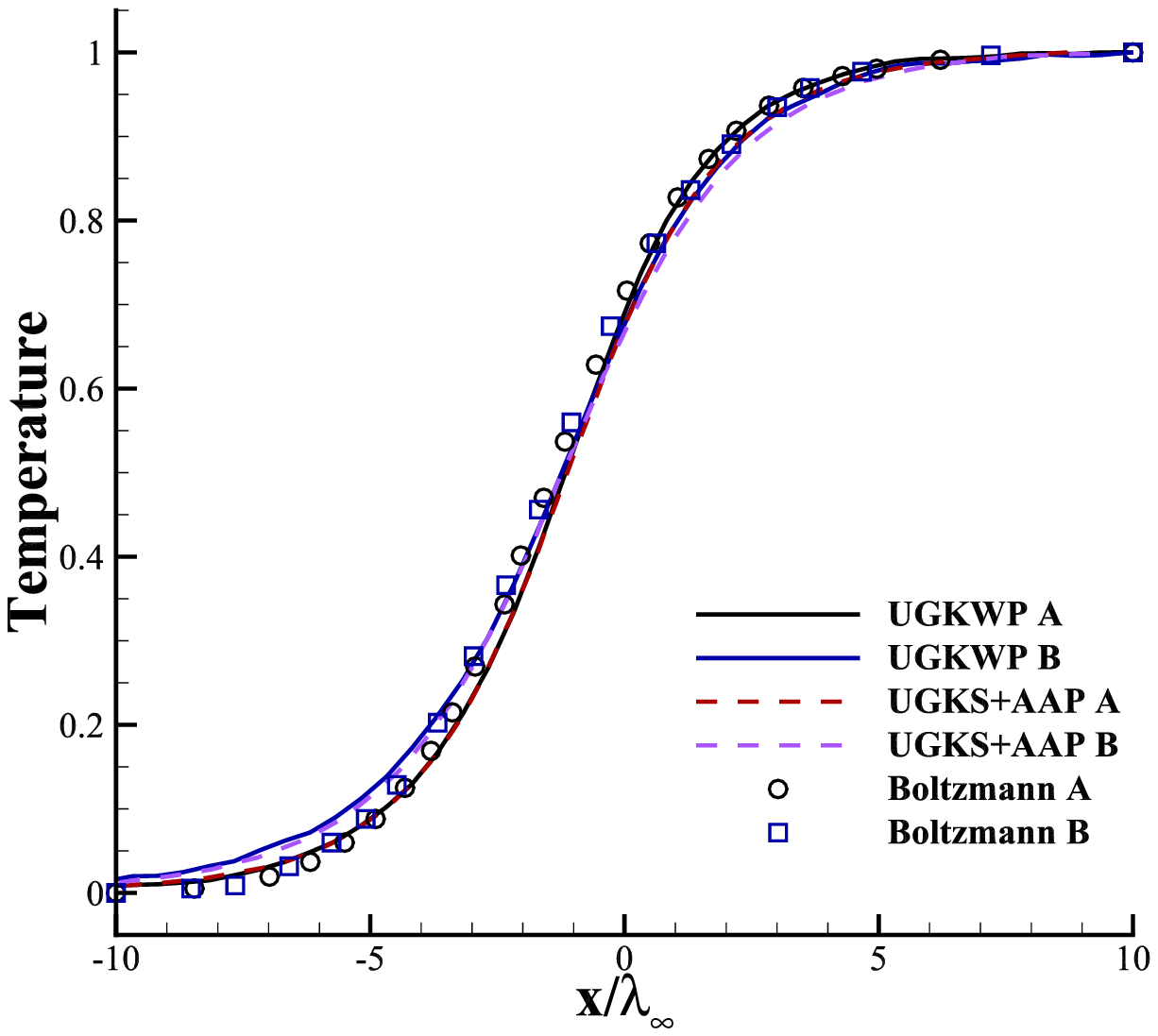}
    	}
	\caption{\label{ma1.5xb0.1ratio2_aap} Shock structure in binary gas mixture (${\rm{Ma}}^{-}=1.5$, $m_B/m_A=0.5$, $\chi^{-}_B=0.1$): (a) Number density and (b) temperature.}
\end{figure}

\begin{figure}[H]
	\centering
	\subfigure[]{
			\includegraphics[width=0.32 \textwidth]{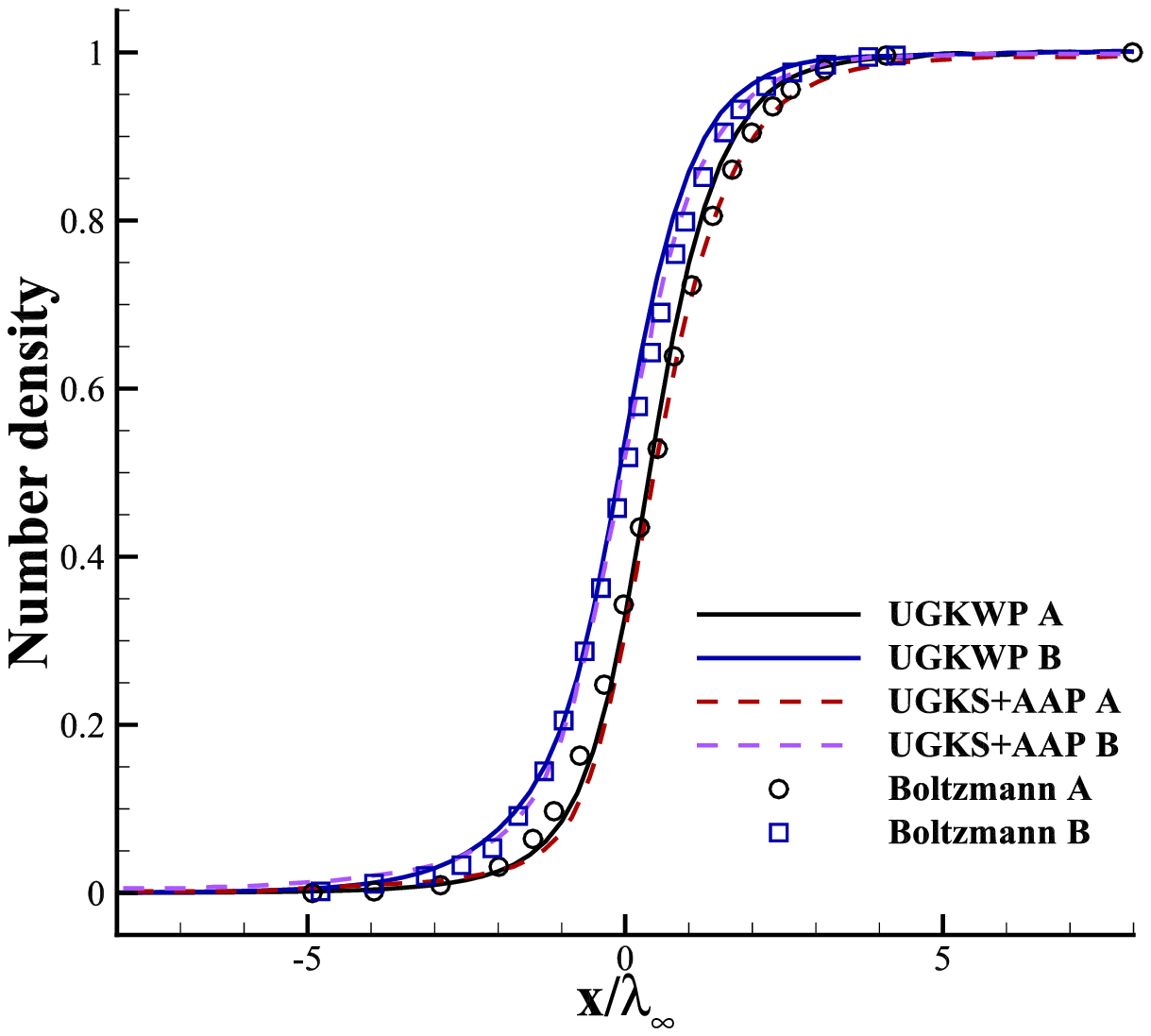}
		}
    \subfigure[]{
    		\includegraphics[width=0.32 \textwidth]{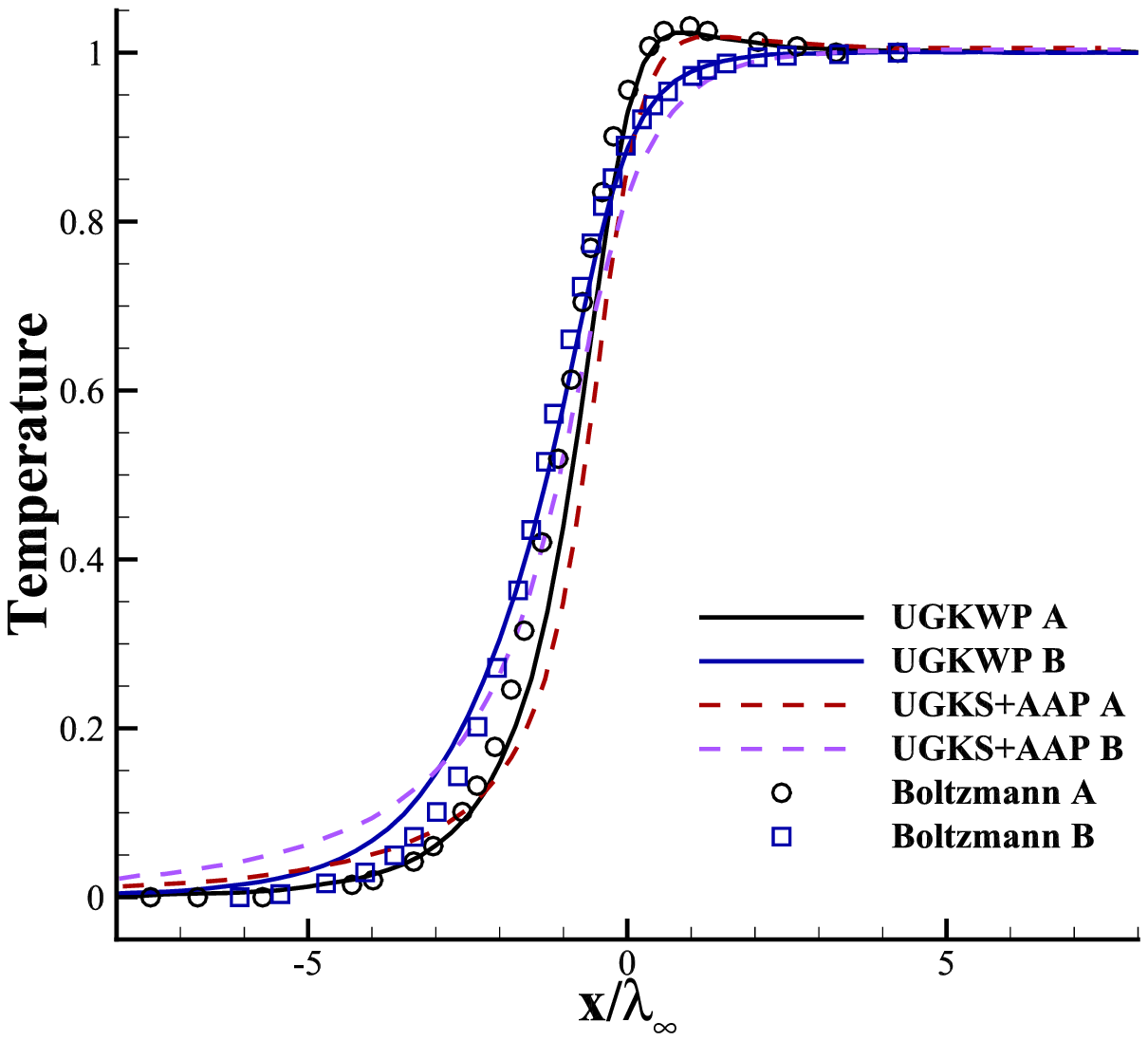}
    	}
	\caption{\label{ma3xb0.9ratio2_aap} Shock structure in binary gas mixture (${\rm{Ma}}^{-}=3$, $m_B/m_A=0.5$, $\chi^{-}_B=0.9$): (a) Number density and (b) temperature.}
\end{figure}

\begin{figure}[H]
	\centering
	\subfigure[]{
			\includegraphics[width=0.32 \textwidth]{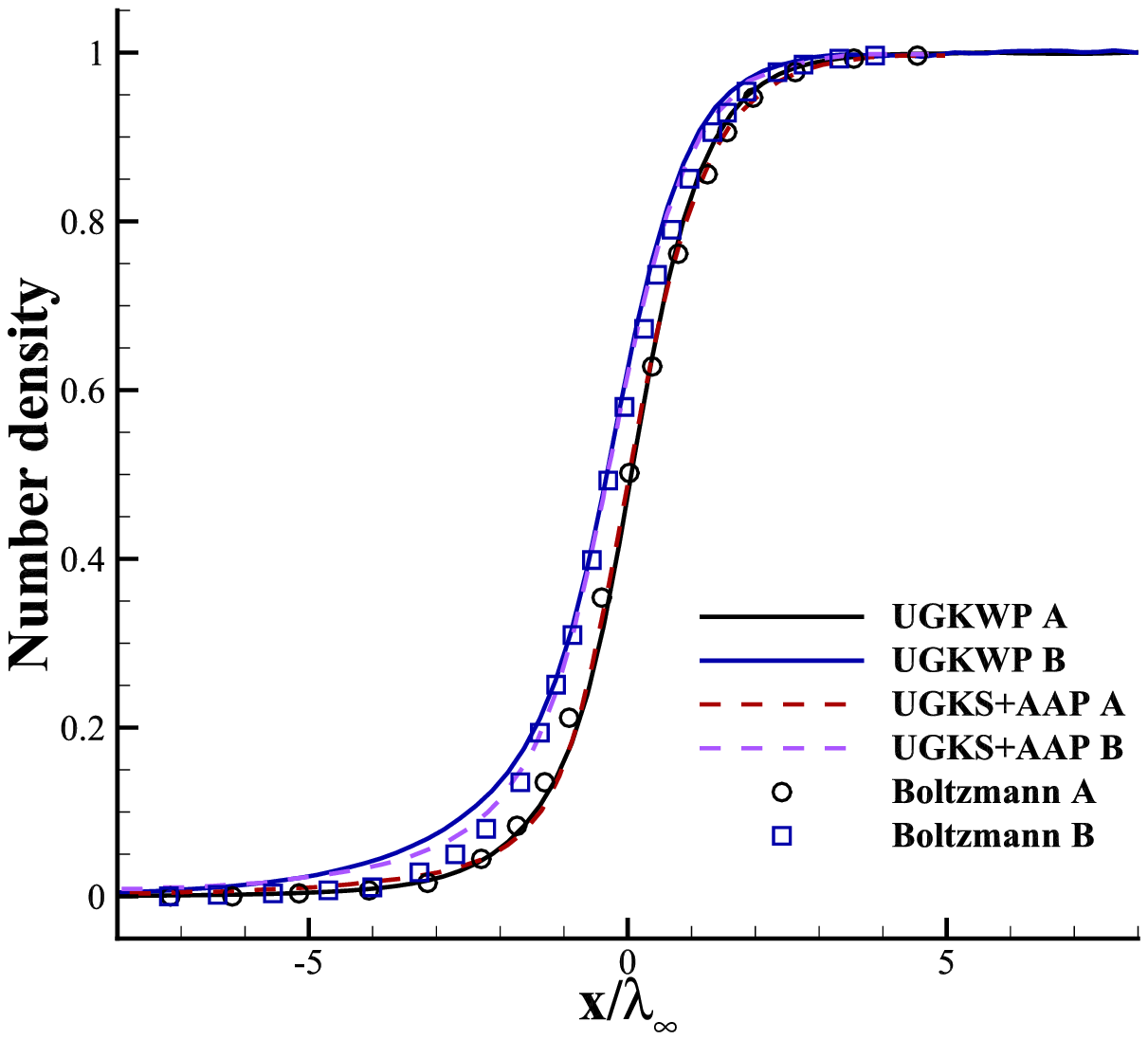}
		}
    \subfigure[]{
    		\includegraphics[width=0.32 \textwidth]{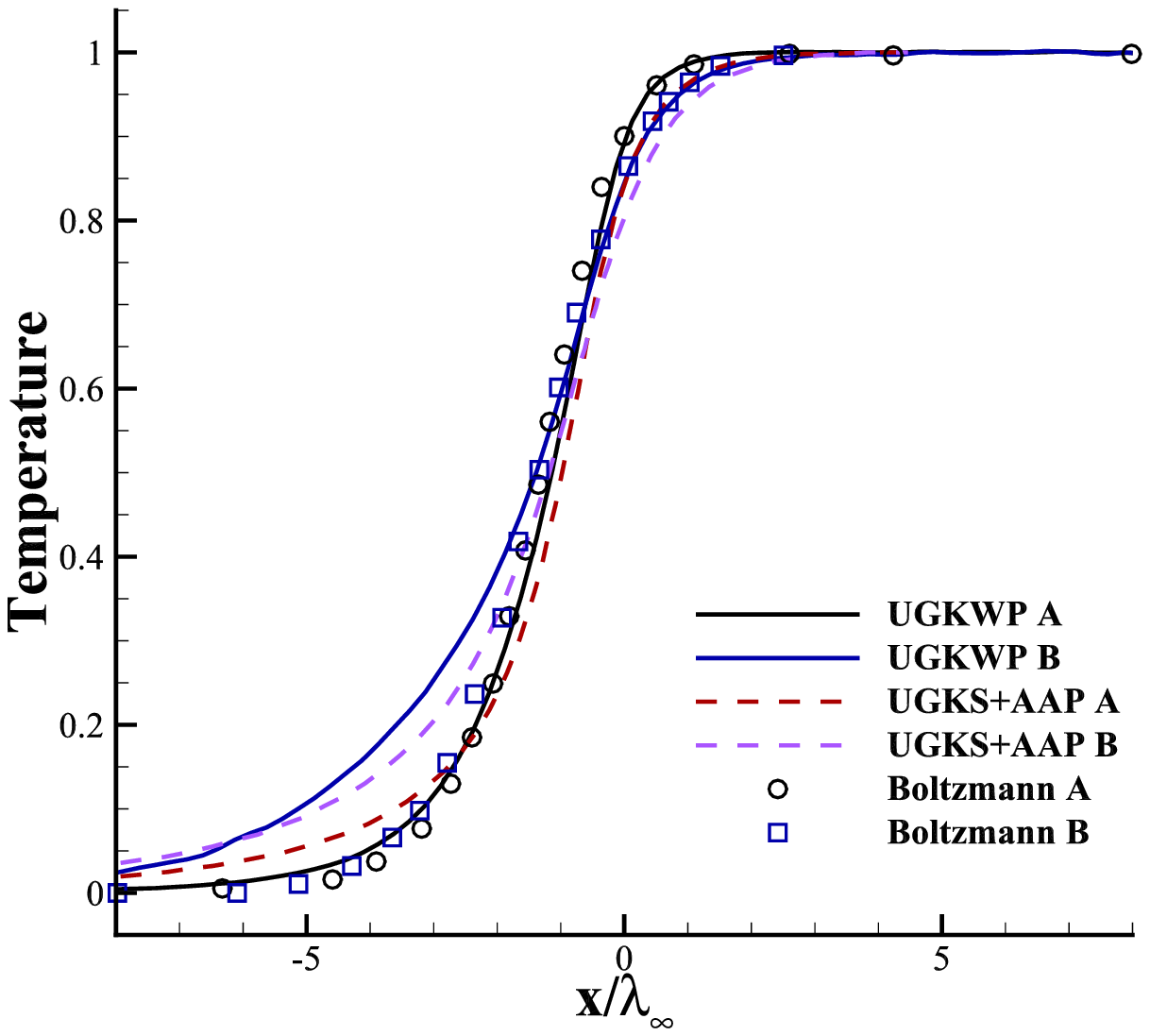}
    	}
	\caption{\label{ma3xb0.1ratio2_aap} Shock structure in binary gas mixture (${\rm{Ma}}^{-}=3$, $m_B/m_A=0.5$, $\chi^{-}_B=0.1$): (a) Number density and (b) temperature.}
\end{figure}

\begin{figure}[H]
	\centering
	\subfigure[]{
			\includegraphics[width=0.32 \textwidth]{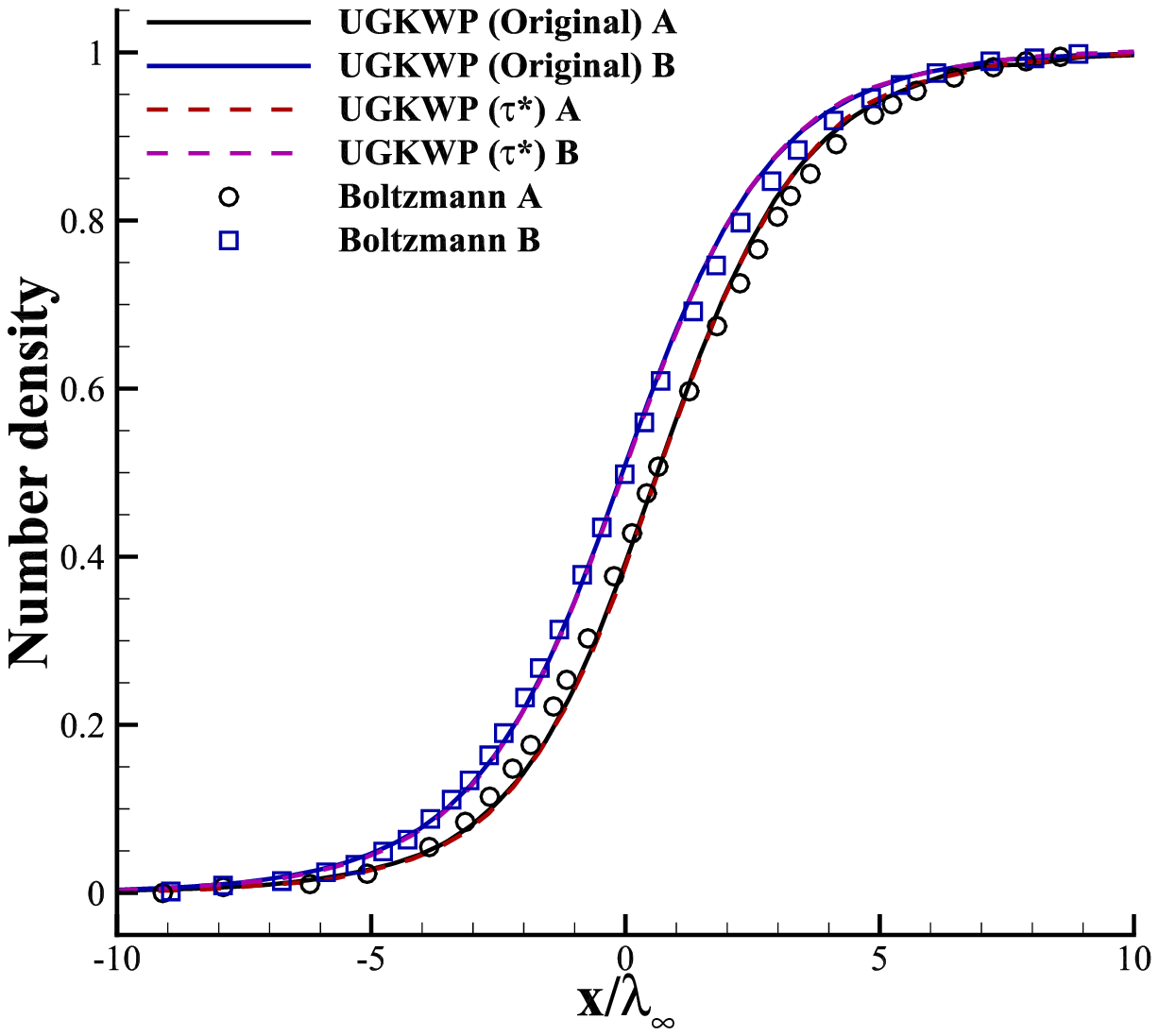}
		}
    \subfigure[]{
    		\includegraphics[width=0.32 \textwidth]{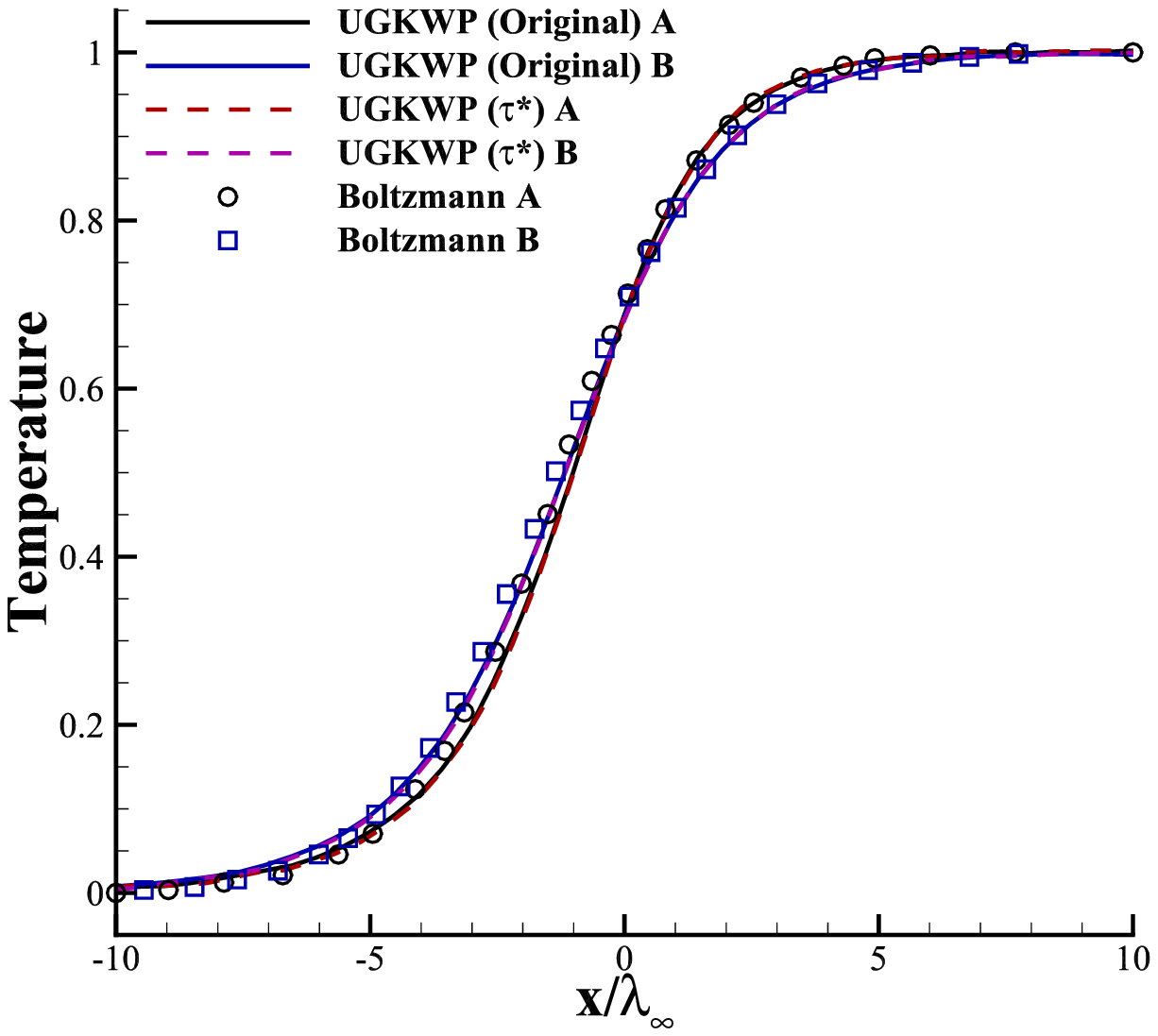}
    	}
	\caption{\label{ma1.5xb0.9ratio2} Shock structure in binary gas mixture (${\rm{Ma}}^{-}=1.5$, $m_B/m_A=0.5$, $\chi^{-}_B=0.9$): (a) Number density and (b) temperature.}
\end{figure}

\begin{figure}[H]
	\centering
	\subfigure[]{
			\includegraphics[width=0.32 \textwidth]{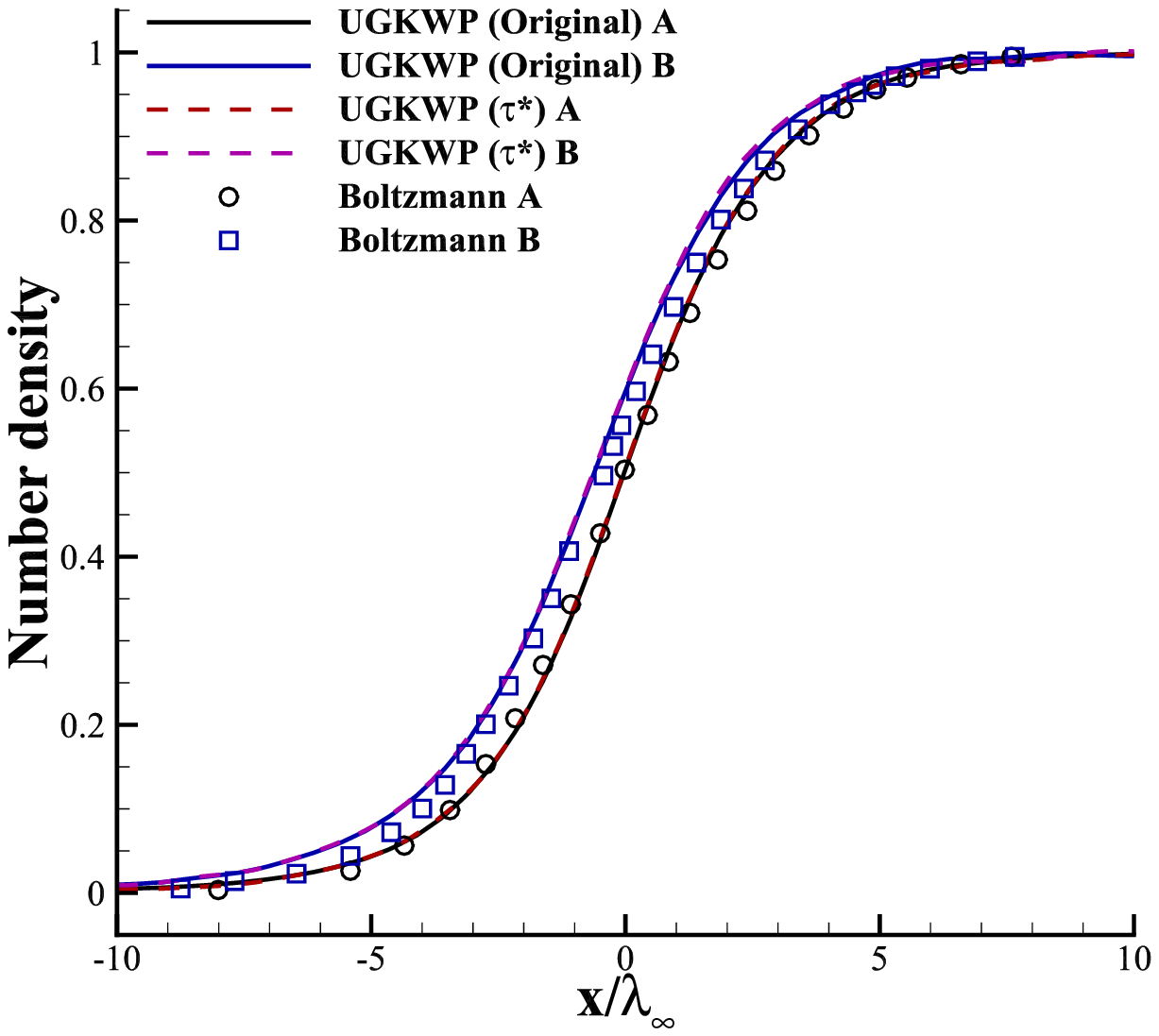}
		}
    \subfigure[]{
    		\includegraphics[width=0.32 \textwidth]{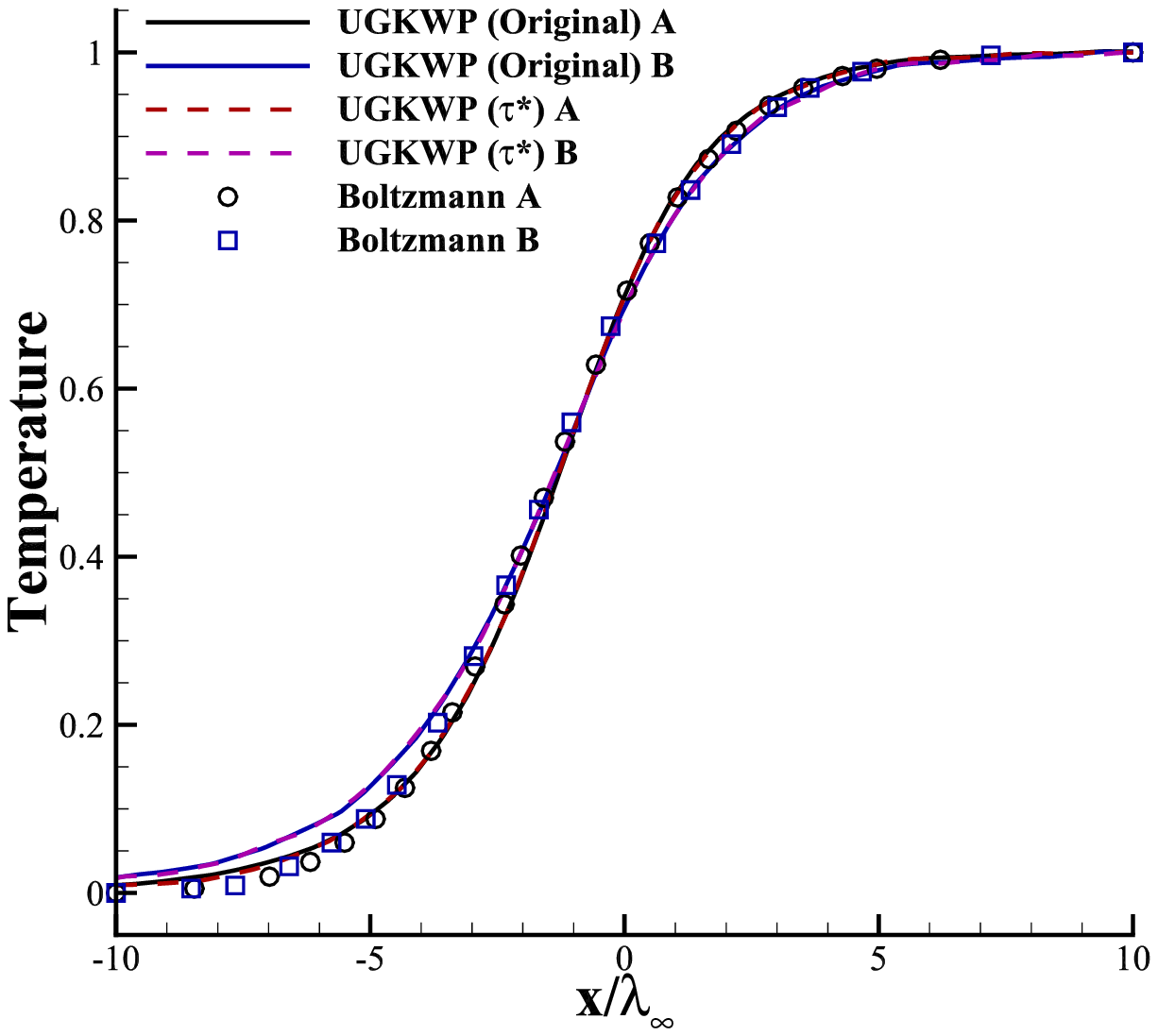}
    	}
	\caption{\label{ma1.5xb0.1ratio2} Shock structure in binary gas mixture (${\rm{Ma}}^{-}=1.5$, $m_B/m_A=0.5$, $\chi^{-}_B=0.1$): (a) Number density and (b) temperature.}
\end{figure}

\begin{figure}[H]
	\centering
	\subfigure[]{
			\includegraphics[width=0.32 \textwidth]{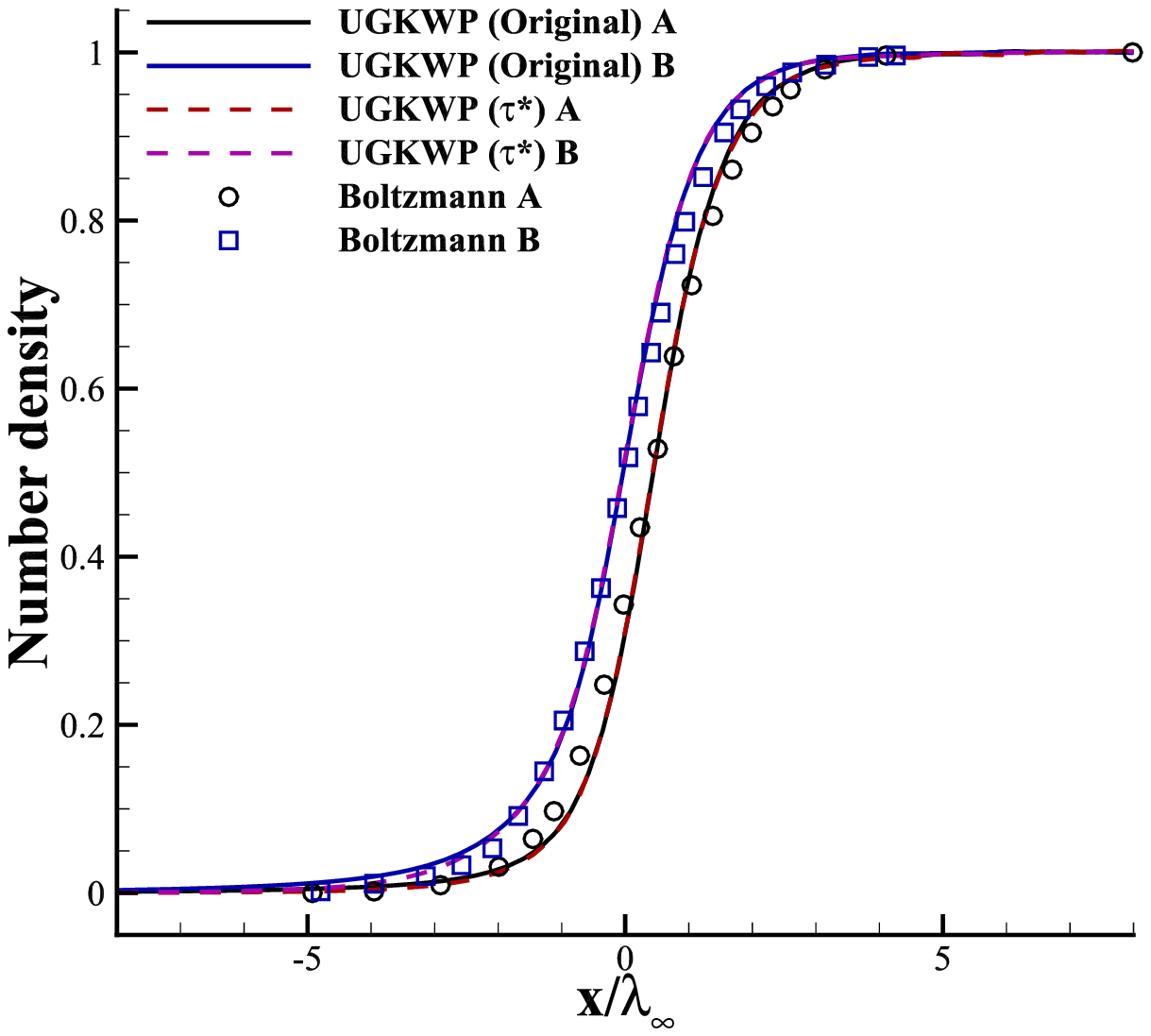}
		}
    \subfigure[]{
    		\includegraphics[width=0.32 \textwidth]{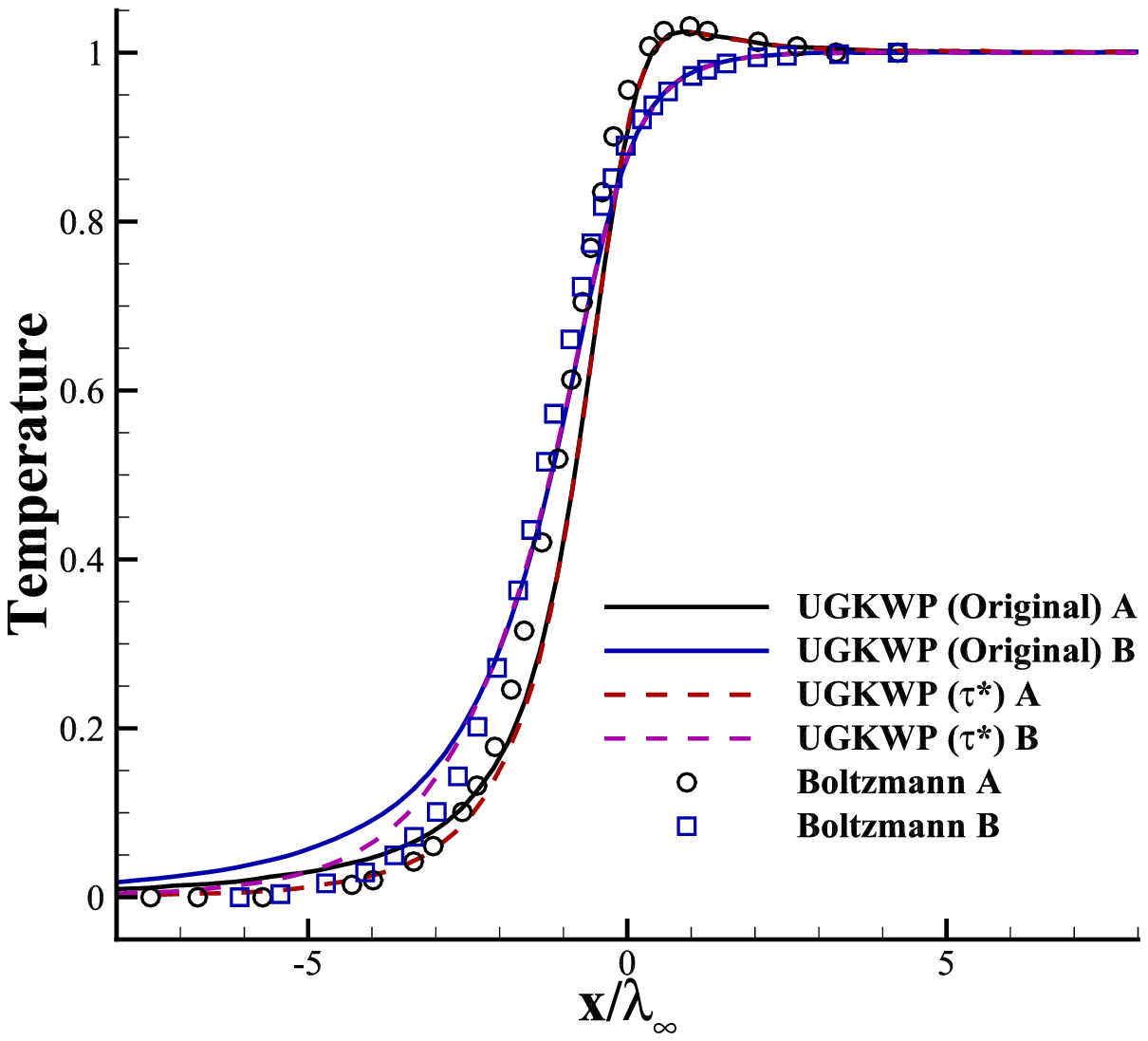}
    	}
	\caption{\label{ma3xb0.9ratio2} Shock structure in binary gas mixture (${\rm{Ma}}^{-}=3$, $m_B/m_A=0.5$, $\chi^{-}_B=0.9$): (a) Number density and (b) temperature.}
\end{figure}

\begin{figure}[H]
	\centering
	\subfigure[]{
			\includegraphics[width=0.32 \textwidth]{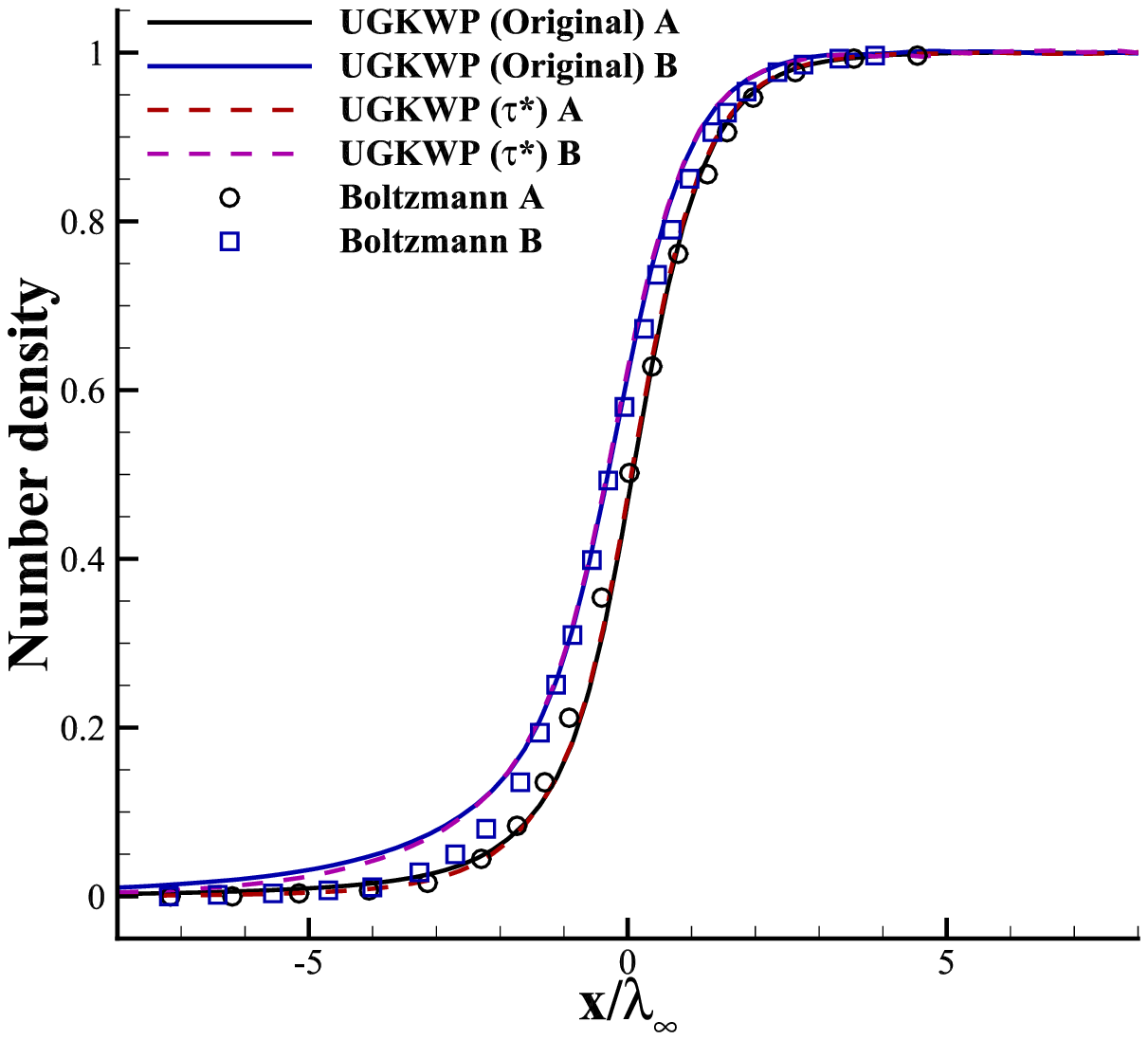}
		}
    \subfigure[]{
    		\includegraphics[width=0.32 \textwidth]{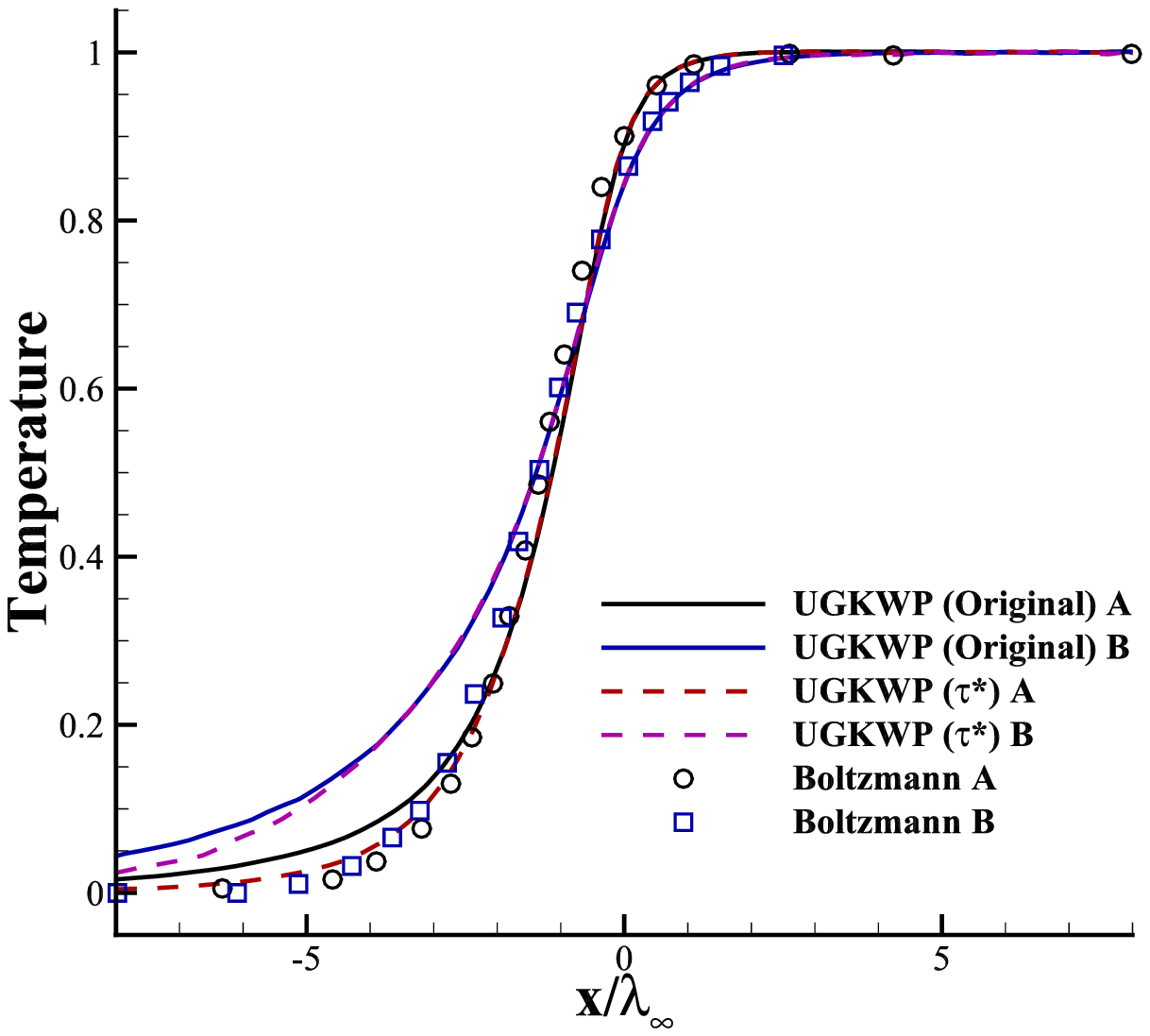}
    	}
	\caption{\label{ma3xb0.1ratio2} Shock structure in binary gas mixture (${\rm{Ma}}^{-}=3$, $m_B/m_A=0.5$, $\chi^{-}_B=0.1$): (a) Number density and (b) temperature.}
\end{figure}

As shown from Fig.~\ref{ma1.5xb0.9ratio2_aap} to Fig.~\ref{ma3xb0.1ratio2_aap}, the proposed UGKWP method is compared with the AAP model-based UGKS~\cite{ugks-aap}. At the post-shock location, the temperature profile of the proposed UGKWP method matches better with the reference result, and this advantage is more significant at higher ${\rm{Ma}}$. Referring to Refs.\cite{todorova2,shock-model}, it is mainly because the ${\rm{Pr}}$ of current method is more accurate, and when using different methods to correct ${\rm{Pr}}$ the approach of Shakhov model gives better results. Another advantage lies in the temperature profile at the pre-shock location. Because in the BGK-type kinetic model the relaxation time is not a function of particle speed, the negative-velocity high-speed particle at the pre-shock location has a lower collision frequency than the real physics. As a result, the temperature there is always higher than the Boltzmann solution. Benefiting from a further microscopic model in Eq.~\eqref{eq:taustar}, the current method can give a better result. A more detailed comparison whether using $\tau^{\ast}$ or not is shown from Fig.~\ref{ma1.5xb0.9ratio2} to Fig.~\ref{ma3xb0.1ratio2}. The improvement is much more obvious at higher ${\rm{Ma}}$ as well.

\subsection{Mass diffusion}\label{sec:diffusion}
In Ref.\cite{dsmc}, this case primarily verified that the DSMC can recover the Navier-Stokes diffusion coefficient in the continuum flow regime. In this section, the one-dimensional mass diffusion in an argon-neon (Ar-Ne) mixture is simulated and compared with the DSMC in a wide ${\rm{Kn}}$ range. The gases properties can be referred to Tab.~\ref{arne}.

\begin{table}[H]
\centering
\caption{\label{arne} Properties of Ar and Ne~\cite{dsmc}}
\begin{tabular}{|c|c|c|}
\hline
    Gas                                                                                                   &Ar                   &Ne                   \\ \hline
	Molecular mass $m$ $\left({\rm{kg}}\right)$                                                           &$6.63\times10^{-26}$ &$3.35\times10^{-26}$ \\
    Gas constant $R$ $\left({\rm{\frac{J}{kg\cdot K}}}\right)$                                            &$208.242$            &$412.134$            \\
    Viscosity index $\omega$                                                                              &$0.81$               &$0.66$               \\
    Scattering parameter $\alpha$ in the variable soft sphere model                                       &$1.40$               &$1.31$               \\
    Viscosity coefficient $\mu_{\rm{ref}}$ $\left({\rm{Pa\cdot s}}\right)$ at $T_{\rm{ref}}=273 {\rm{K}}$ &$2.117\times10^{-5}$ &$2.975\times10^{-5}$ \\
    Prandtl number ${\rm{Pr}}$                                                                            &$2/3$                &$2/3$ \\
\hline
\end{tabular}
\end{table}

In this case, two reservoirs containing pure stationary Ar and Ne gases at $T=273{\rm{K}}$ are established at the two ends of a spatial interval of $L=1{\rm{m}}$. According to Ref.\cite{dsmc}, $200$ cells are set within the interval, and the number density of both reservoirs are set to be $2.8\times10^{20}$, $2.8\times10^{19}$ and $2.8\times10^{18}$ from continuum to rarefied condition. Using the following HS model, their corresponding Knudsen numbers are $0.0061$, $0.061$ and $0.61$ respectively.
\begin{equation}\label{eq:kn}
\begin{aligned}
{\rm{Kn}}=\frac{16}{5L}\sqrt{\frac{m_0}{2\pi k_BT_0}}\frac{\mu_0}{\rho_0}.
\end{aligned}
\end{equation}

\begin{figure}[H]
	\centering
	\subfigure[]{
			\includegraphics[width=0.3 \textwidth]{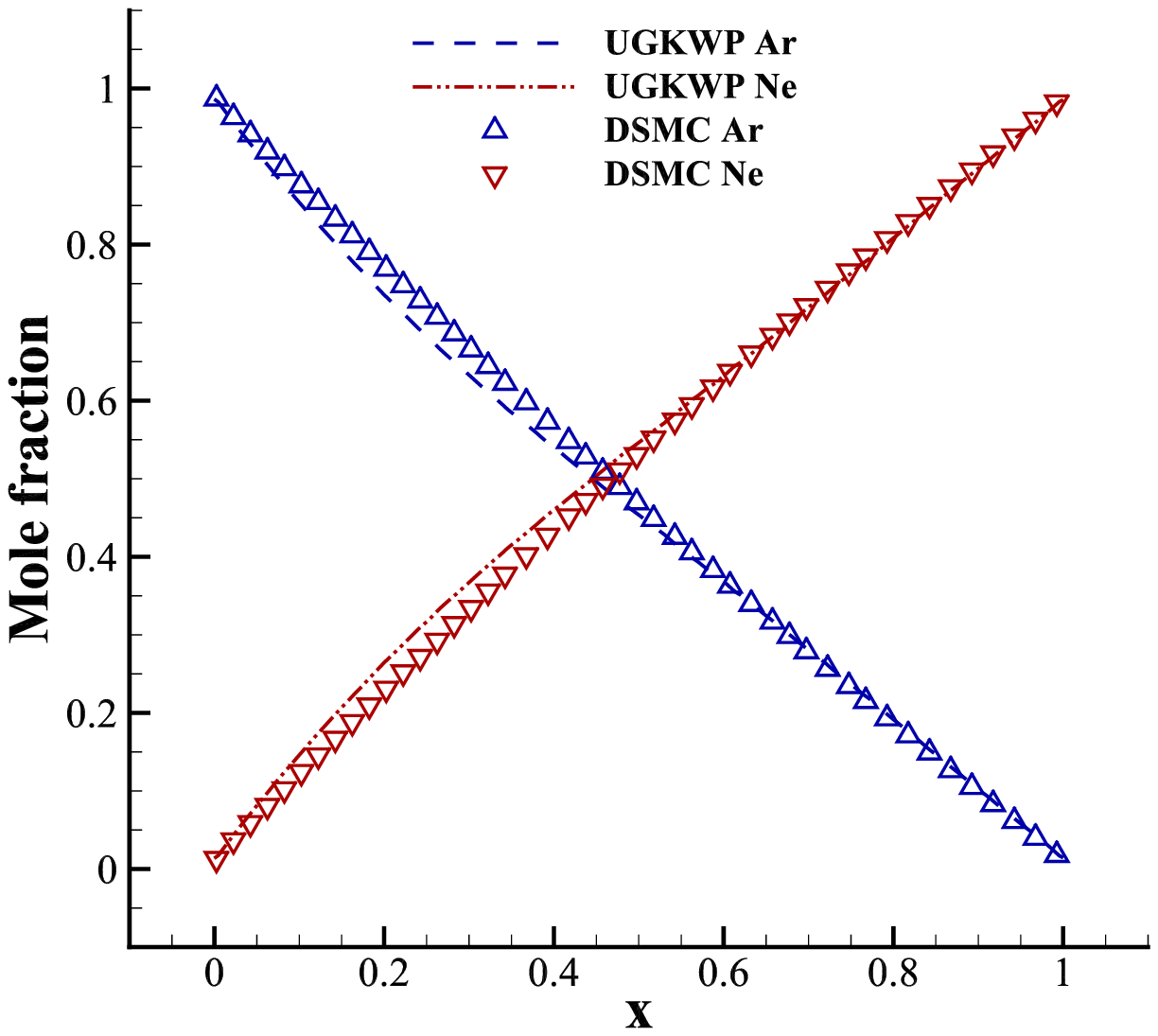}
		}
    \subfigure[]{
    		\includegraphics[width=0.3 \textwidth]{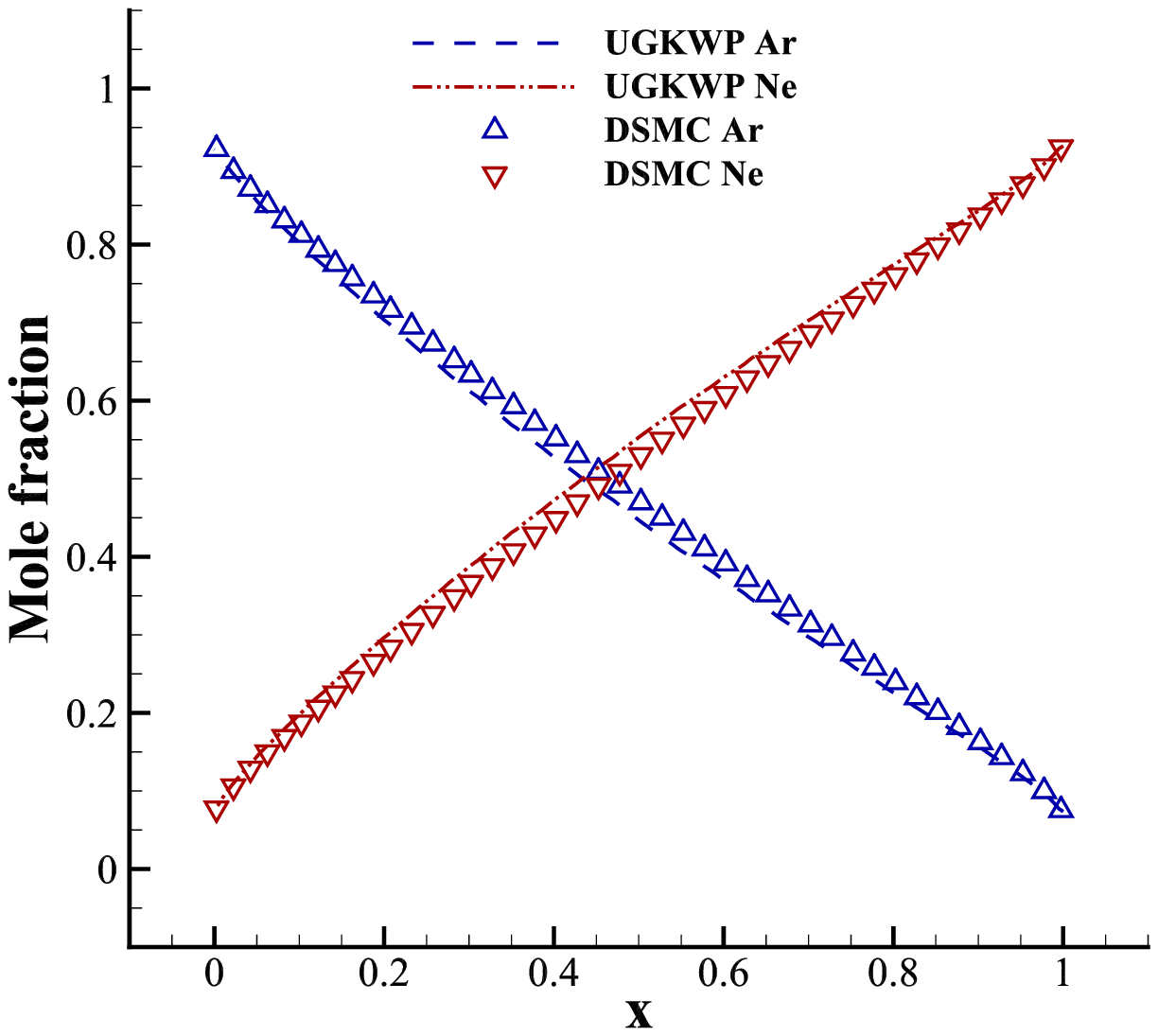}
    	}
    \subfigure[]{
    		\includegraphics[width=0.3 \textwidth]{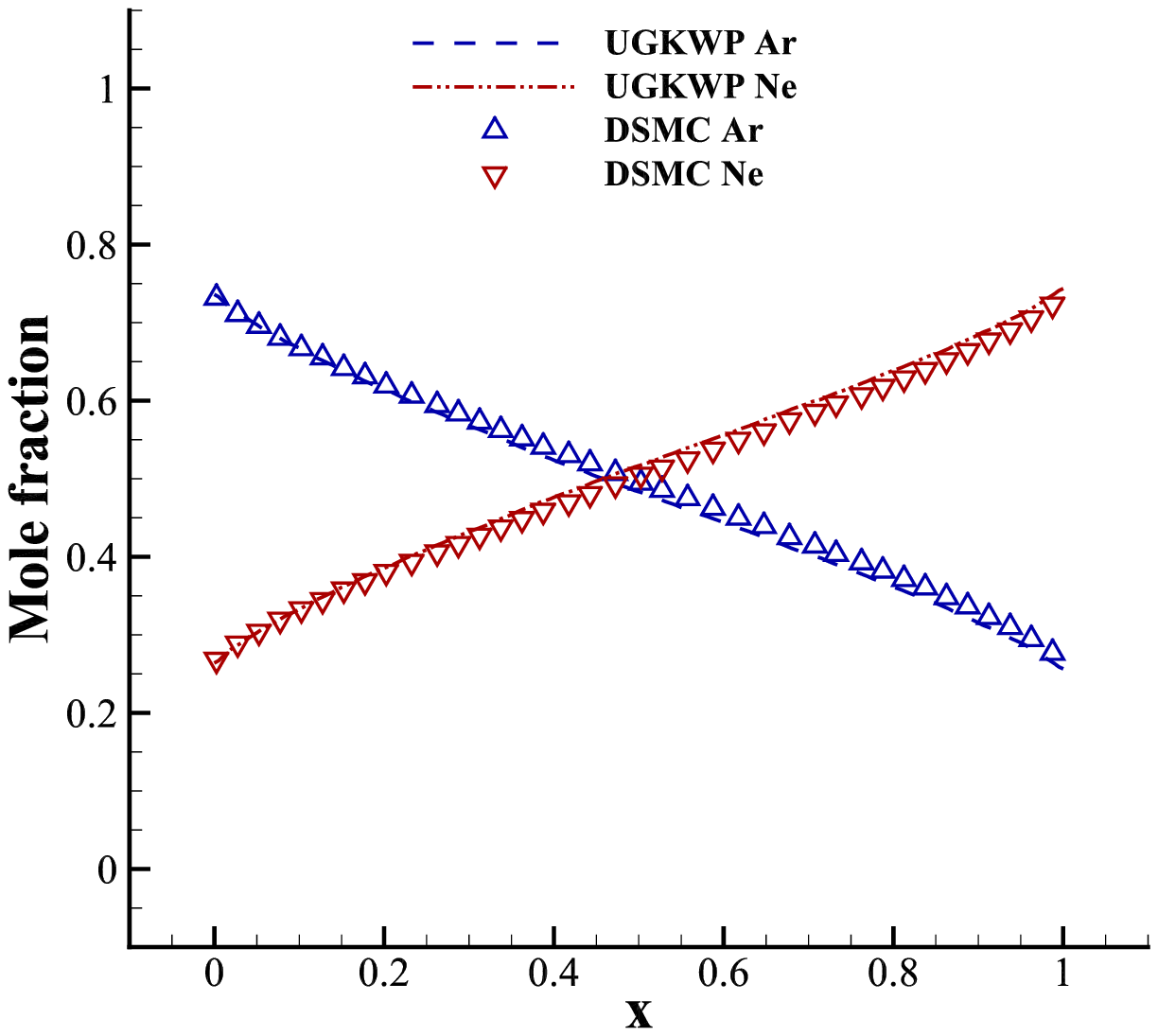}
    	}
	\caption{\label{diffusion-x} Mole fraction profile of mass diffusion case in an Ar-Ne mixture: (a) ${\rm{Kn}}=0.0061$, (b) ${\rm{Kn}}=0.061$ and (c) ${\rm{Kn}}=0.61$.}
\end{figure}

\begin{figure}[H]
	\centering
	\subfigure[]{
			\includegraphics[width=0.3 \textwidth]{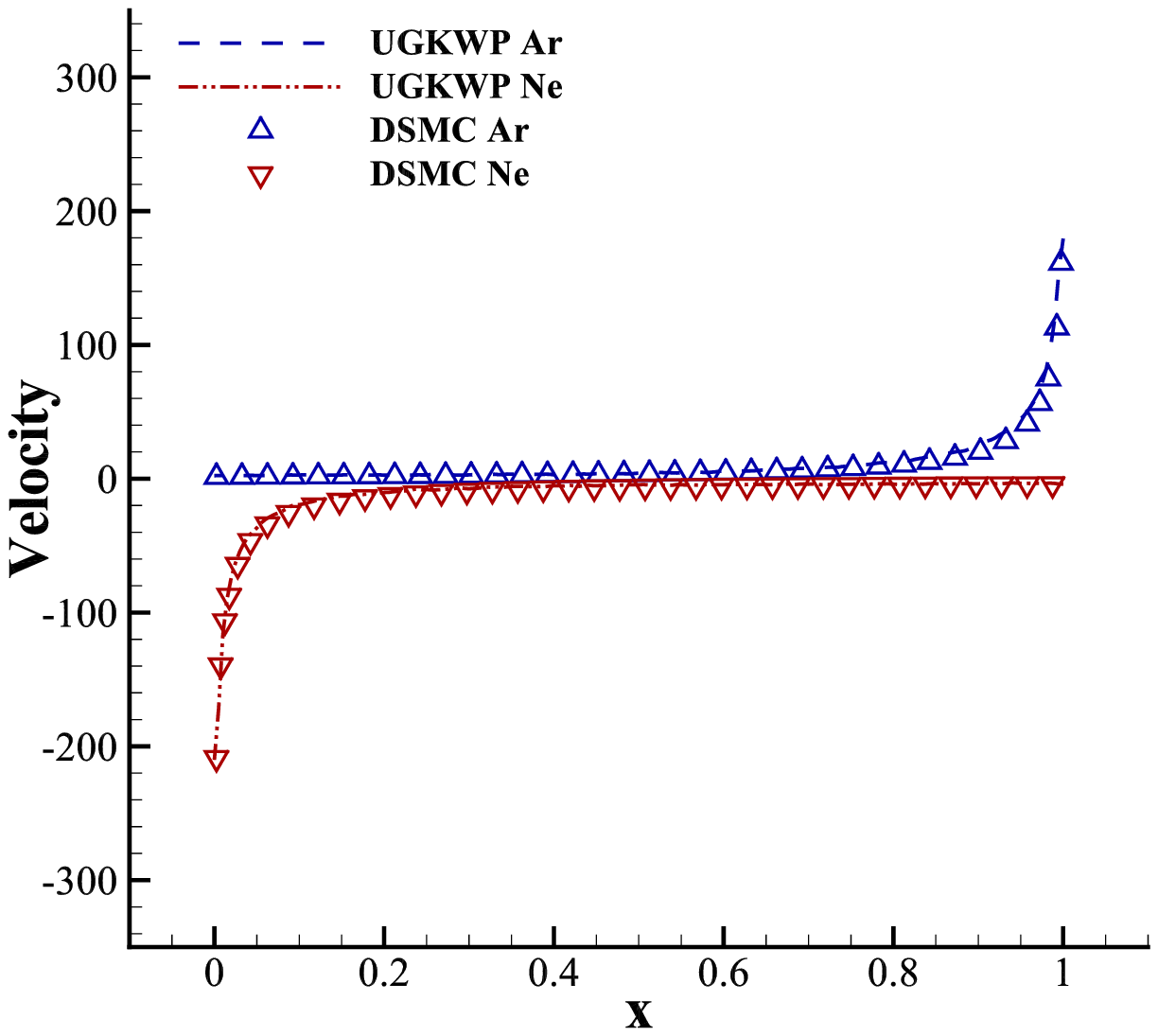}
		}
    \subfigure[]{
    		\includegraphics[width=0.3 \textwidth]{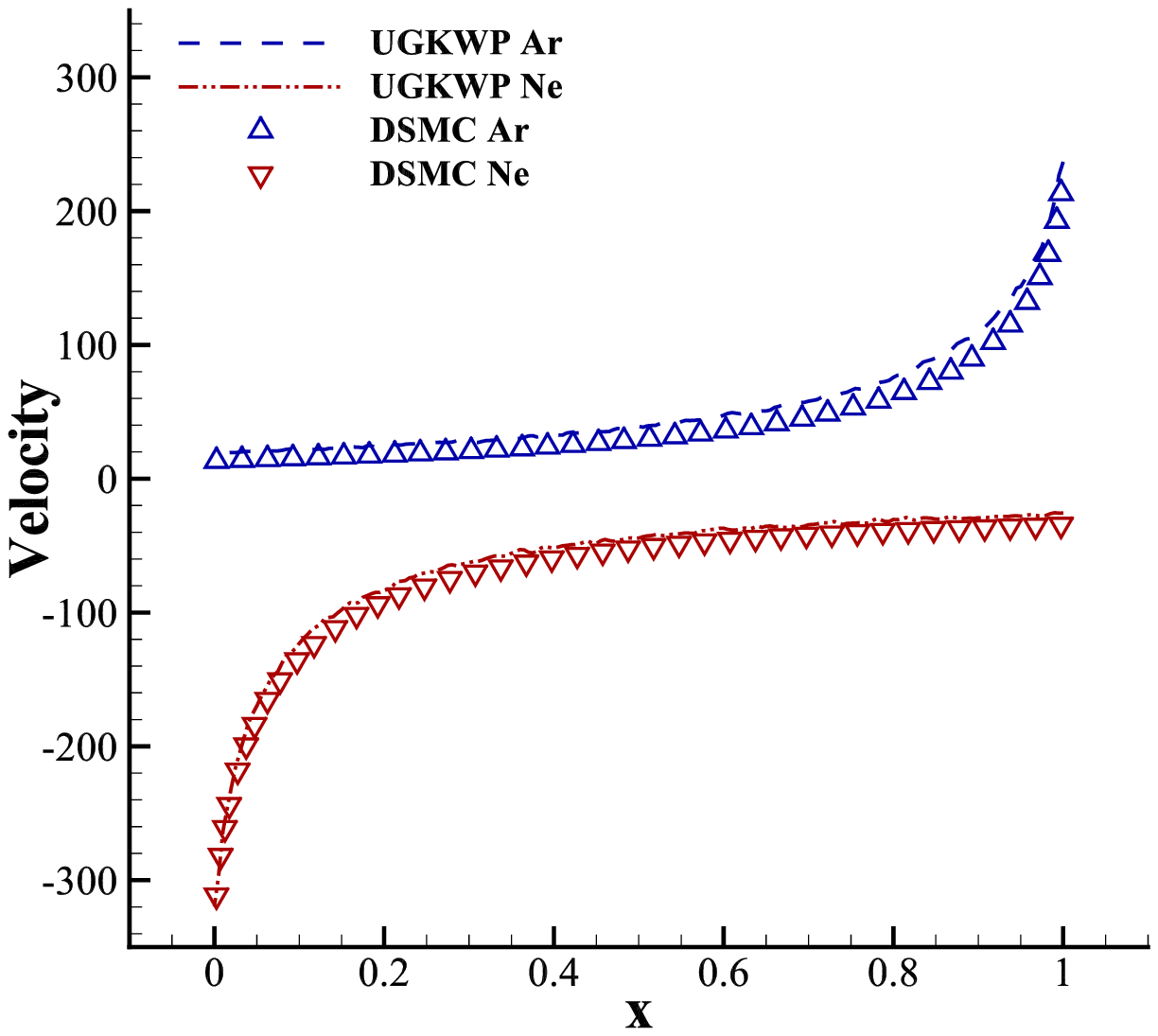}
    	}
    \subfigure[]{
    		\includegraphics[width=0.3 \textwidth]{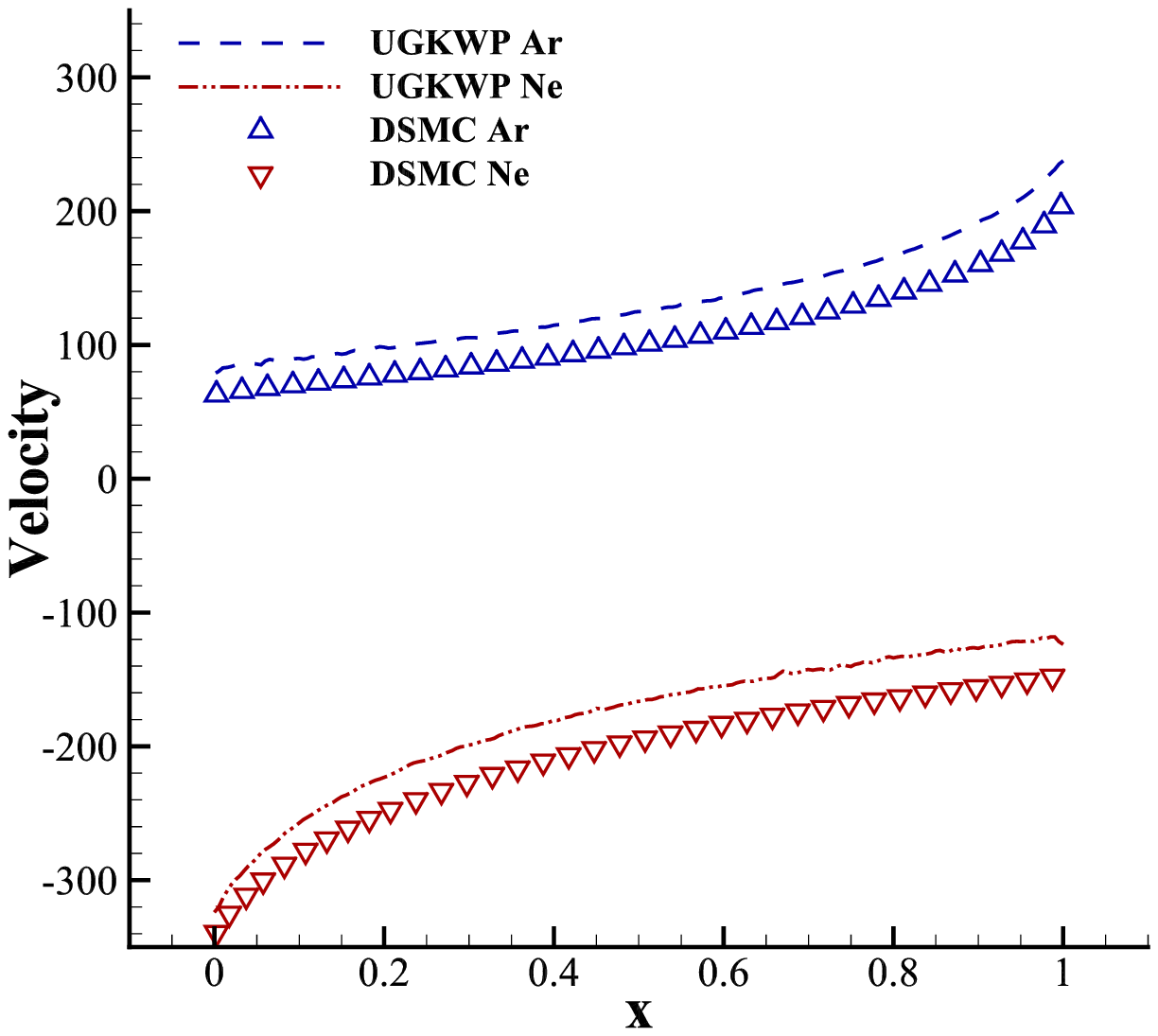}
    	}
	\caption{\label{diffusion-u} Velocity profile of mass diffusion case in an Ar-Ne mixture: (a) ${\rm{Kn}}=0.0061$, (b) ${\rm{Kn}}=0.061$ and (c) ${\rm{Kn}}=0.61$.}
\end{figure}

The mole fraction profile and velocity profile are shown in Fig.~\ref{diffusion-x} and Fig.~\ref{diffusion-u}. Because of $m_{\rm{Ar}}\neq m_{\rm{Ne}}$, the $\chi_{\rm{Ar}}=\chi_{\rm{Ne}}=0.5$ point deviates from $x=0.5$. The ``composition slip'' and ``velocity slip'' at the boundary are more obvious at higher ${\rm{Kn}}$. For comparison, the DSMC results are calculated by DSMC1.FOR code in Ref.\cite{dsmc}. At a lower ${\rm{Kn}}$, the results of proposed UGKWP method match well with those of DSMC because both of them can recover the Navier-Stokes diffusion coefficient in the continuum flow regime. At ${\rm{Kn}}=0.61$, there are deviations between velocity profiles. The reason may be some assumptions in the kinetic model utilized in this study such as $\tau_{\rm{Ar}}=\tau_{\rm{Ne}}$.

\subsection{Couette flow}\label{sec:couette}
\begin{figure}[H]
	\centering
	\subfigure[]{
			\includegraphics[width=0.32 \textwidth]{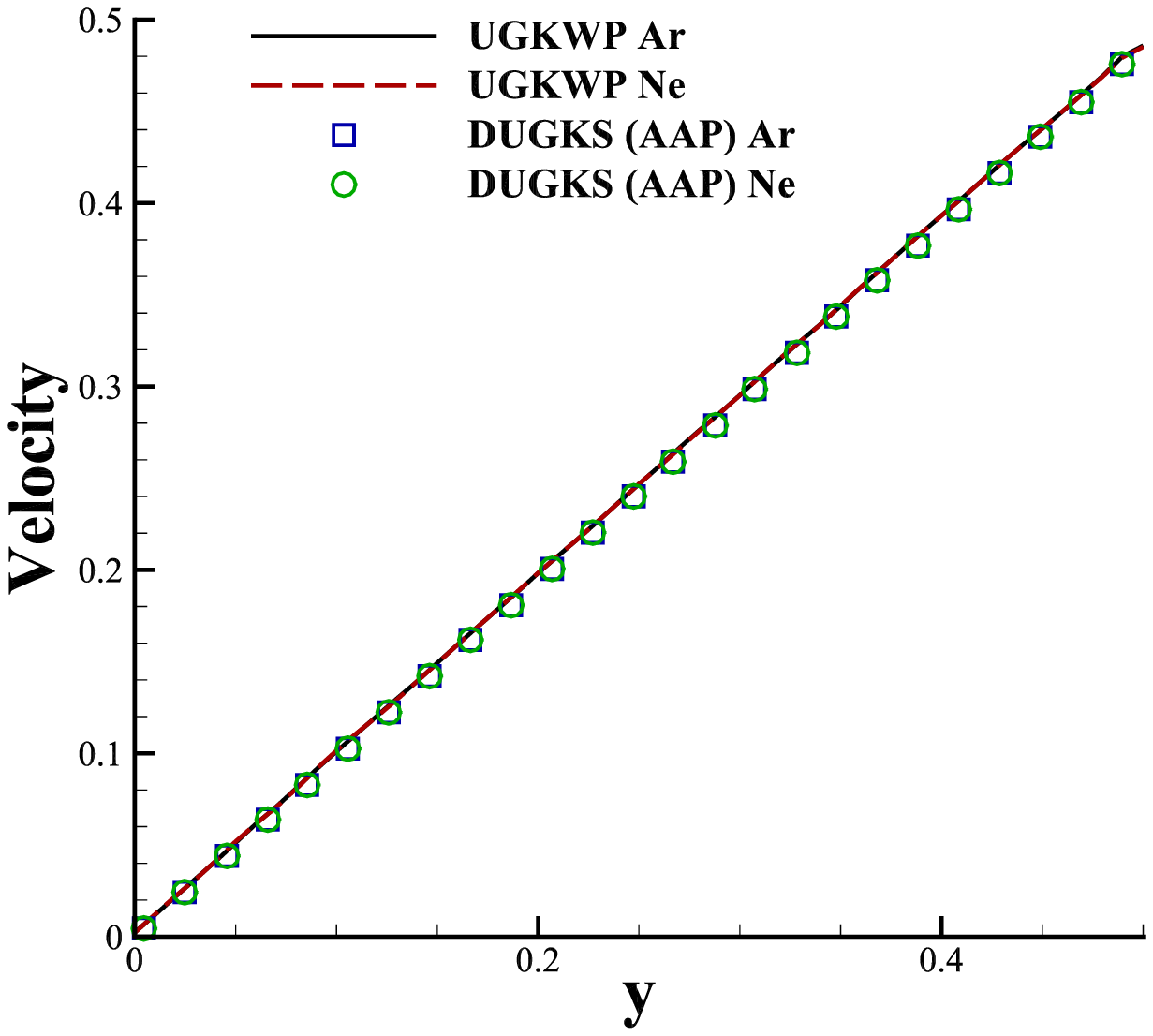}
		}
    \subfigure[]{
    		\includegraphics[width=0.32 \textwidth]{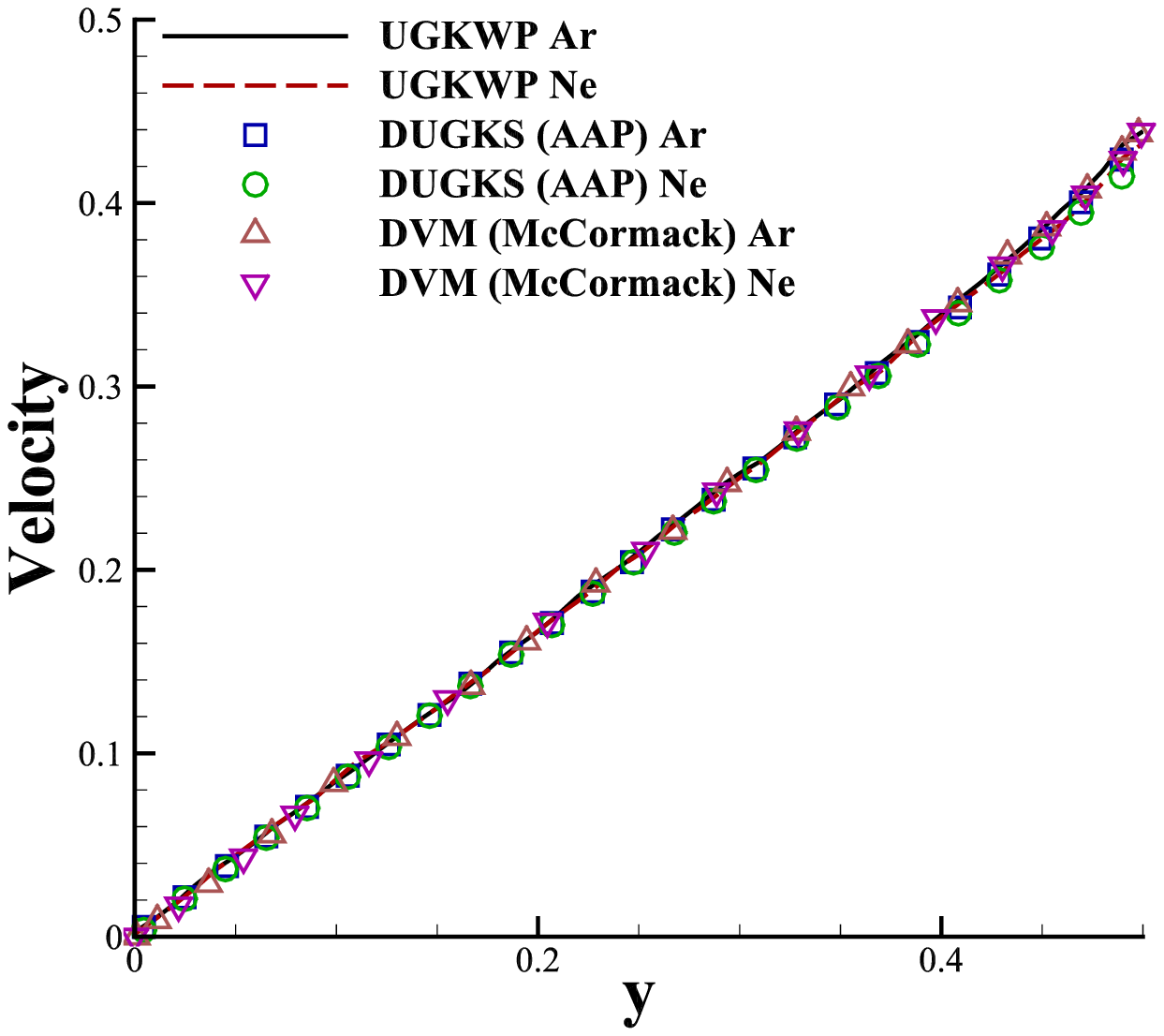}
    	}
    \\
    \subfigure[]{
    		\includegraphics[width=0.32 \textwidth]{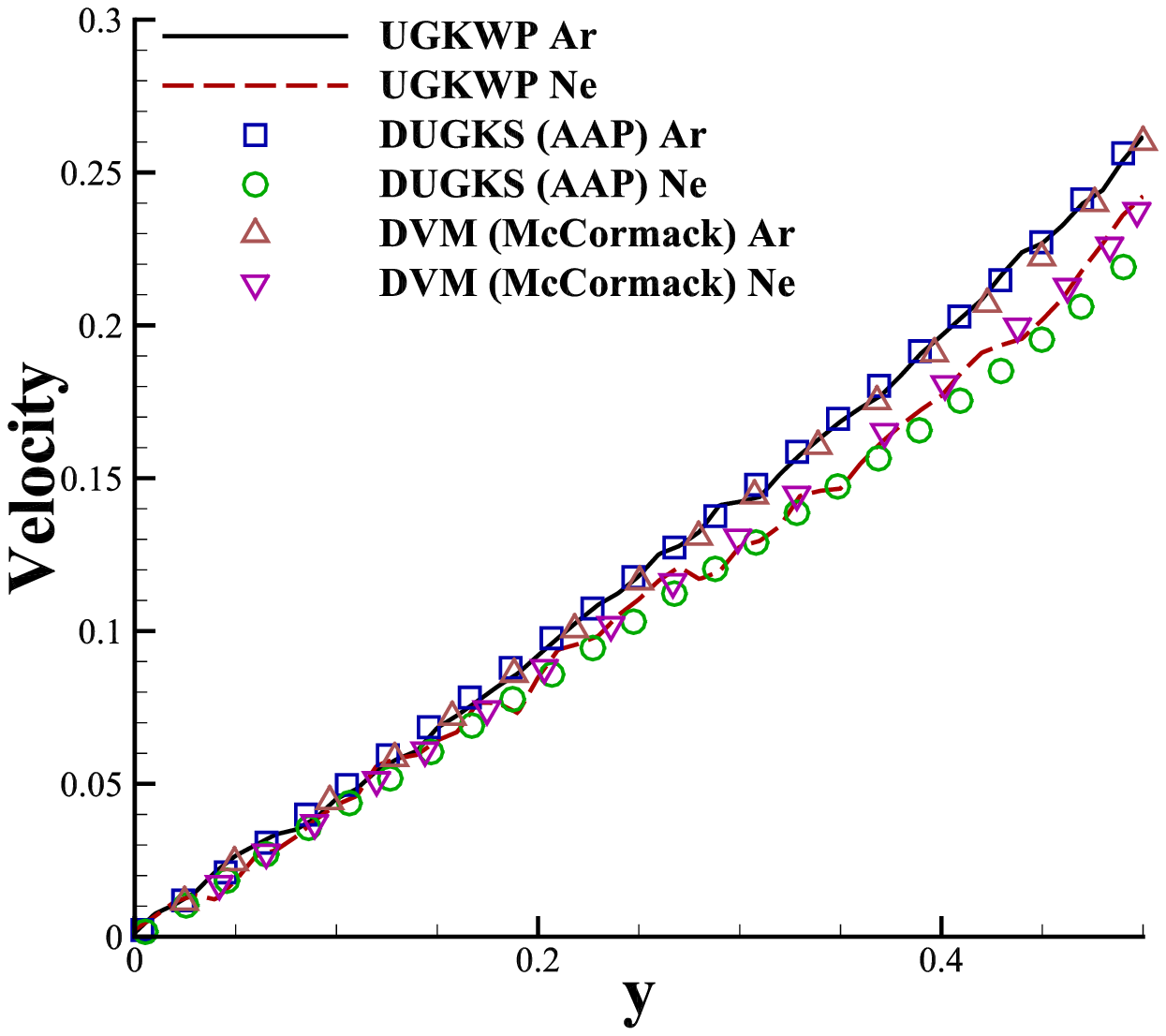}
    	}
    \subfigure[]{
    		\includegraphics[width=0.32 \textwidth]{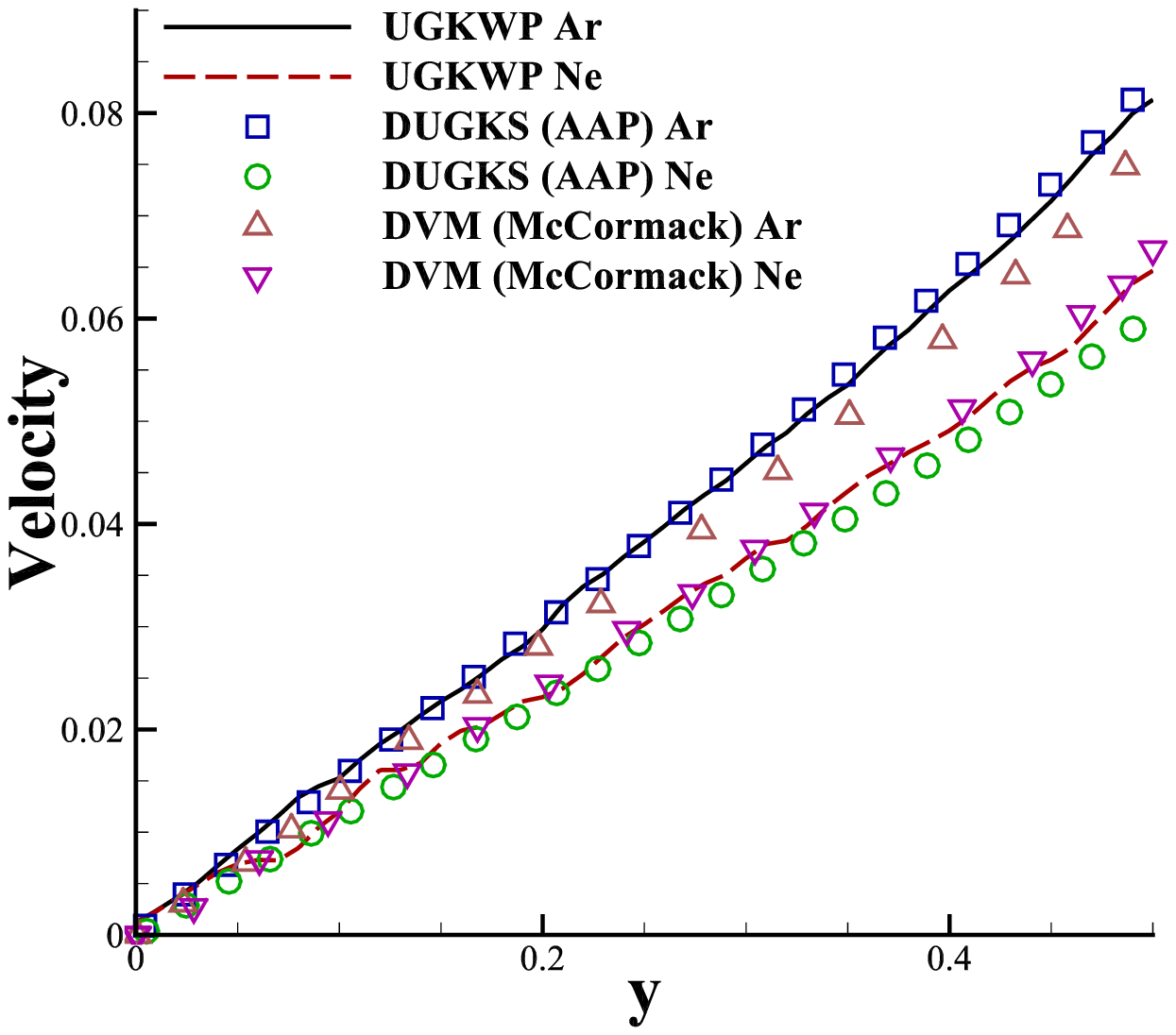}
    	}
	\caption{\label{couette} Velocity profile of Couette flow in an Ar-Ne mixture: (a) $\hat{\delta}_0=100$, (b) $\hat{\delta}_0=10$, (c) $\hat{\delta}_0=1$ and (d) $\hat{\delta}_0=0.1$.}
\end{figure}

The Couette flow between two parallel plates with full diffusion boundary condition and spatial interval $H=1{\rm{m}}$ is simulated in this section. The velocity of top plate is set to be $\left(U_w,0,0\right)^T$ and that of bottom plate is set to be $\left(-U_w,0,0\right)^T$. In the $x$ direction, the periodic boundary condition is applied. For the initial condition, the mole fractions of species are both set to be $0.5$, and their temperature is set to be the same as the wall $T_w$. Ar and Ne are employed in this case and their properties can be referred to Tab.~\ref{arne}, except that $\mu_{\rm{Ar},\rm{ref}}$ and $\mu_{\rm{Ne},\rm{ref}}$ are set to be $2.239\times10^{-5}{\rm{Pa\cdot s}}$ and $3.160\times10^{-5}{\rm{Pa\cdot s}}$ respectively at $T_w$ for comparison with Ref.\cite{dugks-aap2}. Hereby, as Ref.\cite{dugks-aap2}, a characteristic molecular velocity is defined as,
\begin{equation}
\begin{aligned}
v_0=\sqrt{\frac{2k_BT_w}{m_0}},
\nonumber
\end{aligned}
\end{equation}
and a gas rarefaction parameter $\hat{\delta}_0$ is defined as,
\begin{equation}
\begin{aligned}
\hat{\delta}_0=\frac{n_0k_BT_wH}{\mu_0v_0}.
\nonumber
\end{aligned}
\end{equation}
Conditions as $\hat{\delta}_0=100,10,1,0.1$ are considered from continuum to rarefied flow regime, and $U_w$ is set to be $0.1v_0$. The results are shown in Fig.~\ref{couette}, in comparison with results of AAP model-based Discrete UGKS (DUGKS)~\cite{dugks-aap2} and McCormack model-based DVM~\cite{mcc-couette}. The vertical coordinate is normalized by $2U_w$. When the flow gets rarefied, both of diffusion velocity within the flow field and slip velocity at the wall boundary enlarge. The deviation also increases among different methods and models. The velocity profile of the proposed UGKWP method lies in the middle of others.

\subsection{Hypersonic flow around a cylinder}\label{sec:cylinder}
\begin{figure}[H]
	\centering
	\subfigure[]{
			\includegraphics[width=0.22 \textwidth]{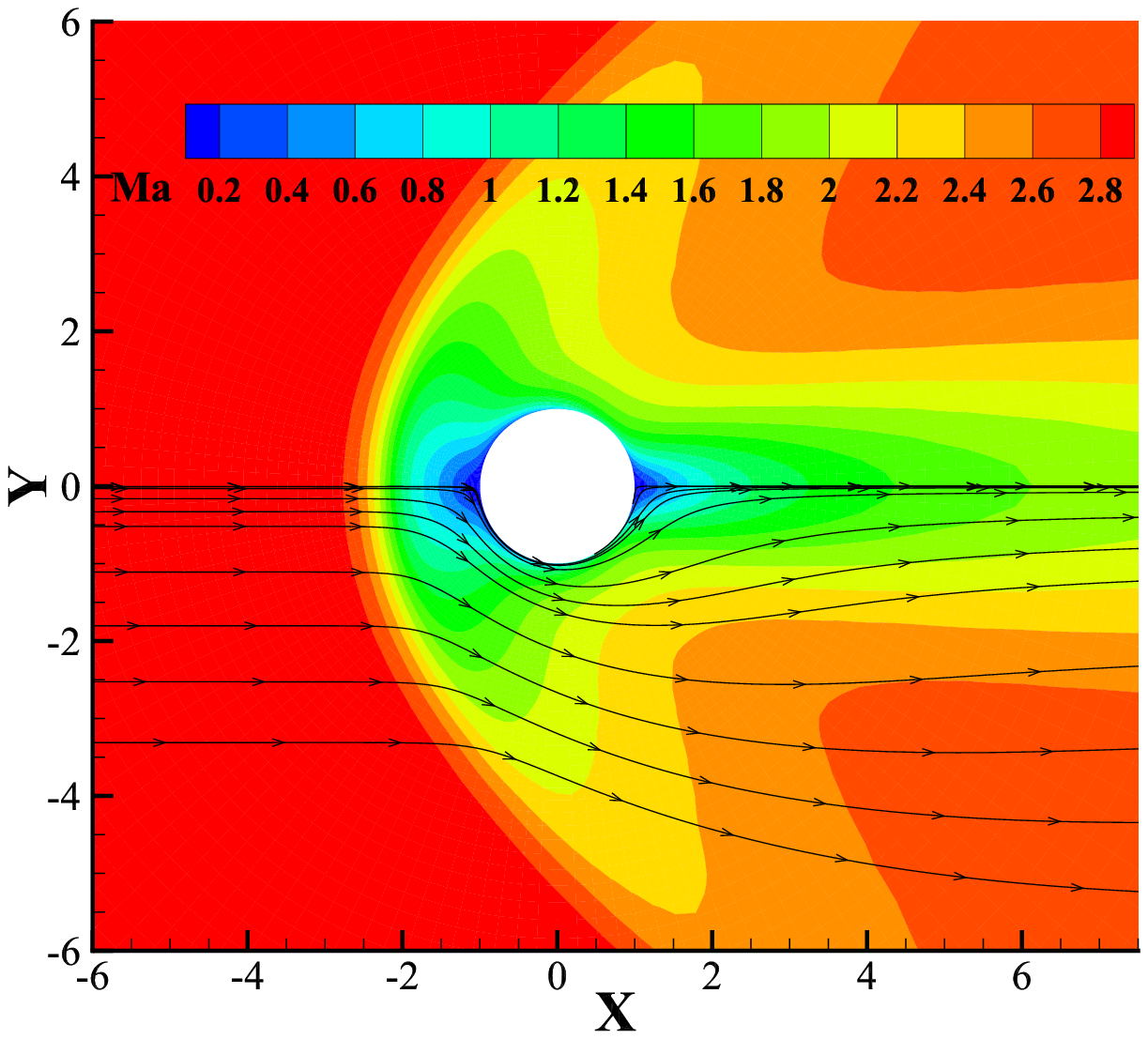}
		}
    \subfigure[]{
    		\includegraphics[width=0.22 \textwidth]{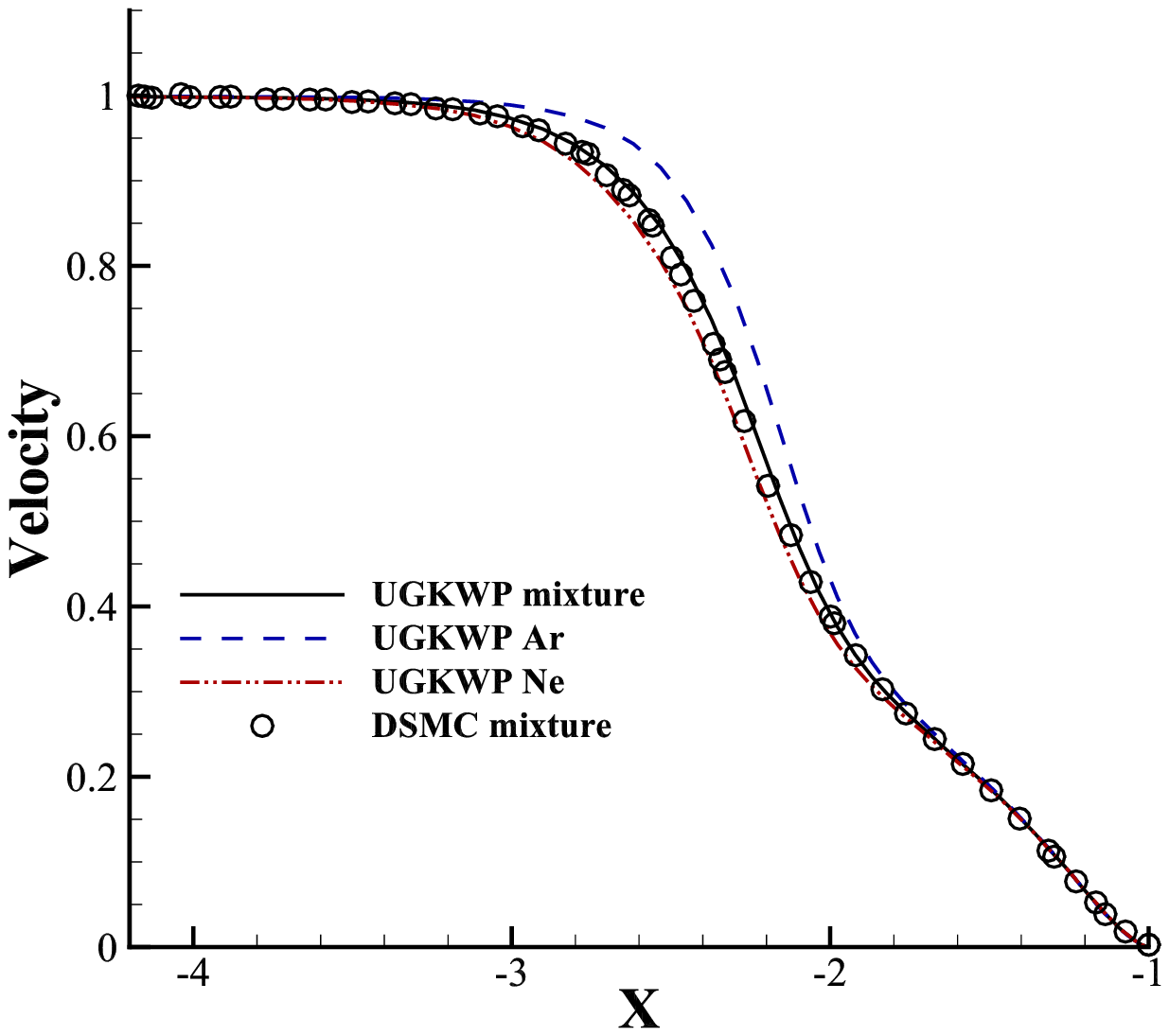}
    	}
    \subfigure[]{
    		\includegraphics[width=0.22 \textwidth]{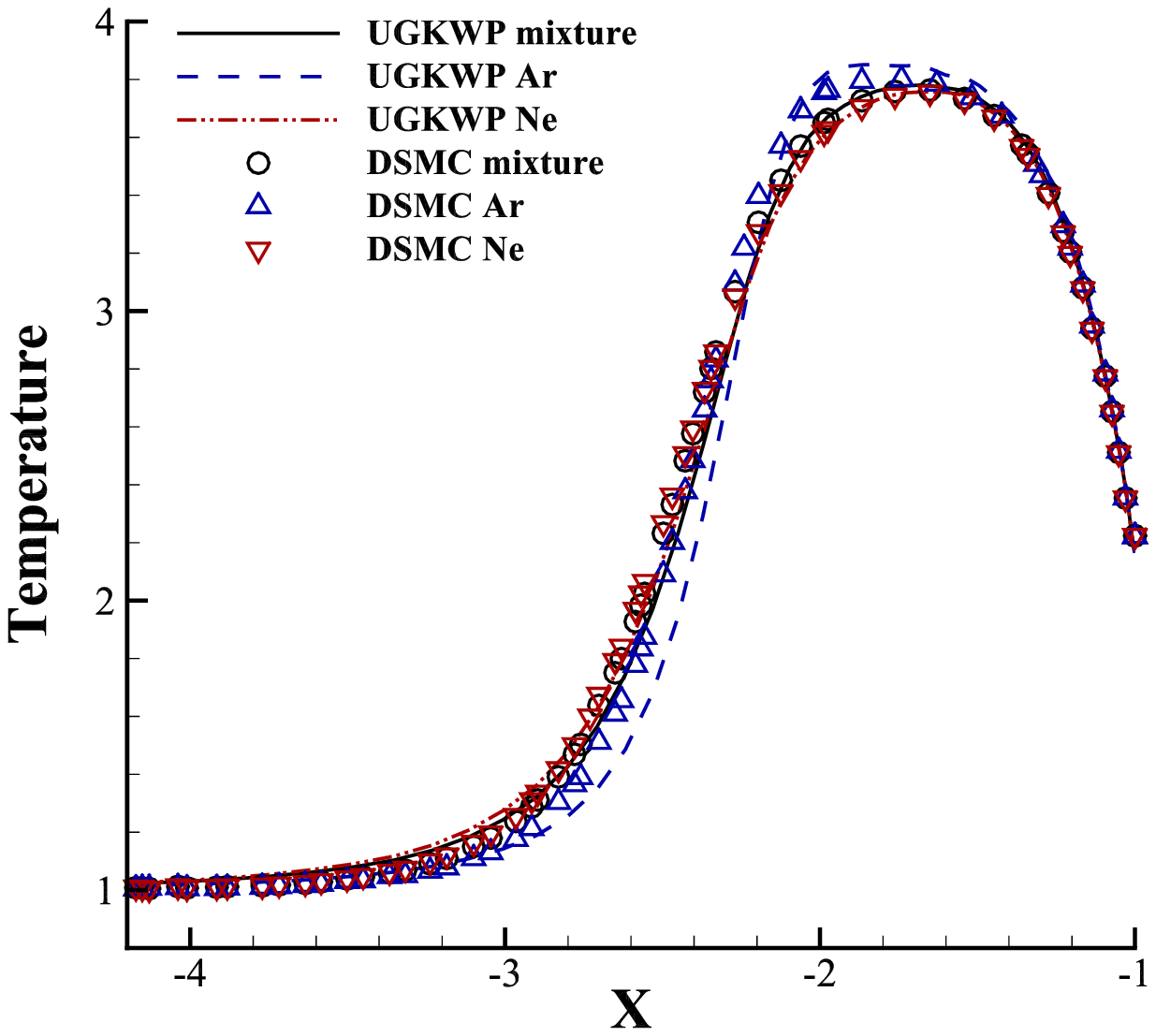}
    	}
    \\
    \subfigure[]{
			\includegraphics[width=0.22 \textwidth]{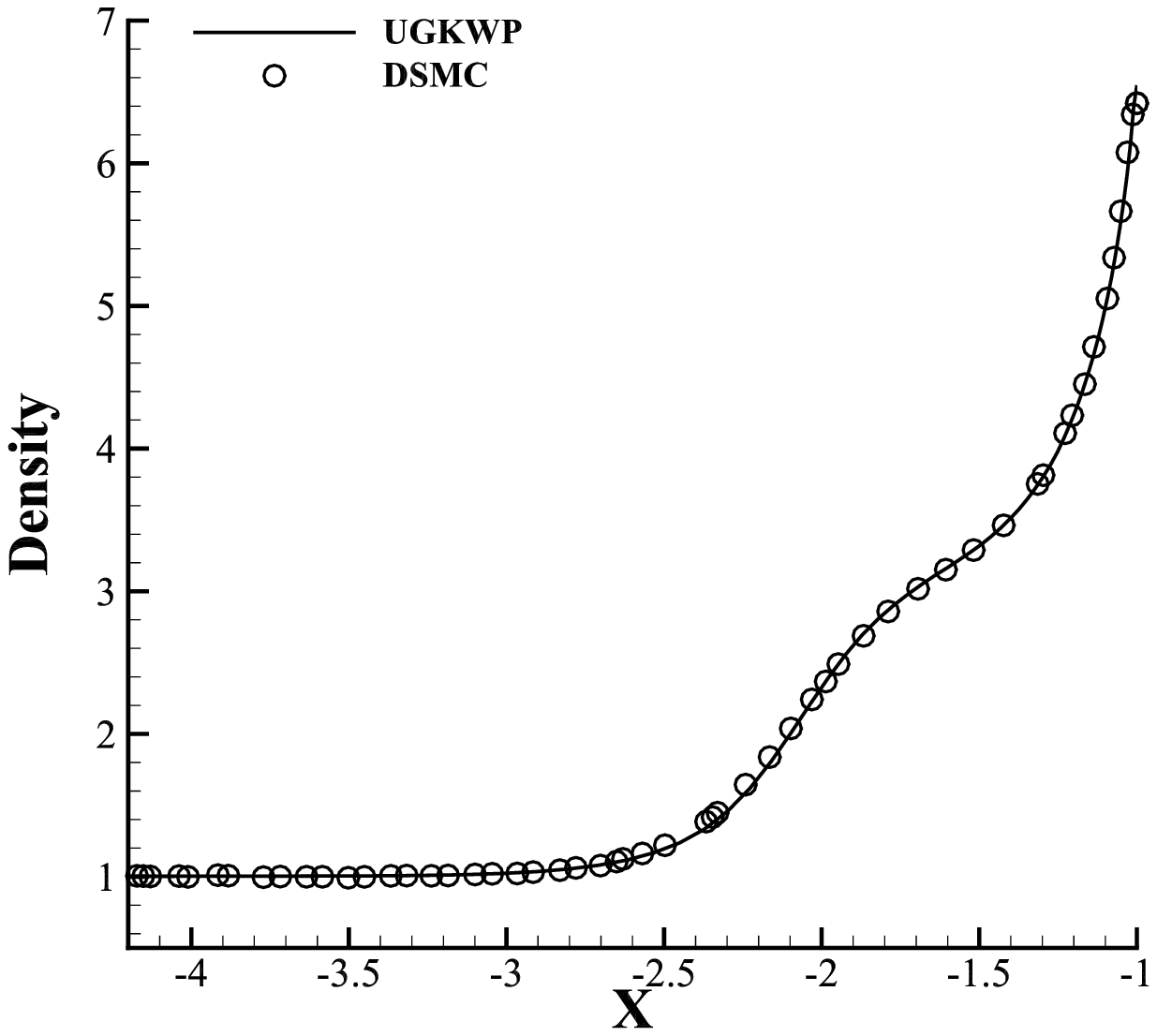}
		}
    \subfigure[]{
    		\includegraphics[width=0.22 \textwidth]{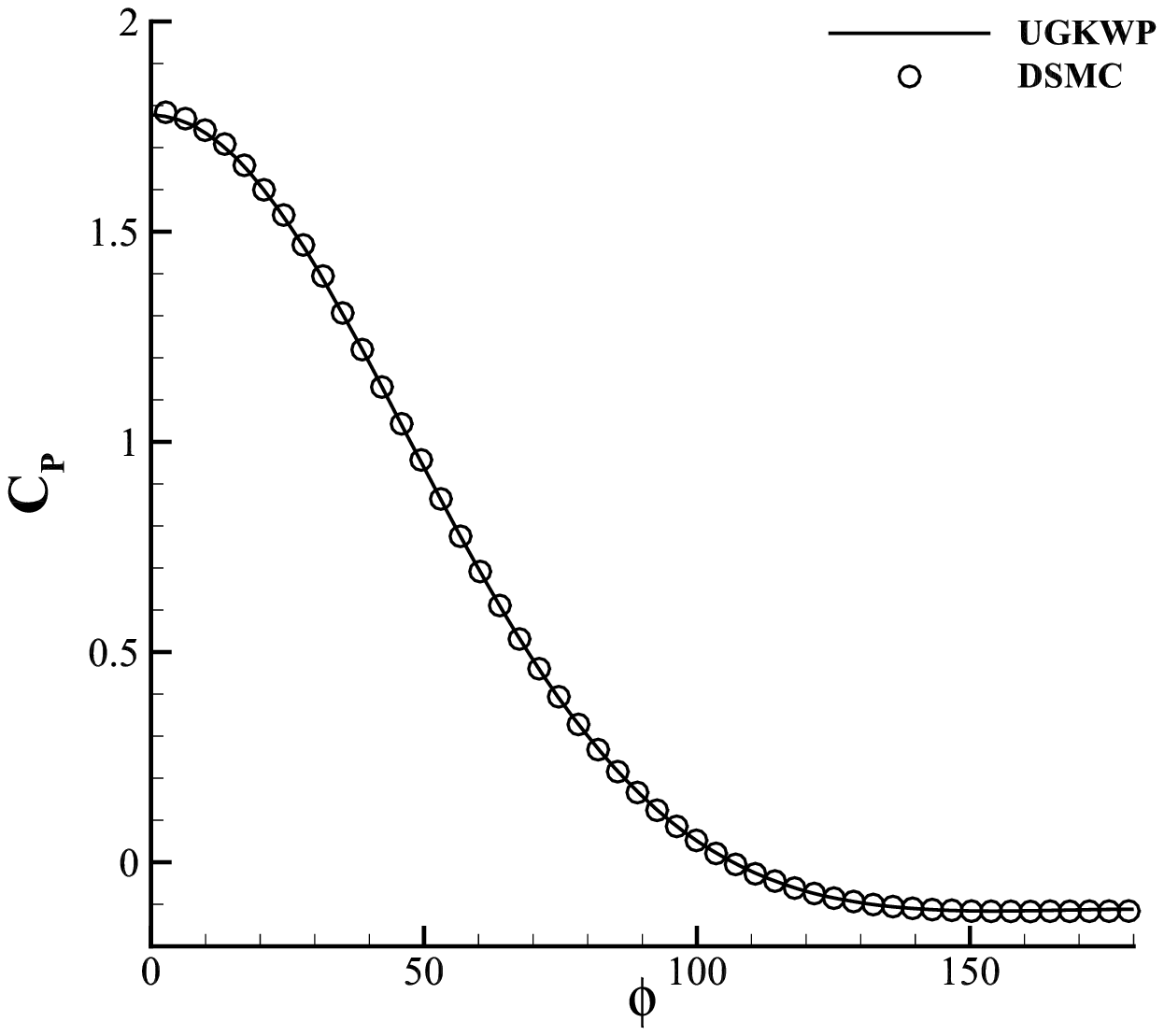}
    	}
    \subfigure[]{
    		\includegraphics[width=0.22 \textwidth]{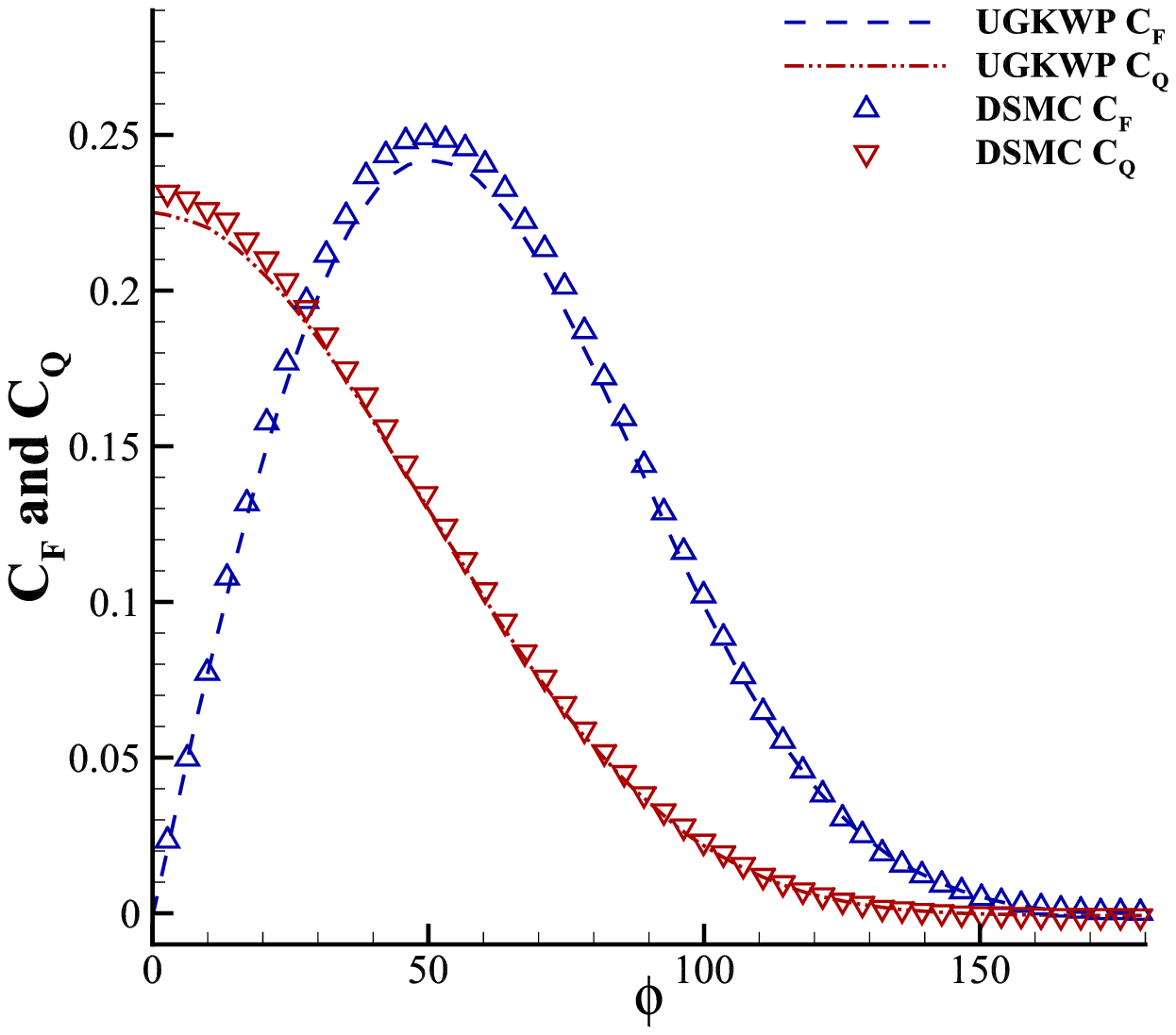}
    	}
	\caption{\label{ma3kn0.1} Hypersonic flow around a cylinder in an Ar-Ne mixture at ${\rm{Ma}}_{\infty}=3$, ${\rm{Kn}}_{\infty}=0.1$: (a) ${\rm{Ma}}$ contour and streamline, (b) velocity along the stagnation line, (c) temperature along the stagnation line, (d) density along the stagnation line, (e) pressure coefficient at the wall, (f) shear stress and heat flux coefficients at the wall.}
\end{figure}

\begin{figure}[H]
	\centering
	\subfigure[]{
			\includegraphics[width=0.22 \textwidth]{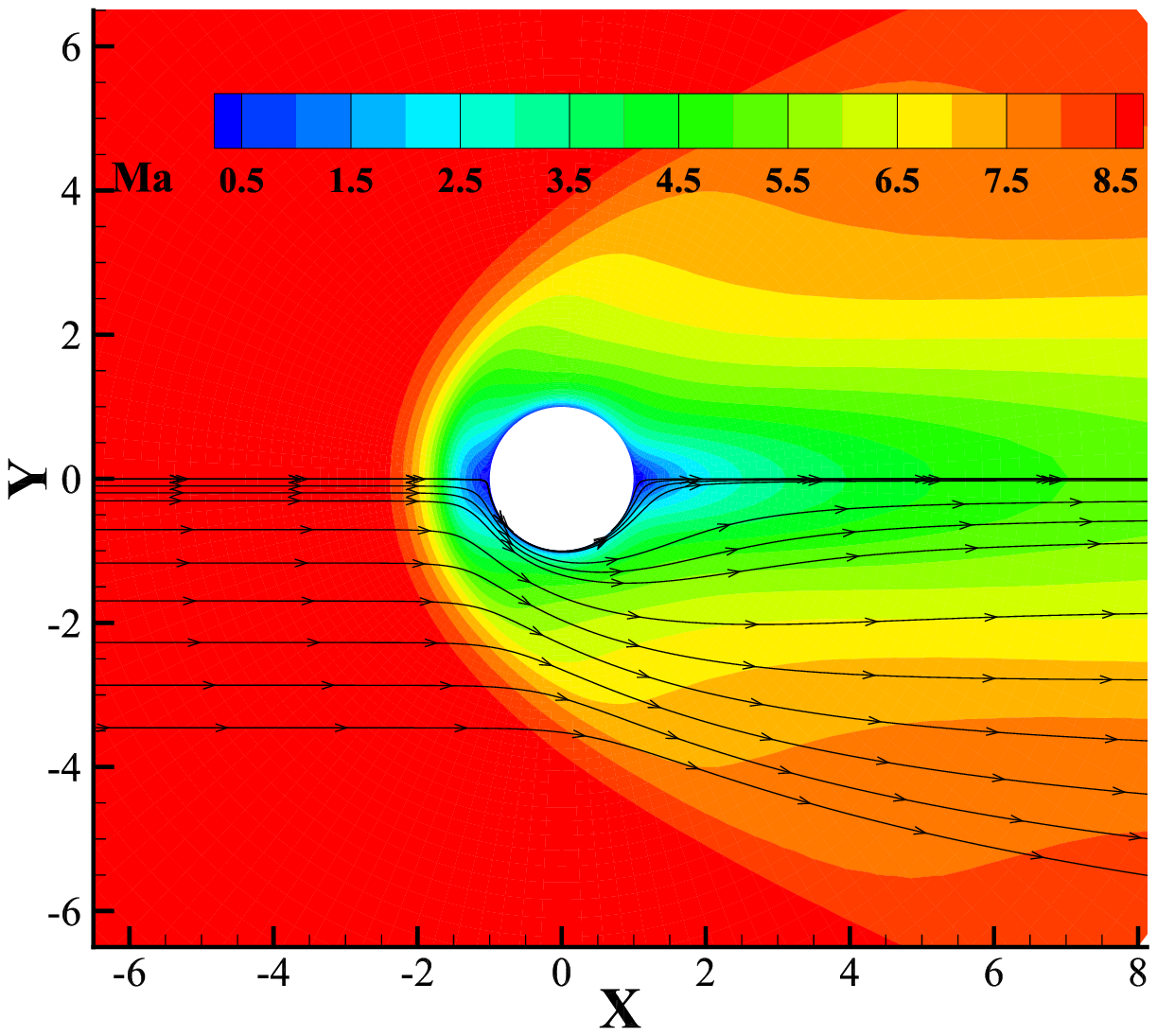}
		}
    \subfigure[]{
    		\includegraphics[width=0.22 \textwidth]{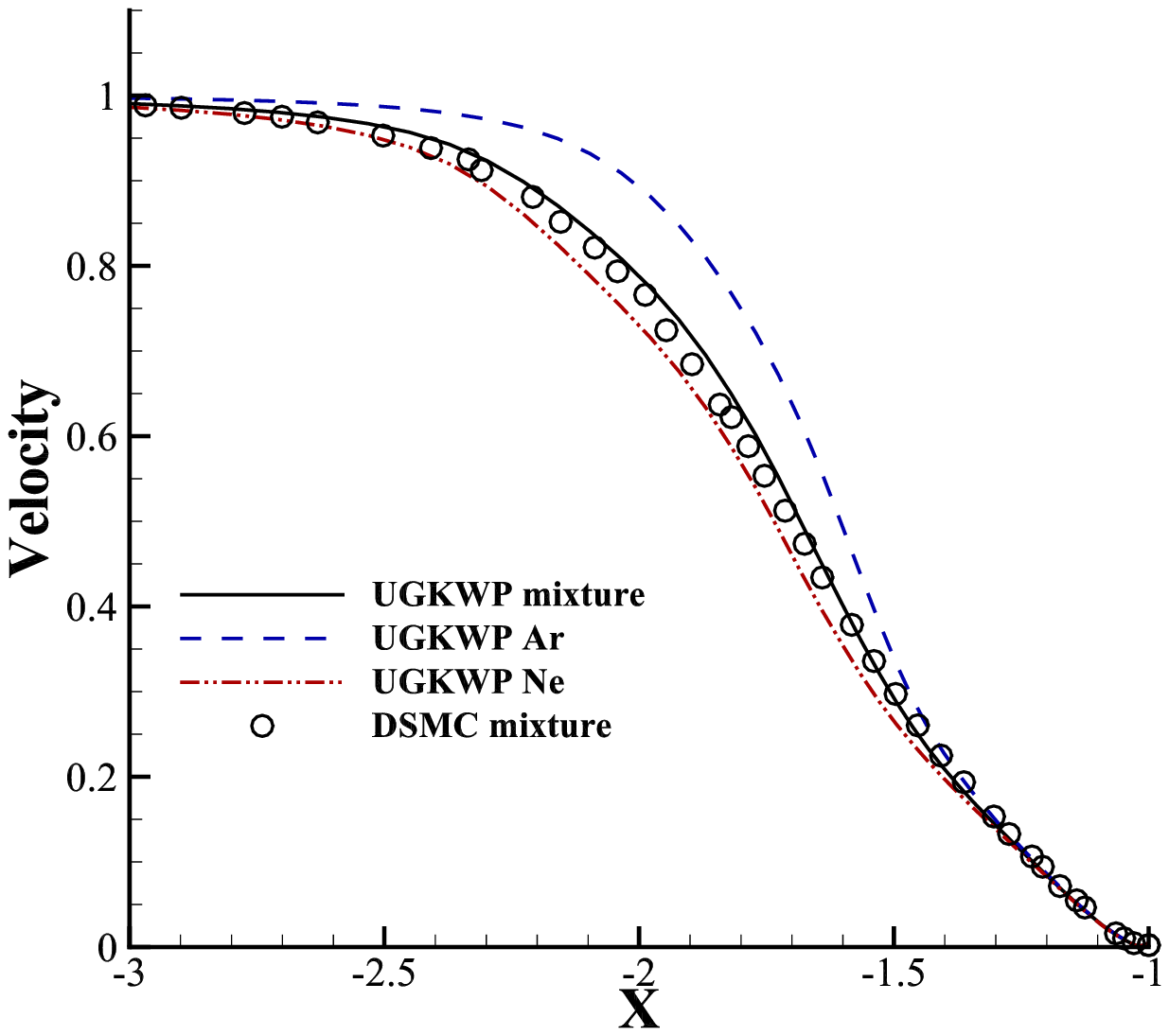}
    	}
    \subfigure[]{
    		\includegraphics[width=0.22 \textwidth]{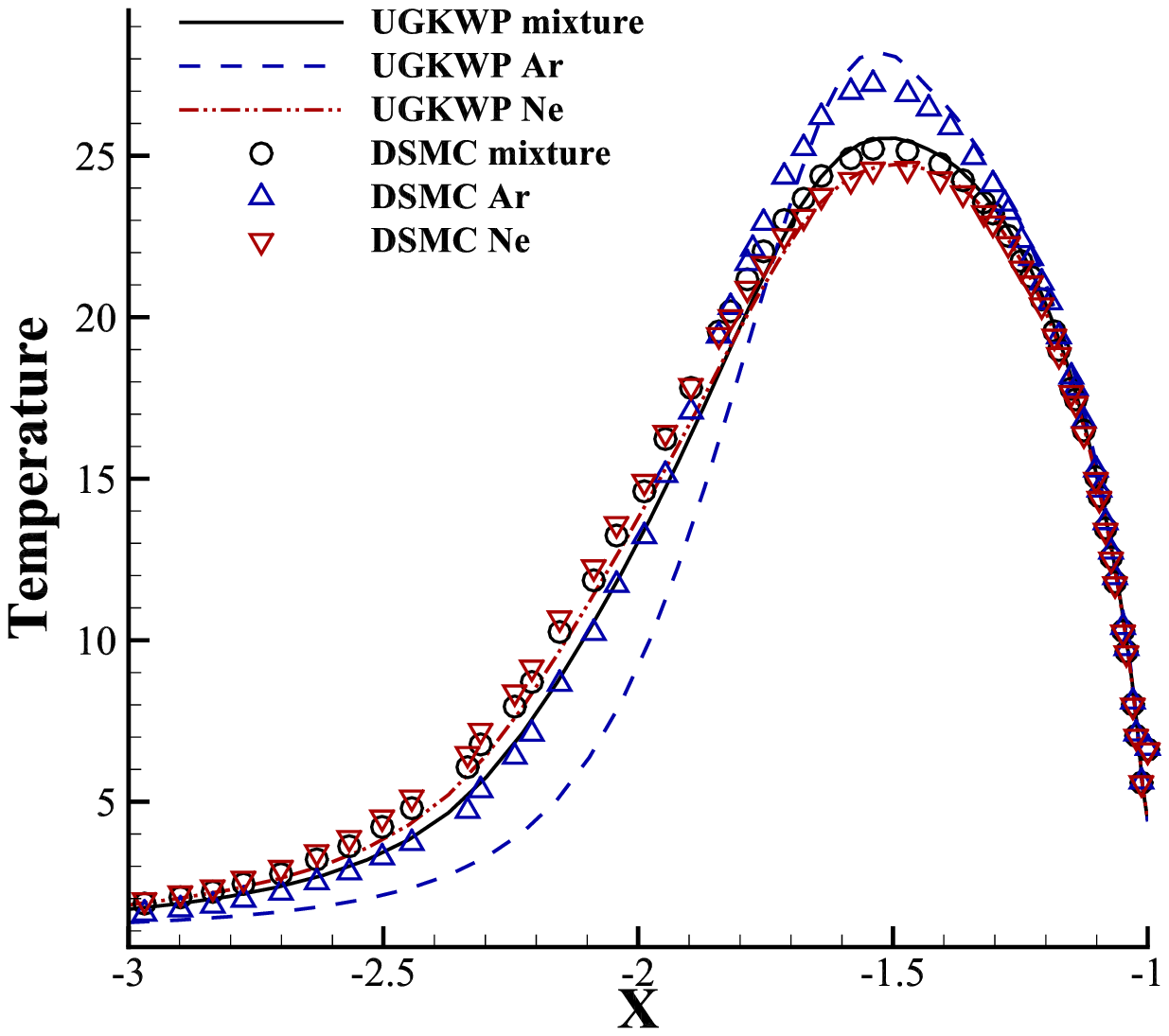}
    	}
    \\
    \subfigure[]{
			\includegraphics[width=0.22 \textwidth]{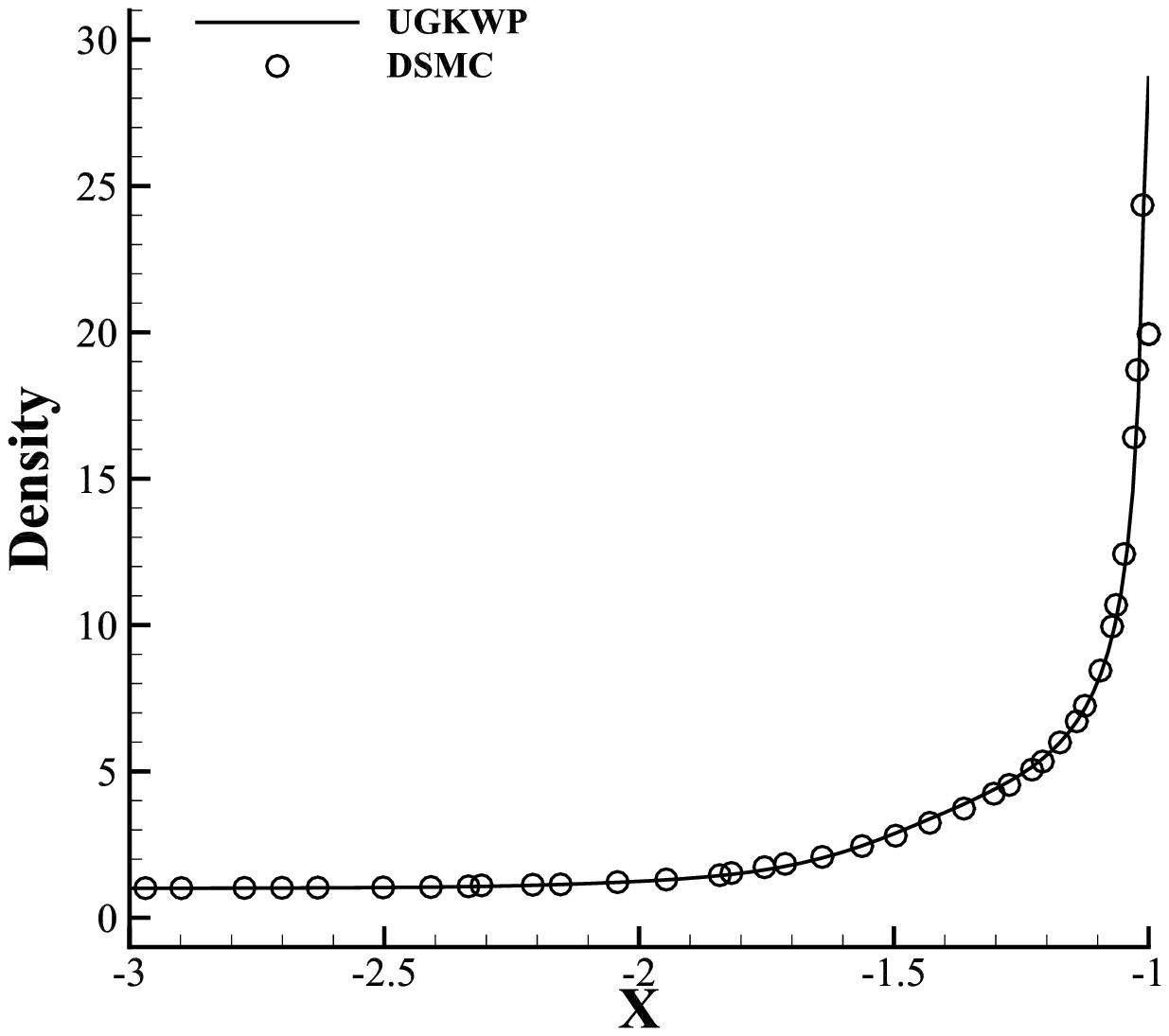}
		}
    \subfigure[]{
    		\includegraphics[width=0.22 \textwidth]{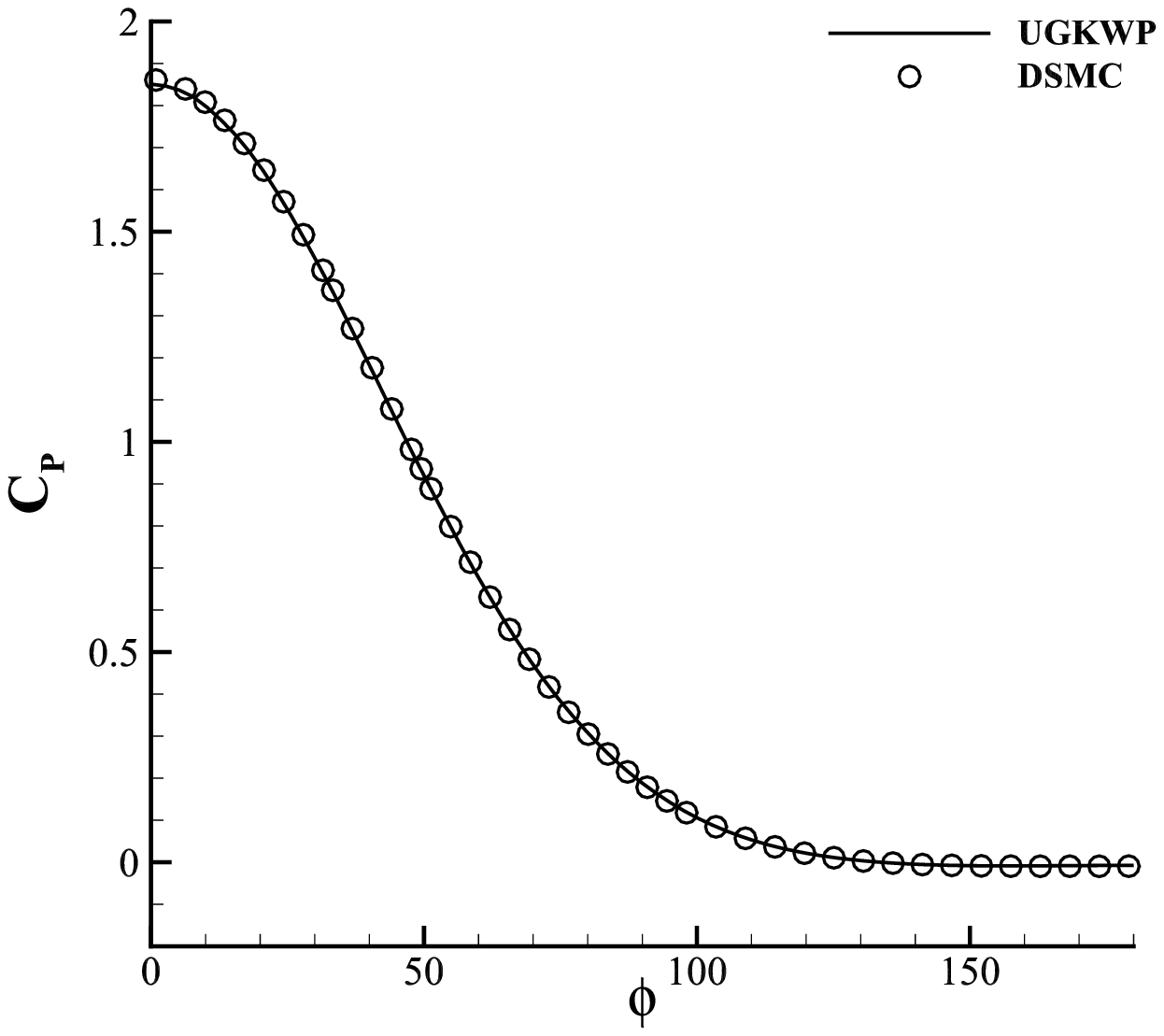}
    	}
    \subfigure[]{
    		\includegraphics[width=0.22 \textwidth]{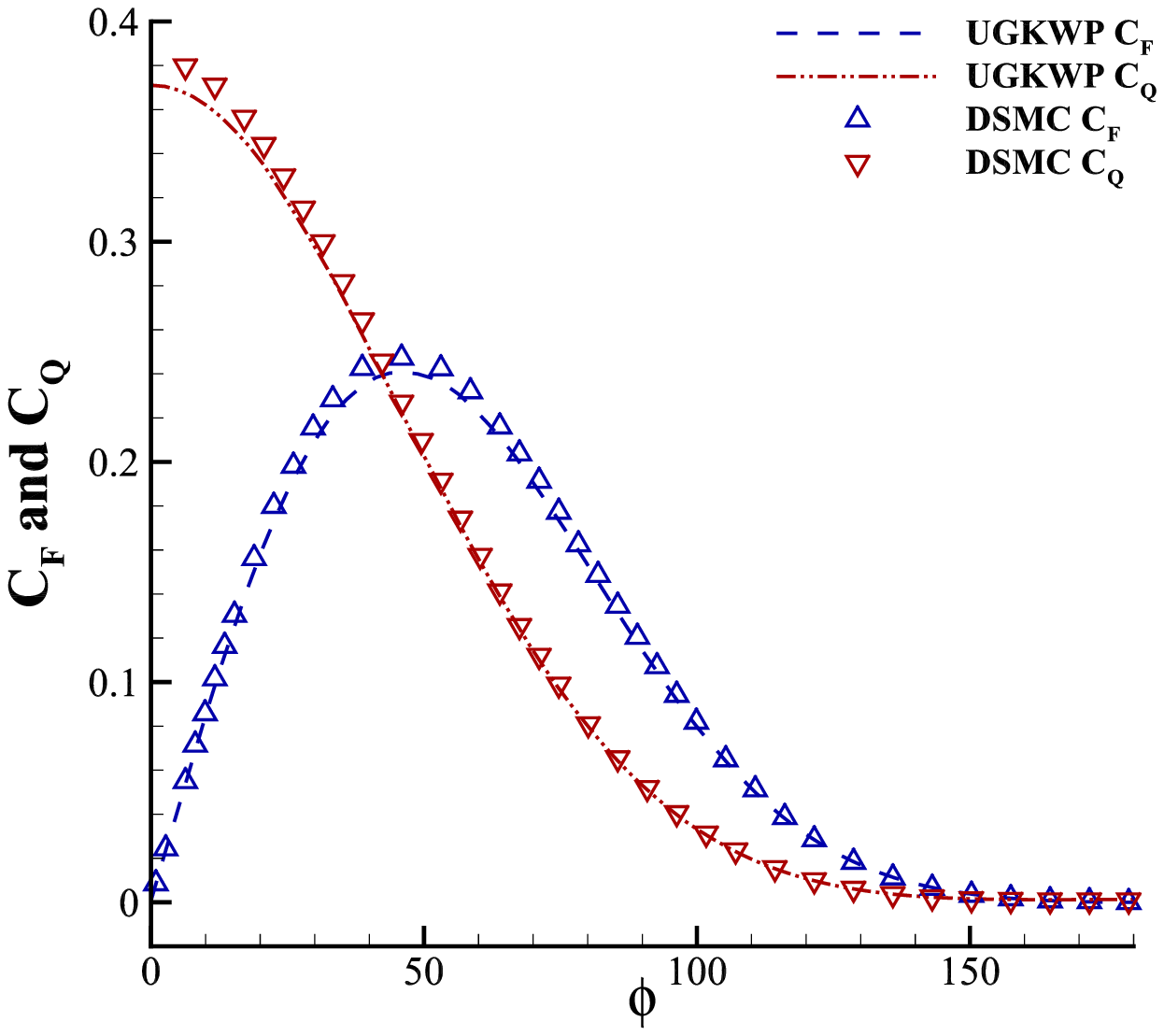}
    	}
	\caption{\label{ma9kn0.1} Hypersonic flow around a cylinder in an Ar-Ne mixture at ${\rm{Ma}}_{\infty}=9$, ${\rm{Kn}}_{\infty}=0.1$: (a) ${\rm{Ma}}$ contour and streamline, (b) velocity along the stagnation line, (c) temperature along the stagnation line, (d) density along the stagnation line, (e) pressure coefficient at the wall, (f) shear stress and heat flux coefficients at the wall.}
\end{figure}

\begin{figure}[H]
	\centering
	\subfigure[]{
			\includegraphics[width=0.22 \textwidth]{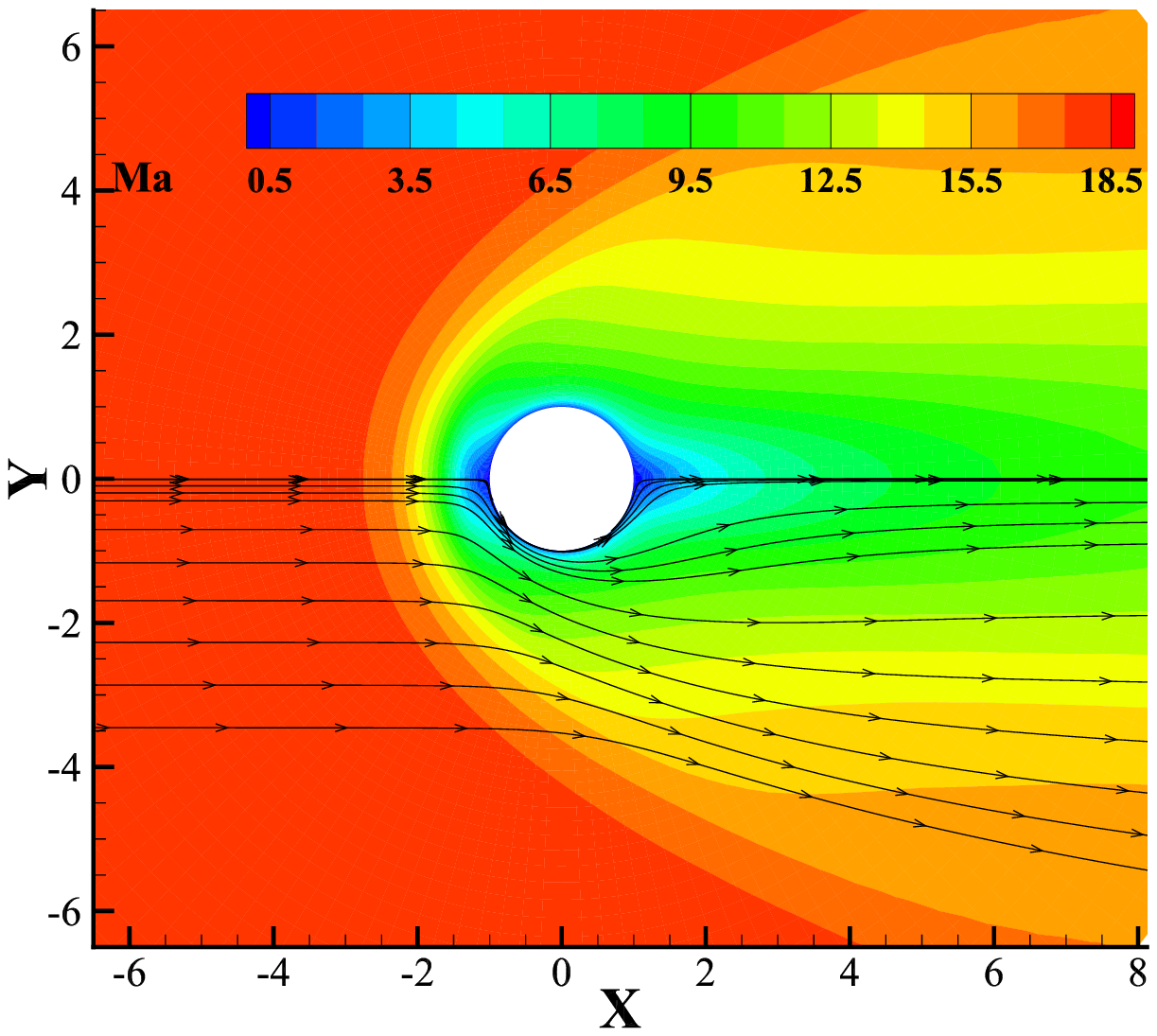}
		}
    \subfigure[]{
    		\includegraphics[width=0.22 \textwidth]{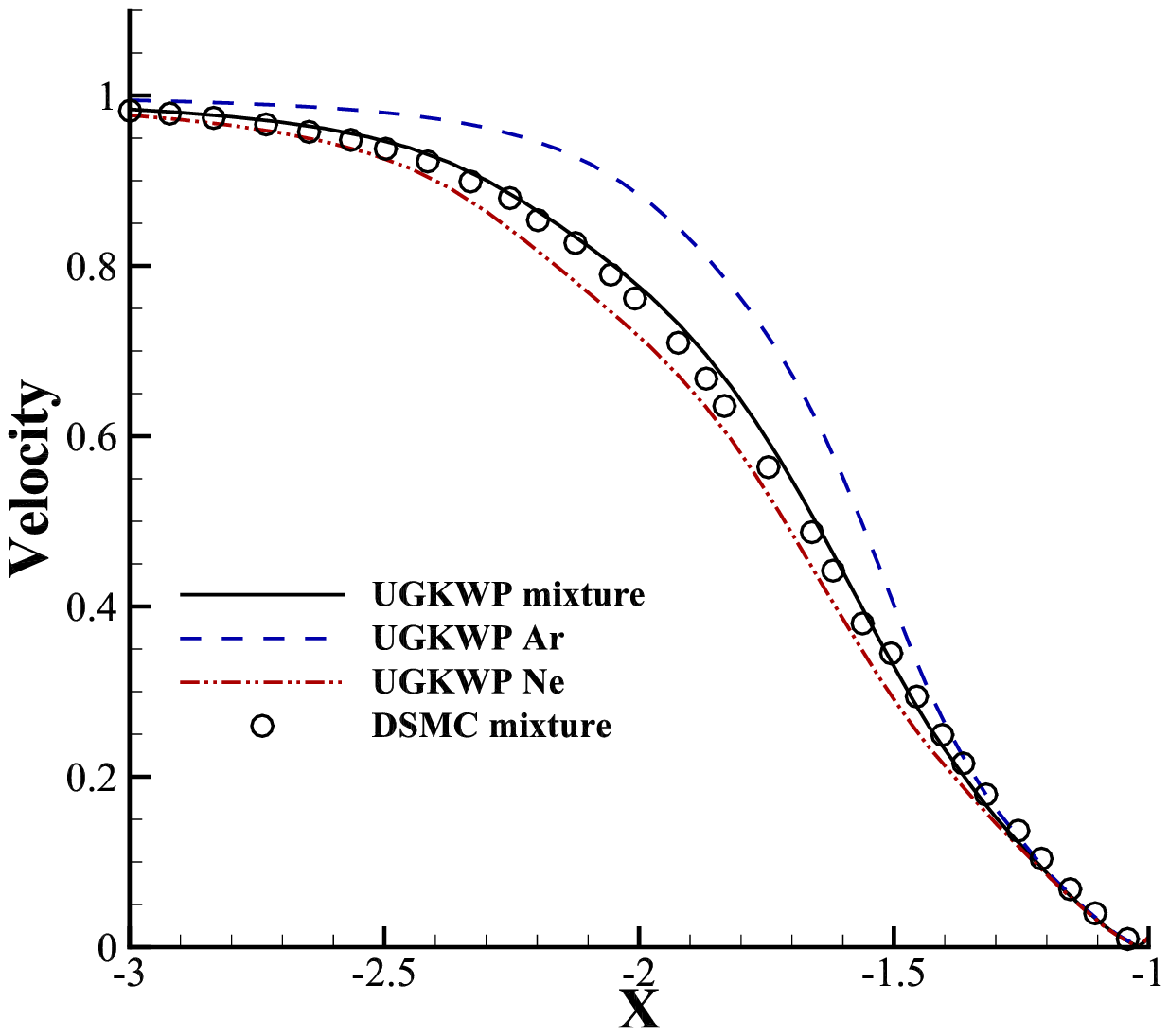}
    	}
    \subfigure[]{
    		\includegraphics[width=0.22 \textwidth]{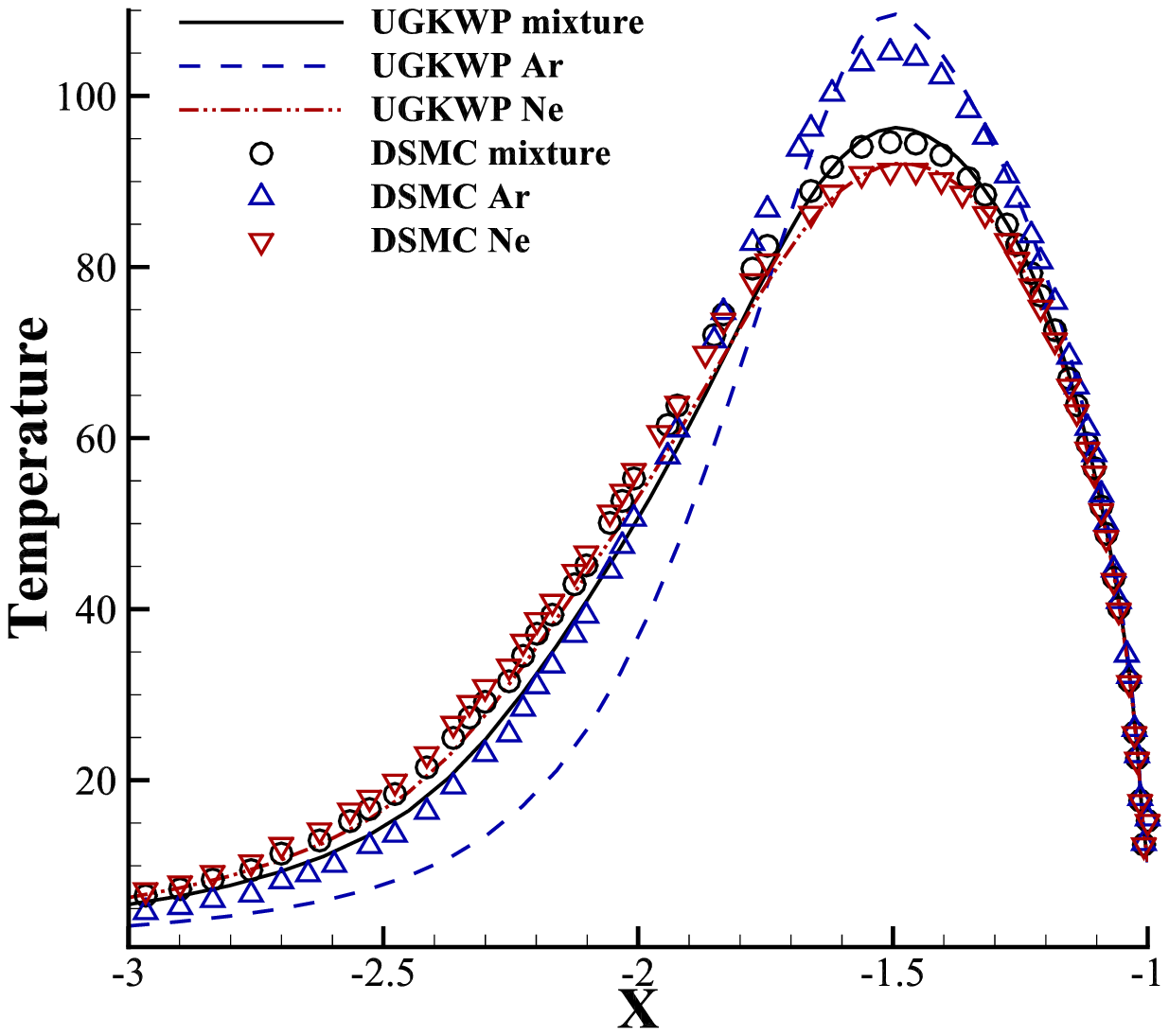}
    	}
    \\
    \subfigure[]{
			\includegraphics[width=0.22 \textwidth]{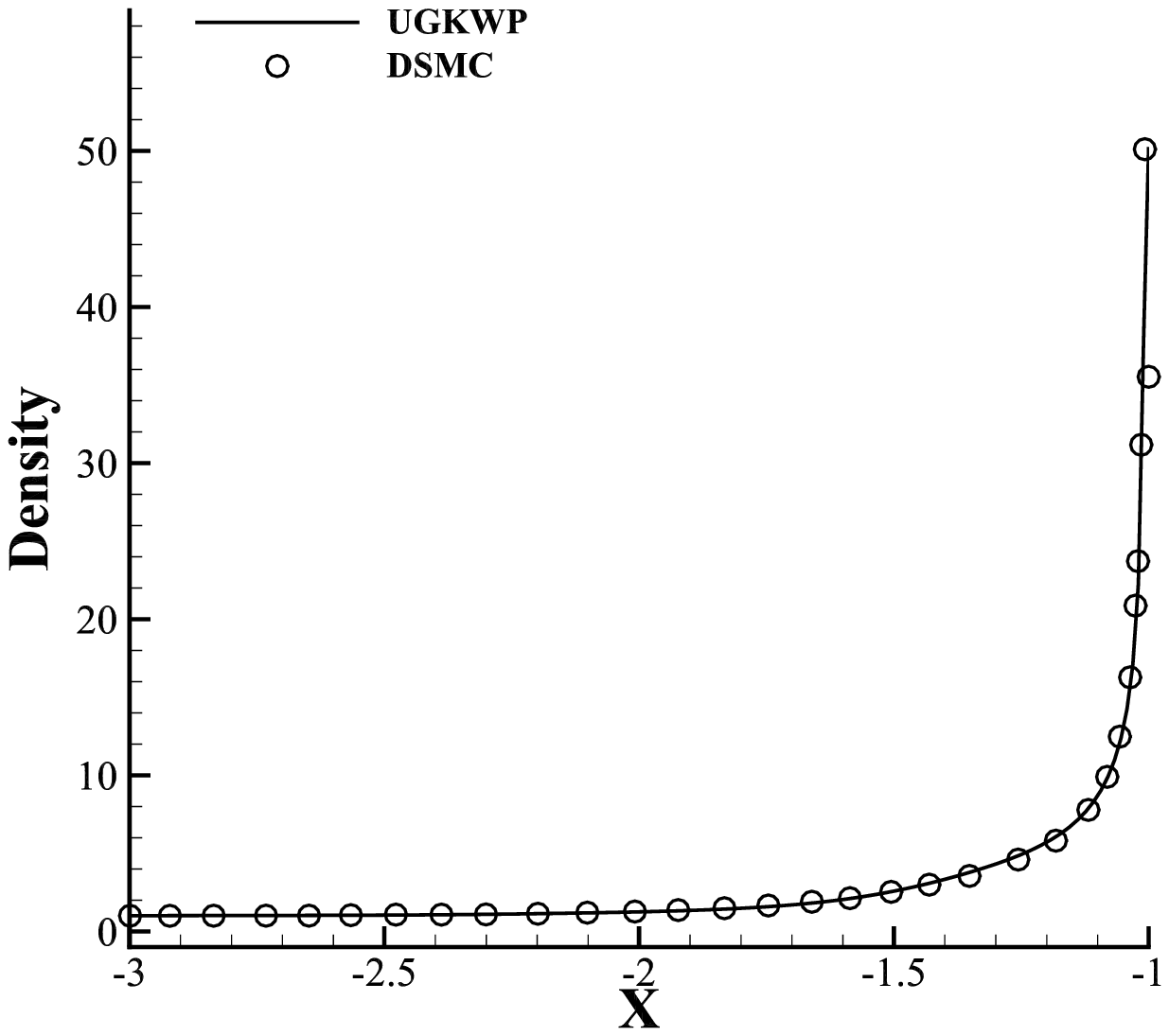}
		}
    \subfigure[]{
    		\includegraphics[width=0.22 \textwidth]{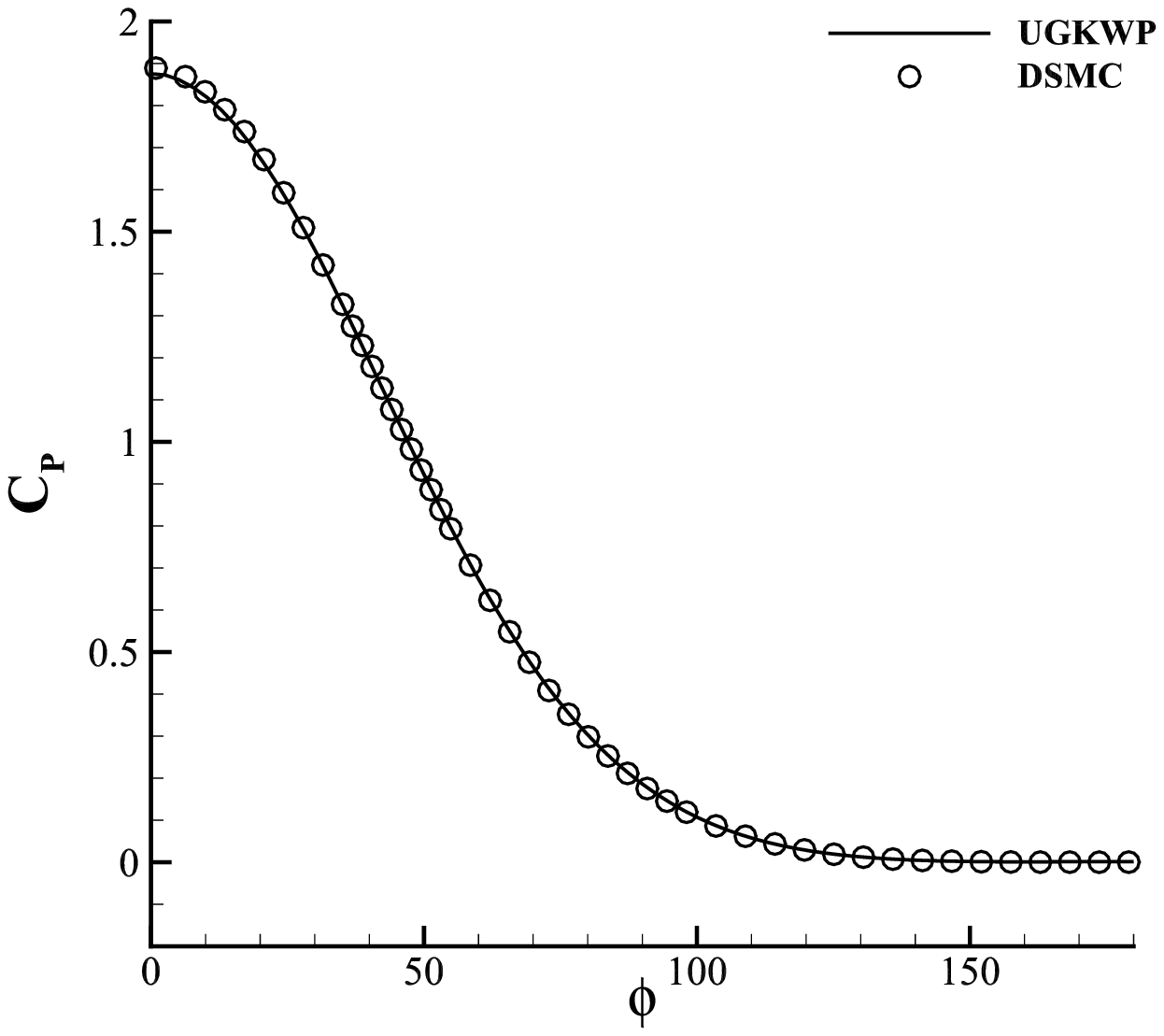}
    	}
    \subfigure[]{
    		\includegraphics[width=0.22 \textwidth]{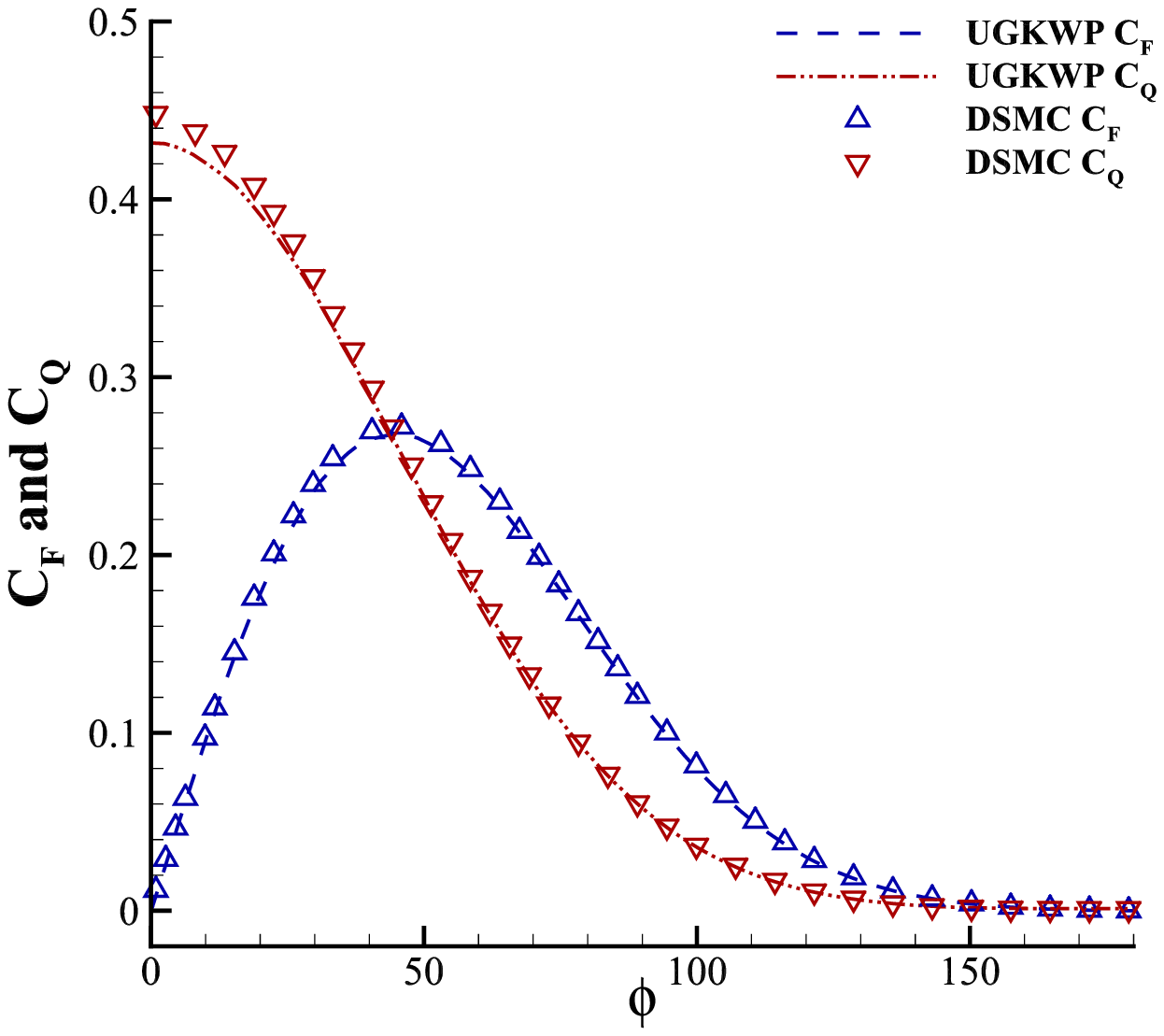}
    	}
	\caption{\label{ma18kn0.1} Hypersonic flow around a cylinder in an Ar-Ne mixture at ${\rm{Ma}}_{\infty}=18$, ${\rm{Kn}}_{\infty}=0.1$: (a) ${\rm{Ma}}$ contour and streamline, (b) velocity along the stagnation line, (c) temperature along the stagnation line, (d) density along the stagnation line, (e) pressure coefficient at the wall, (f) shear stress and heat flux coefficients at the wall.}
\end{figure}

\begin{figure}[H]
	\centering
	\subfigure[]{
			\includegraphics[width=0.22 \textwidth]{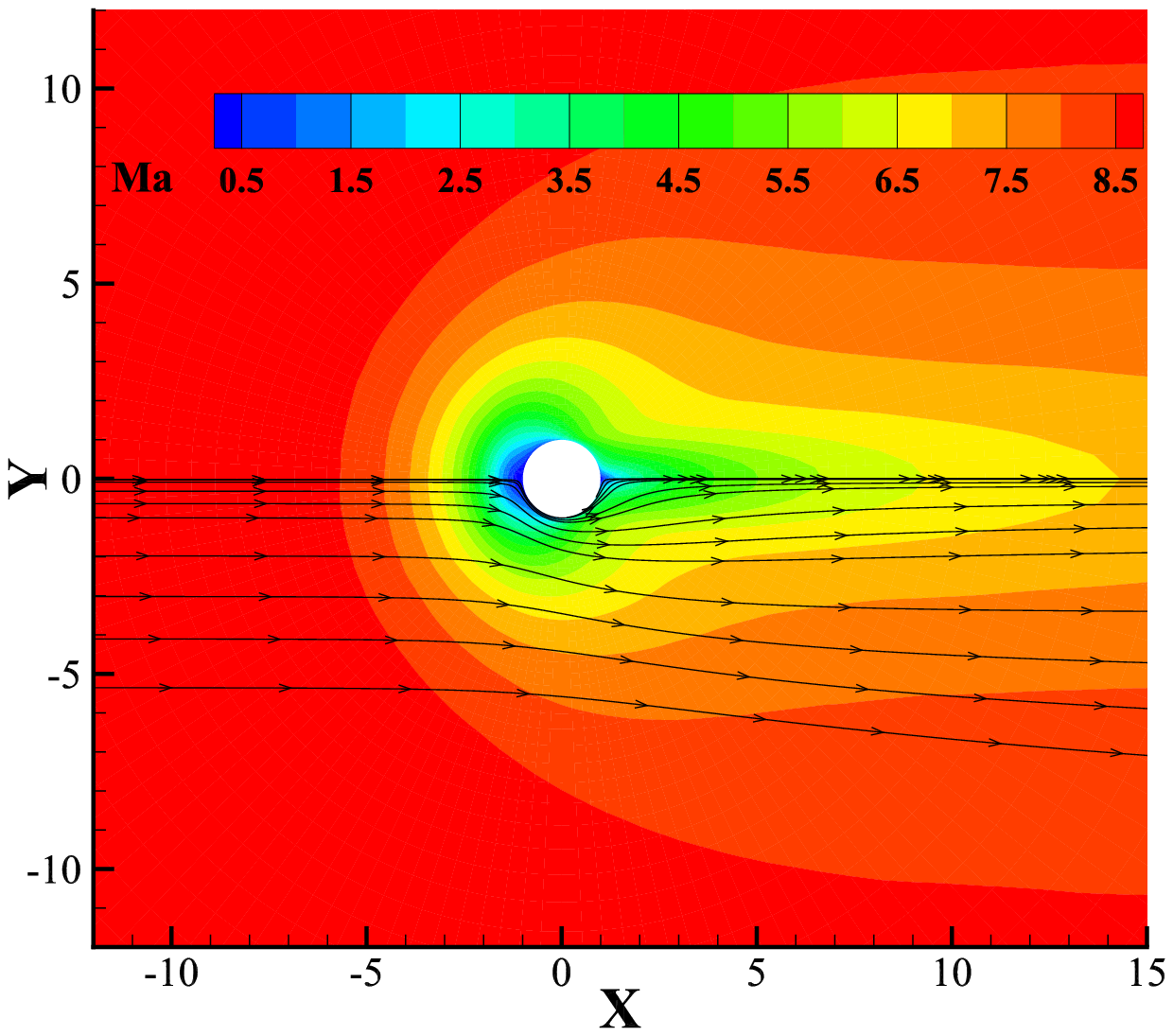}
		}
    \subfigure[]{
    		\includegraphics[width=0.22 \textwidth]{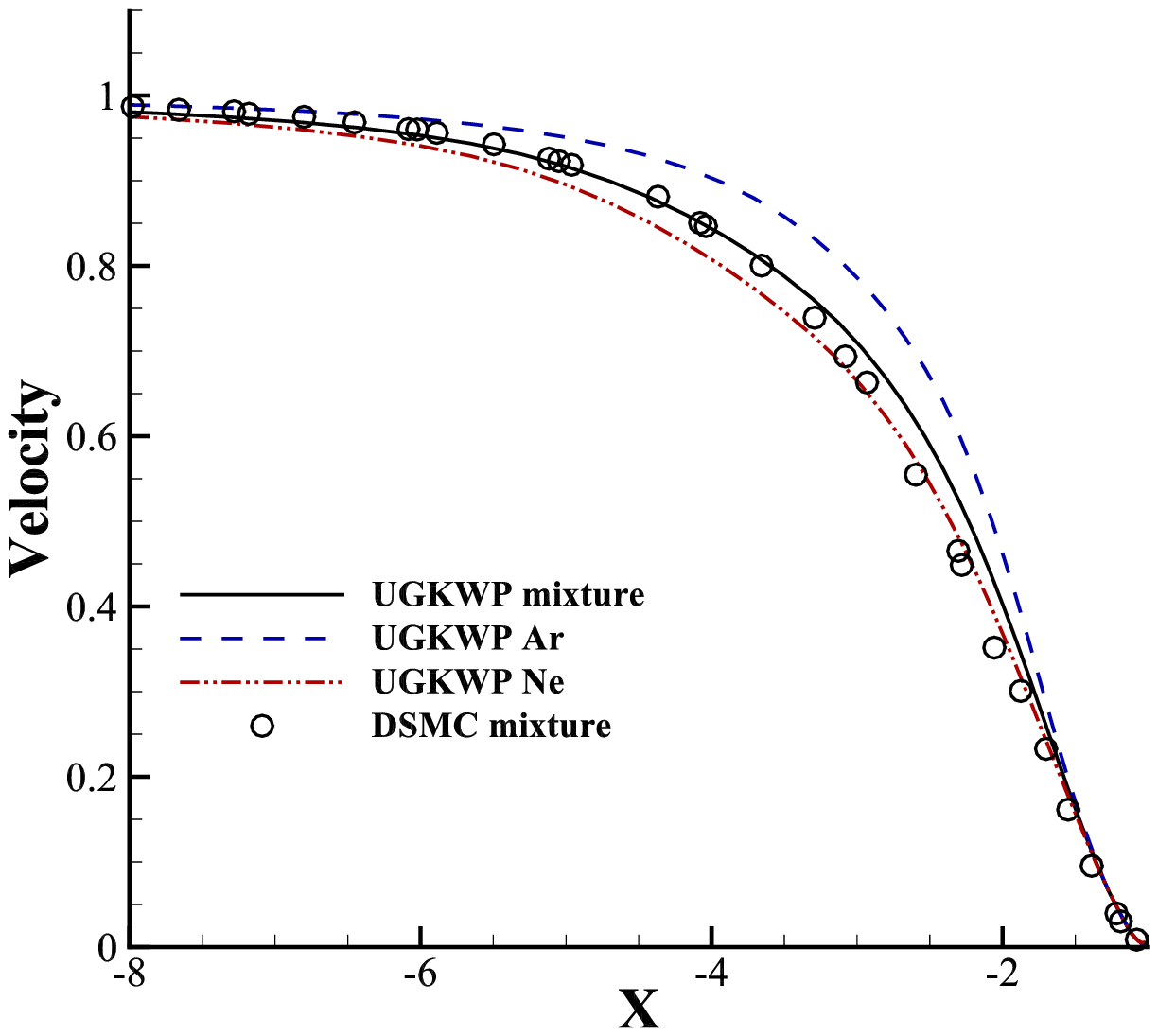}
    	}
    \subfigure[]{
    		\includegraphics[width=0.22 \textwidth]{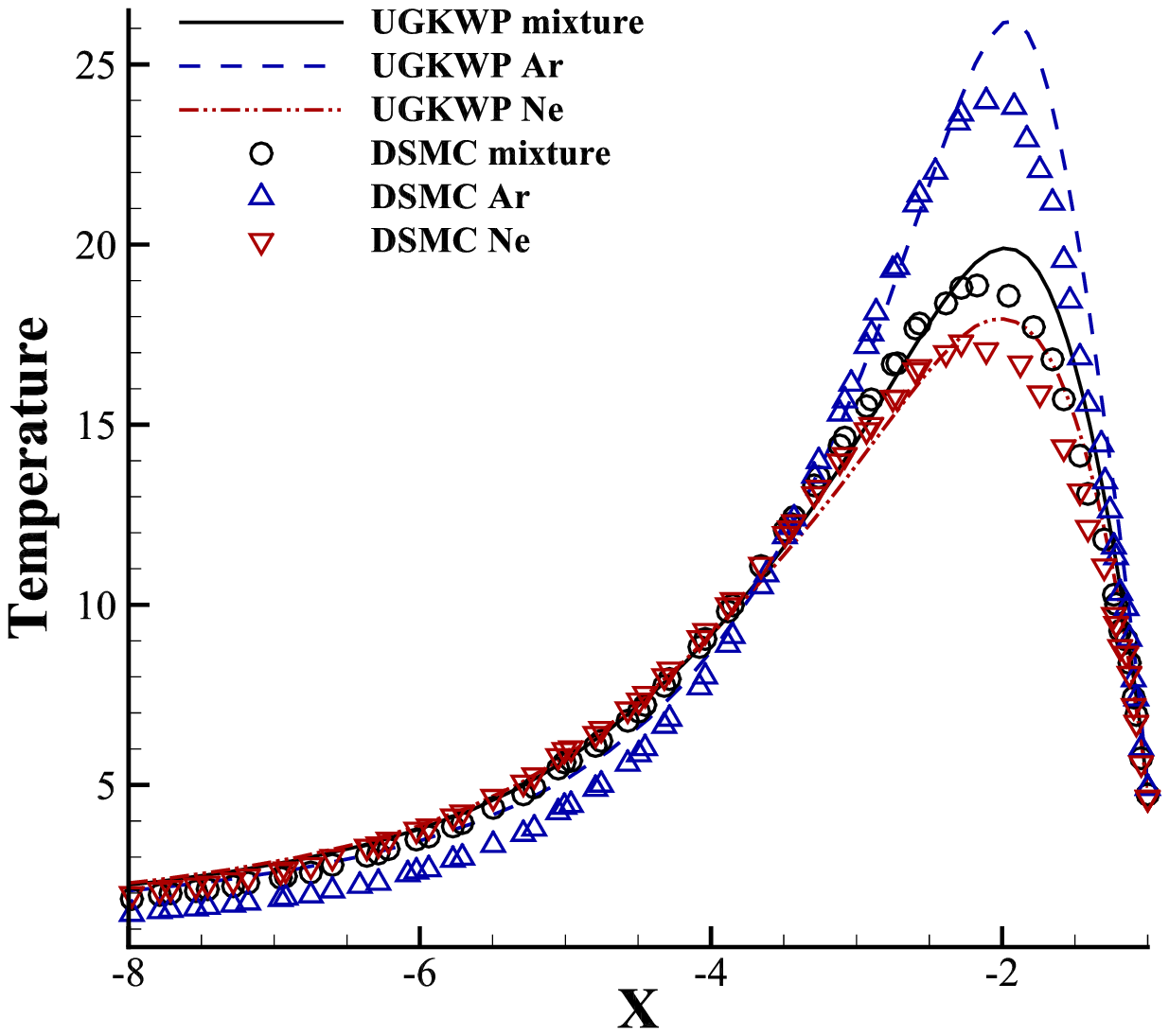}
    	}
    \\
    \subfigure[]{
			\includegraphics[width=0.22 \textwidth]{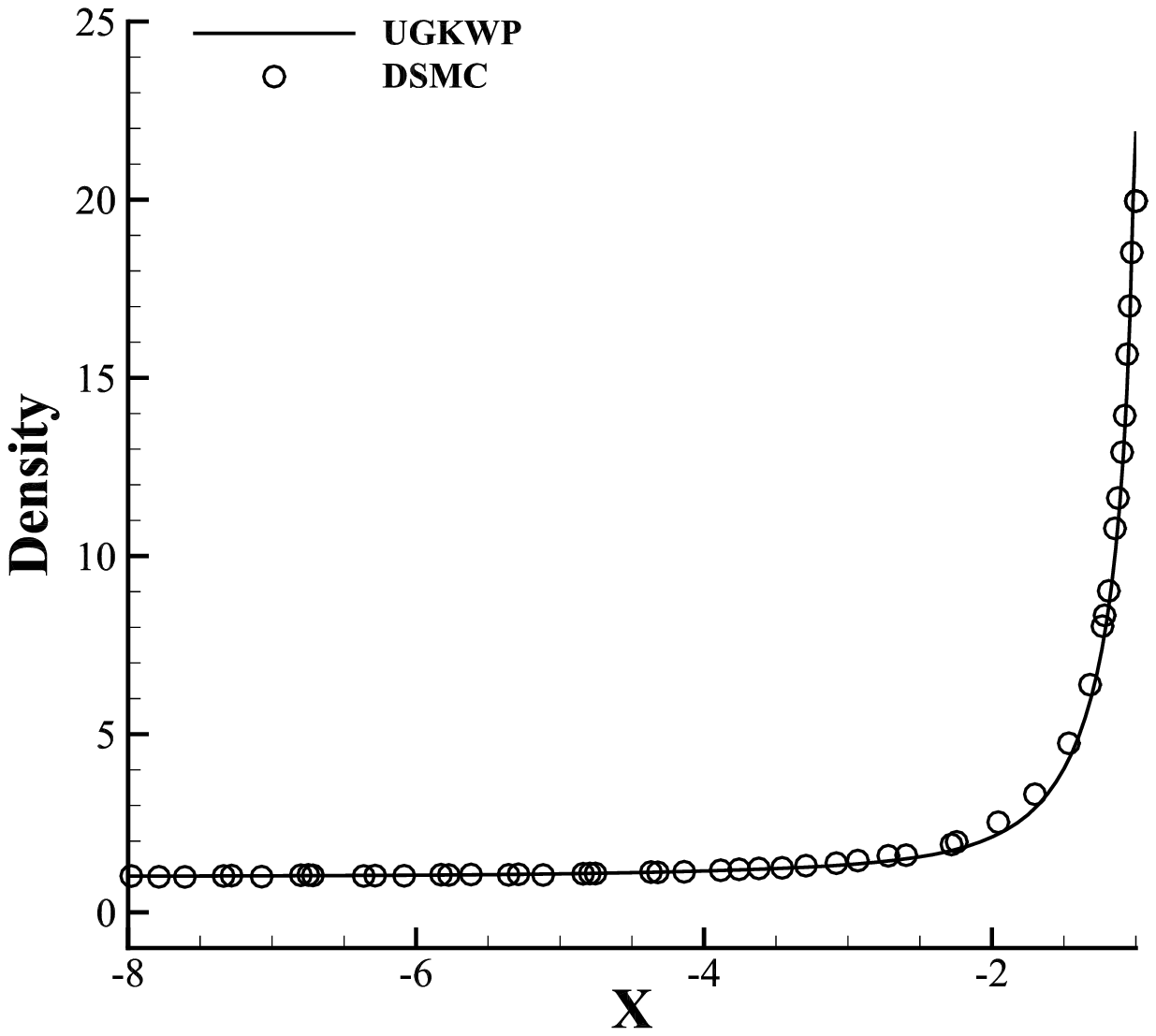}
		}
    \subfigure[]{
    		\includegraphics[width=0.22 \textwidth]{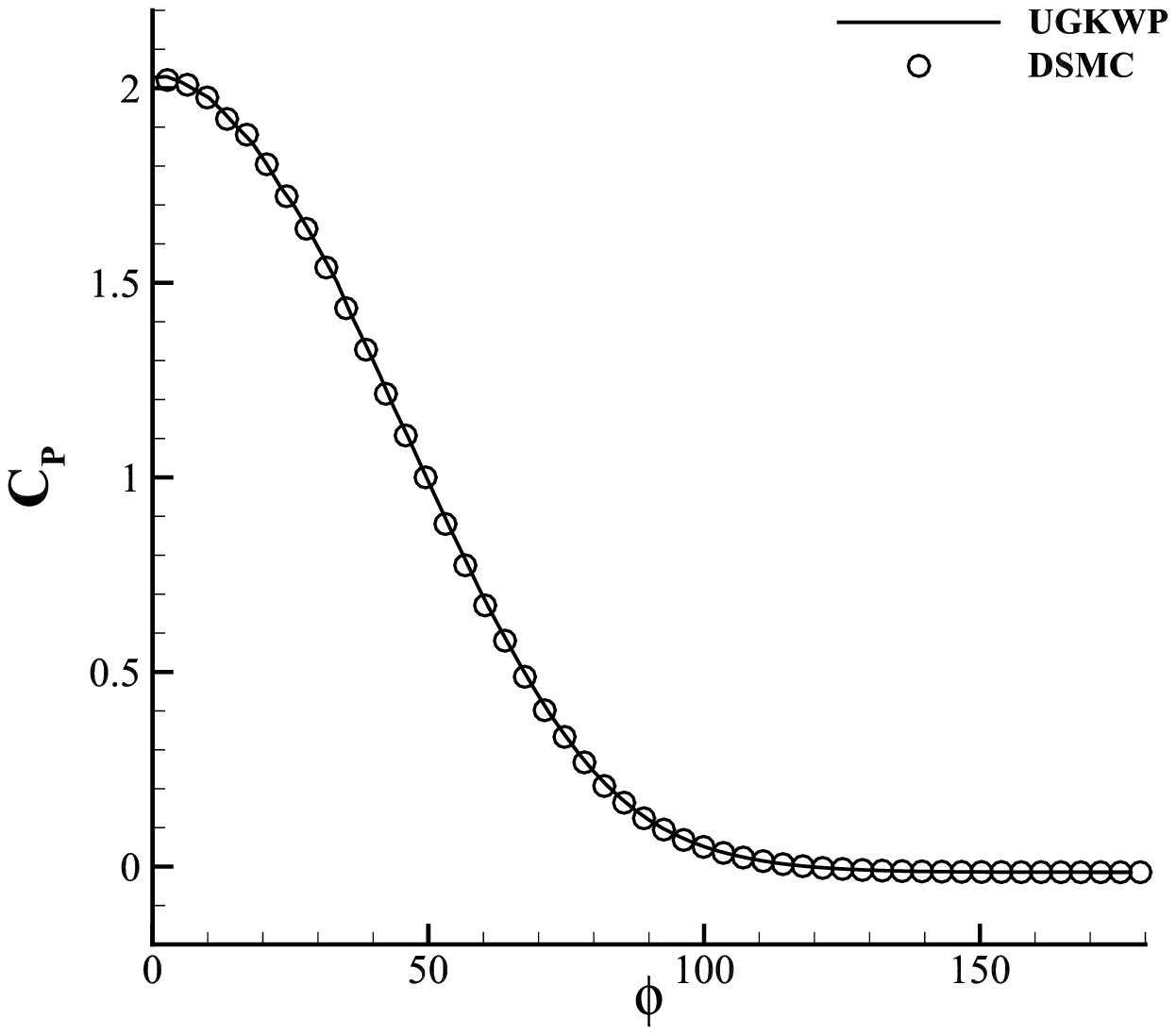}
    	}
    \subfigure[]{
    		\includegraphics[width=0.22 \textwidth]{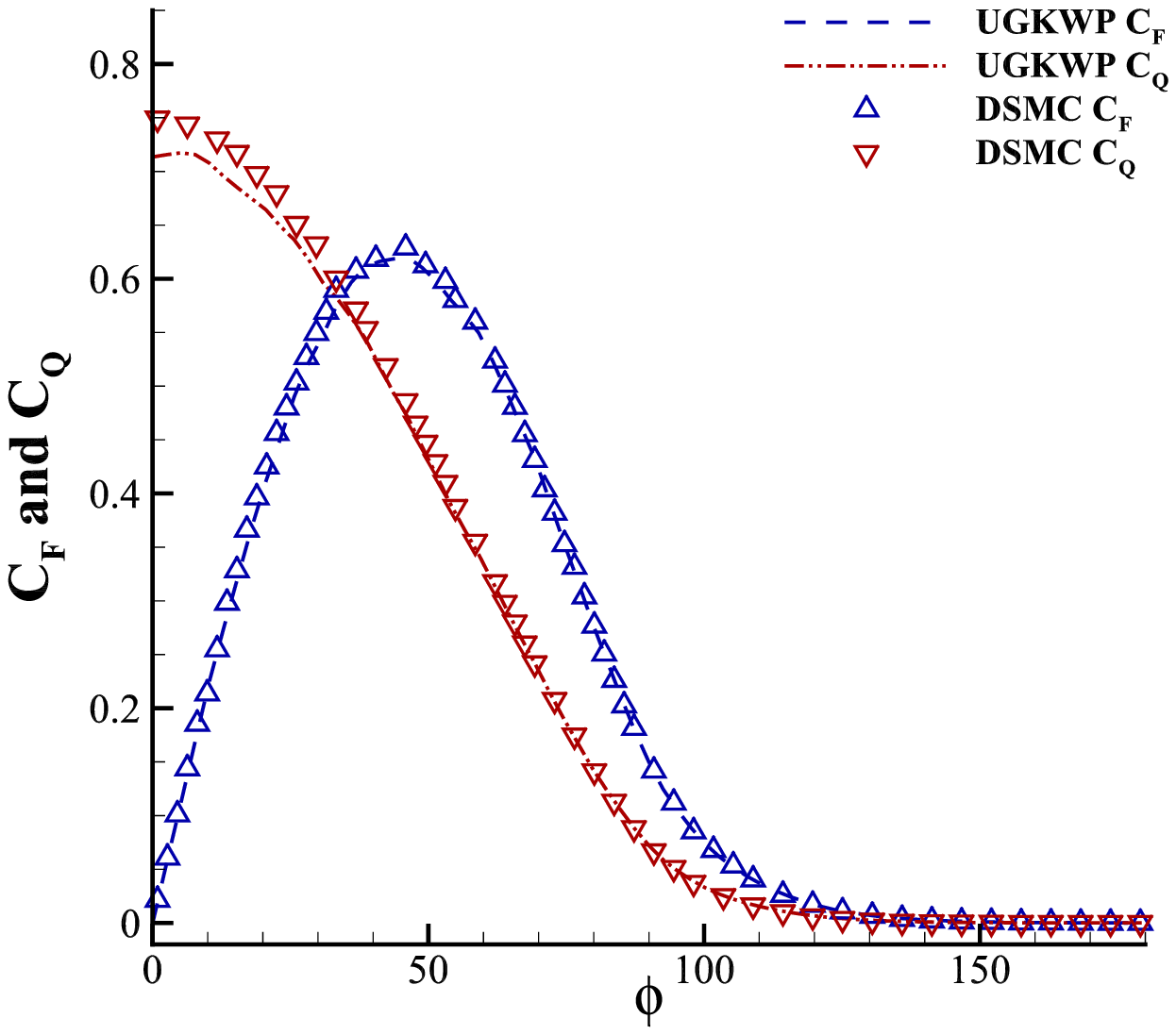}
    	}
	\caption{\label{ma9kn1} Hypersonic flow around a cylinder in an Ar-Ne mixture at ${\rm{Ma}}_{\infty}=9$, ${\rm{Kn}}_{\infty}=1$: (a) ${\rm{Ma}}$ contour and streamline, (b) velocity along the stagnation line, (c) temperature along the stagnation line, (d) density along the stagnation line, (e) pressure coefficient at the wall, (f) shear stress and heat flux coefficients at the wall.}
\end{figure}

\begin{figure}[H]
	\centering
	\subfigure[]{\label{ma9kn0.01-ma}
			\includegraphics[width=0.22 \textwidth]{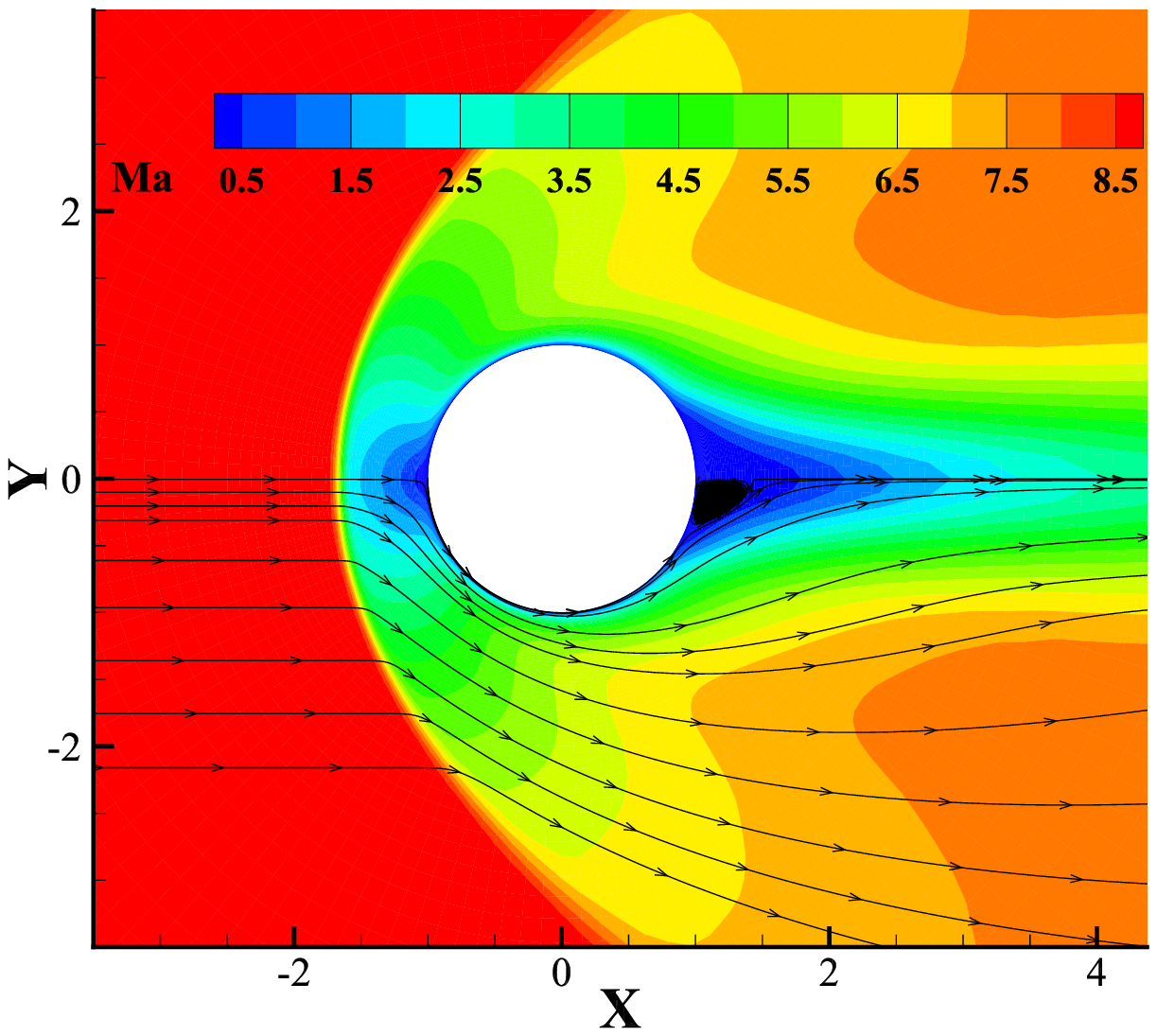}
		}
    \subfigure[]{
    		\includegraphics[width=0.22 \textwidth]{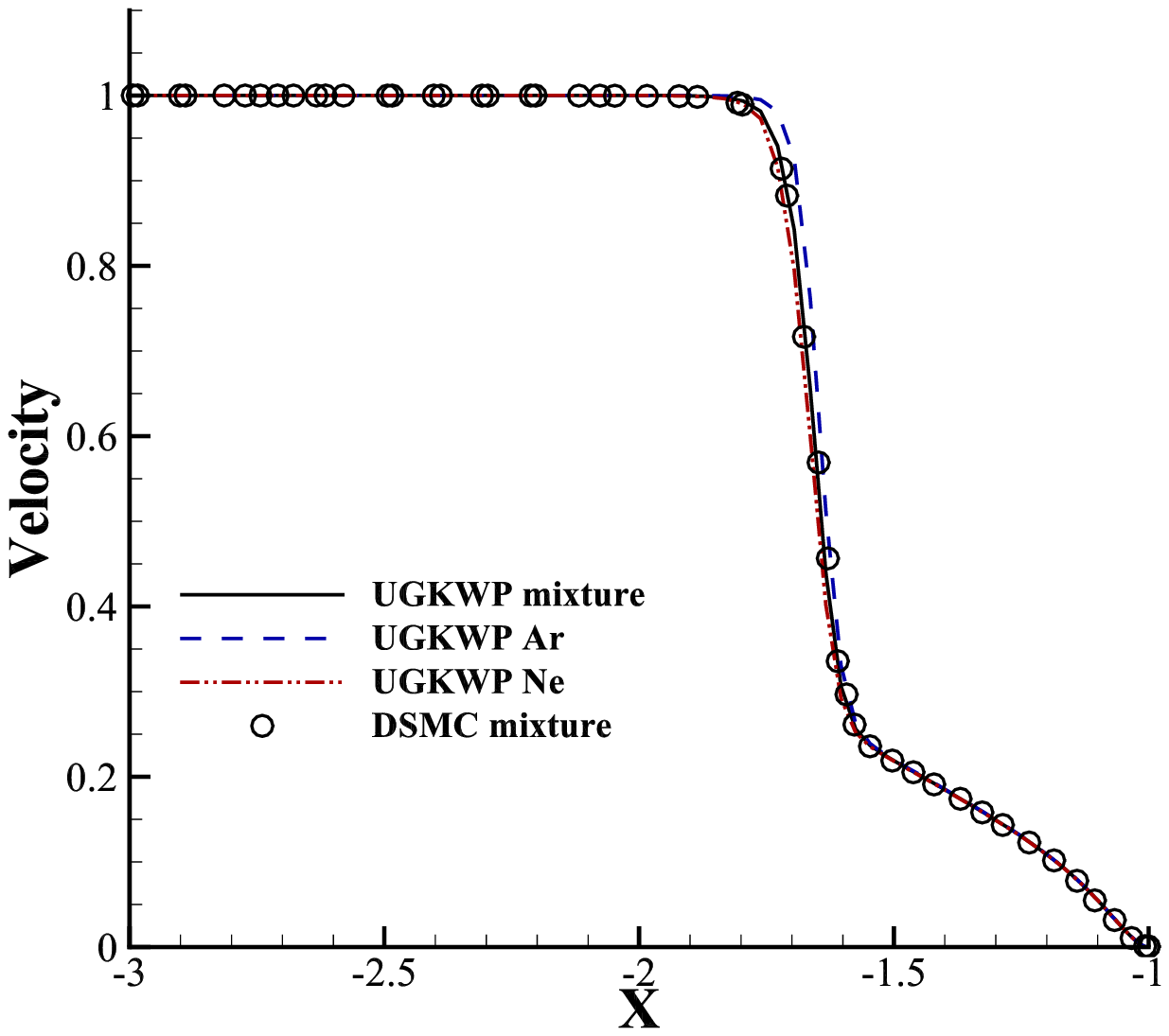}
    	}
    \subfigure[]{\label{ma9kn0.01-t}
    		\includegraphics[width=0.22 \textwidth]{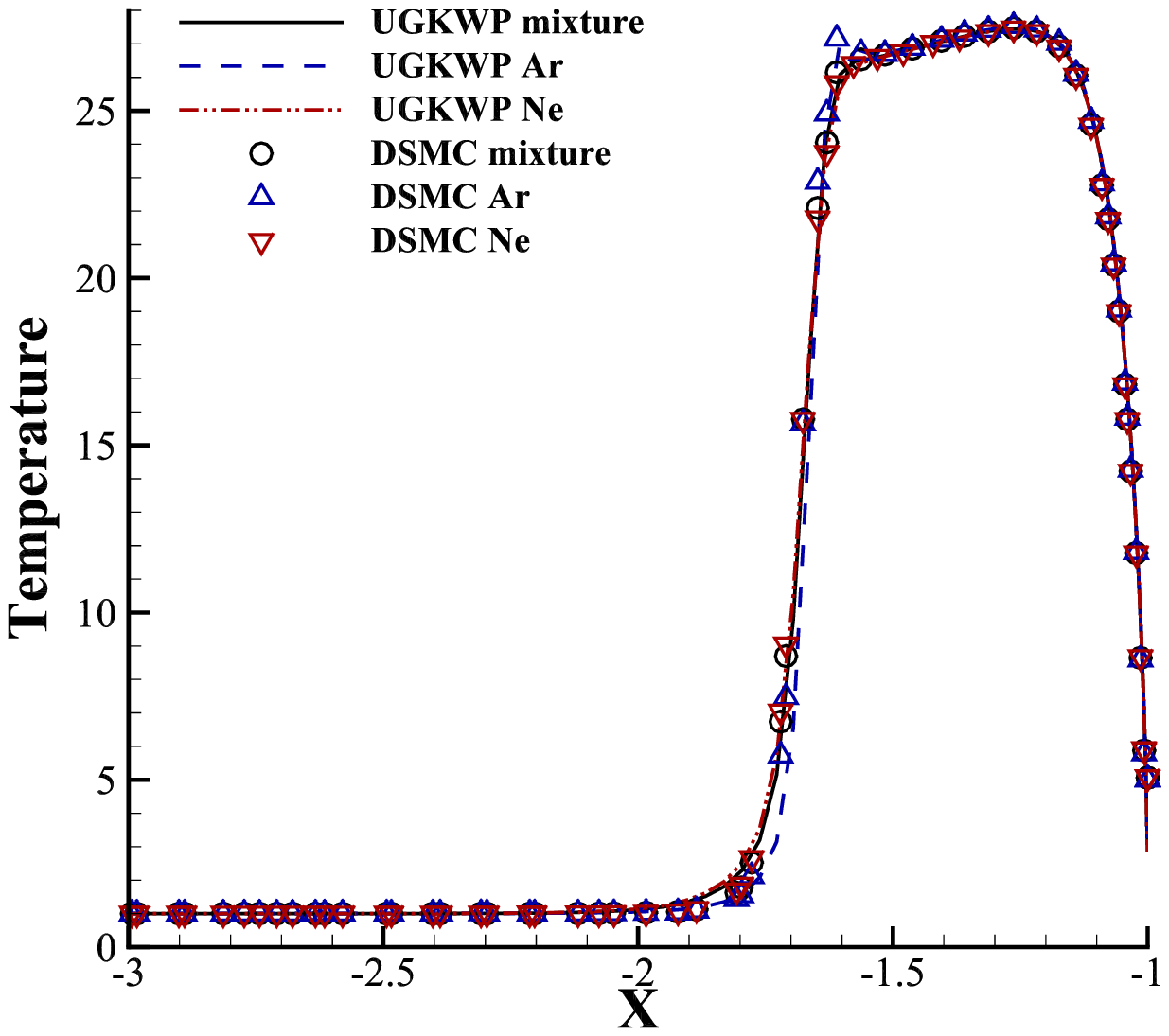}
    	}
    \\
    \subfigure[]{
			\includegraphics[width=0.22 \textwidth]{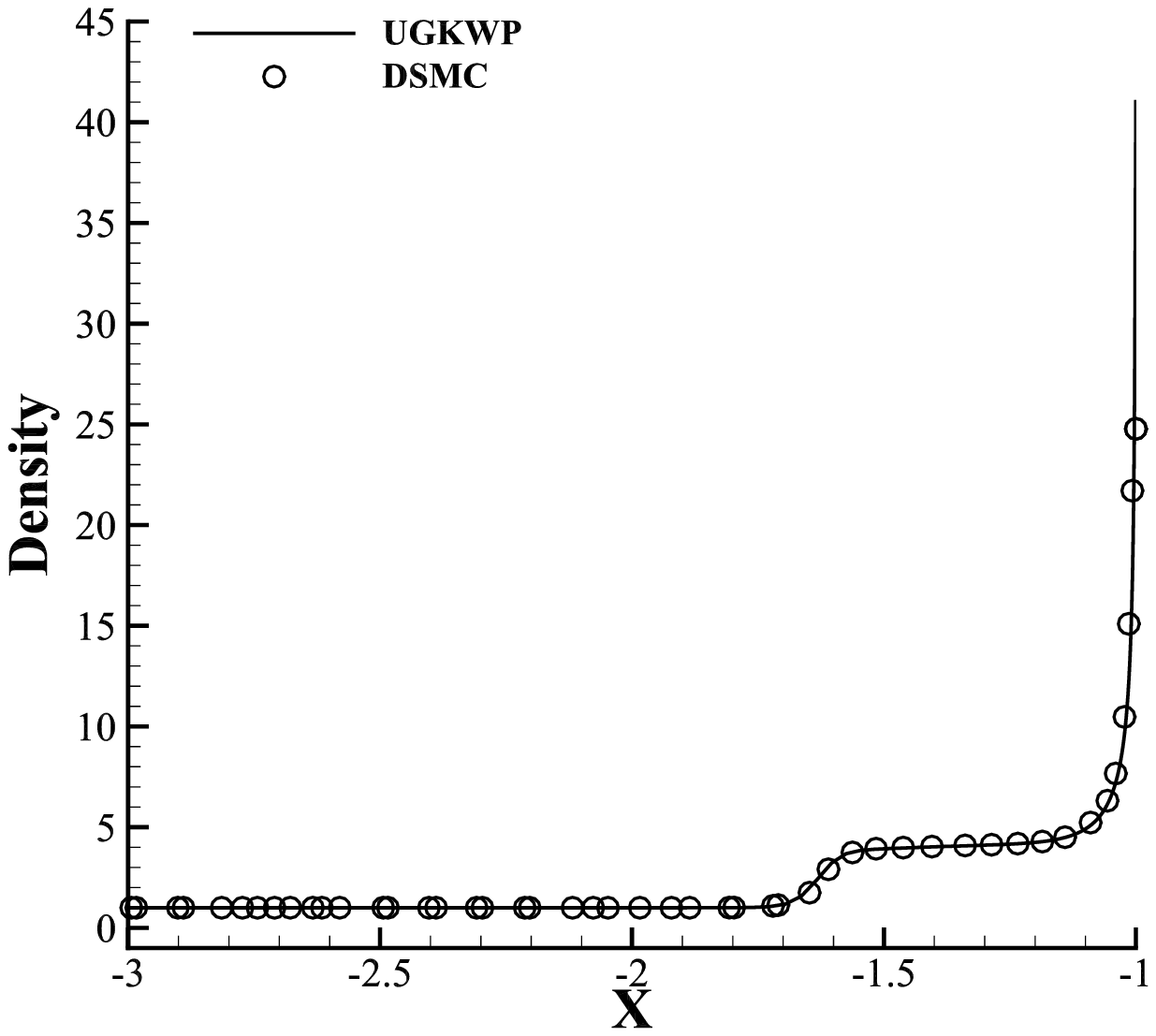}
		}
    \subfigure[]{
    		\includegraphics[width=0.22 \textwidth]{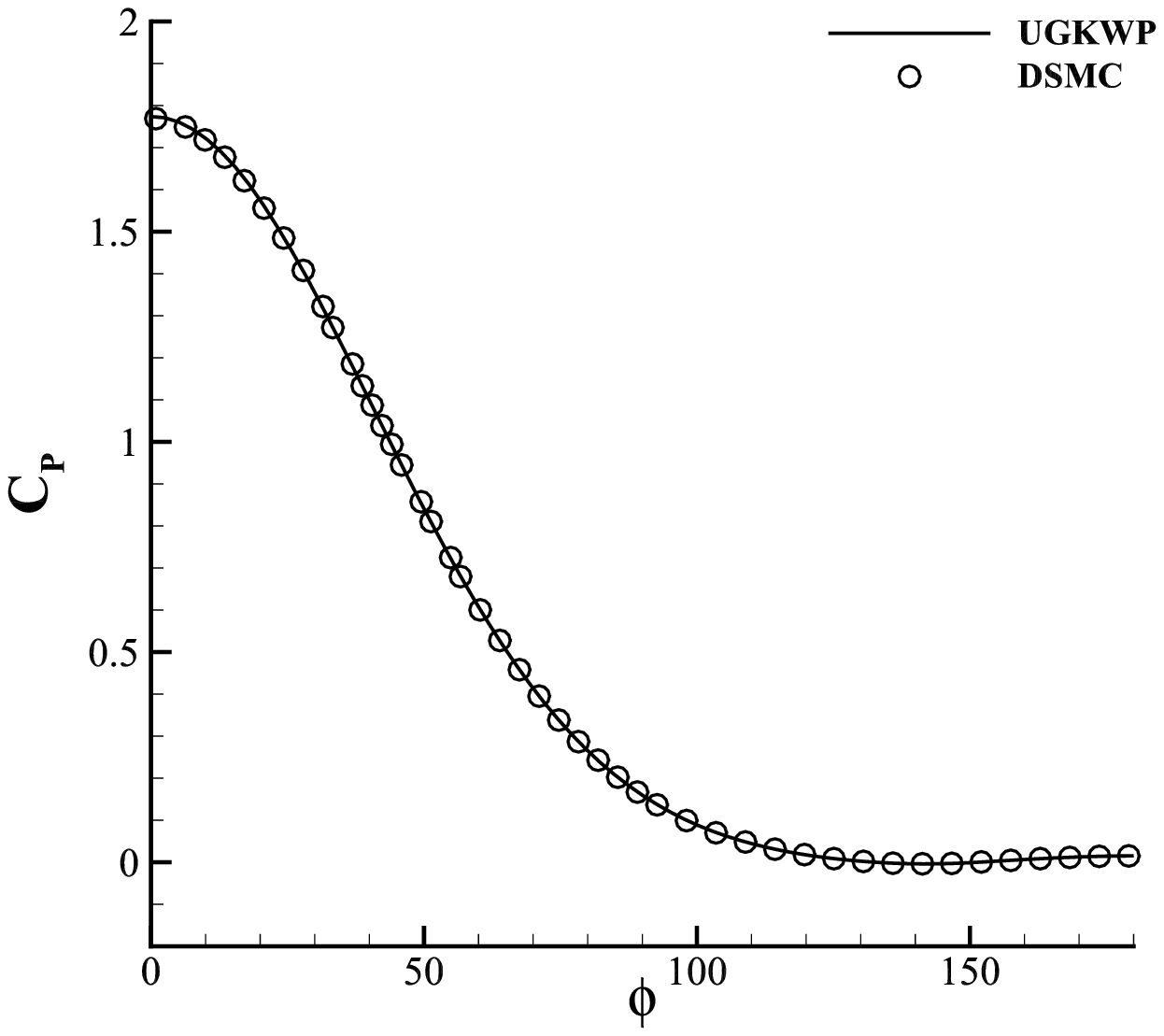}
    	}
    \subfigure[]{\label{ma9kn0.01-cfcq}
    		\includegraphics[width=0.22 \textwidth]{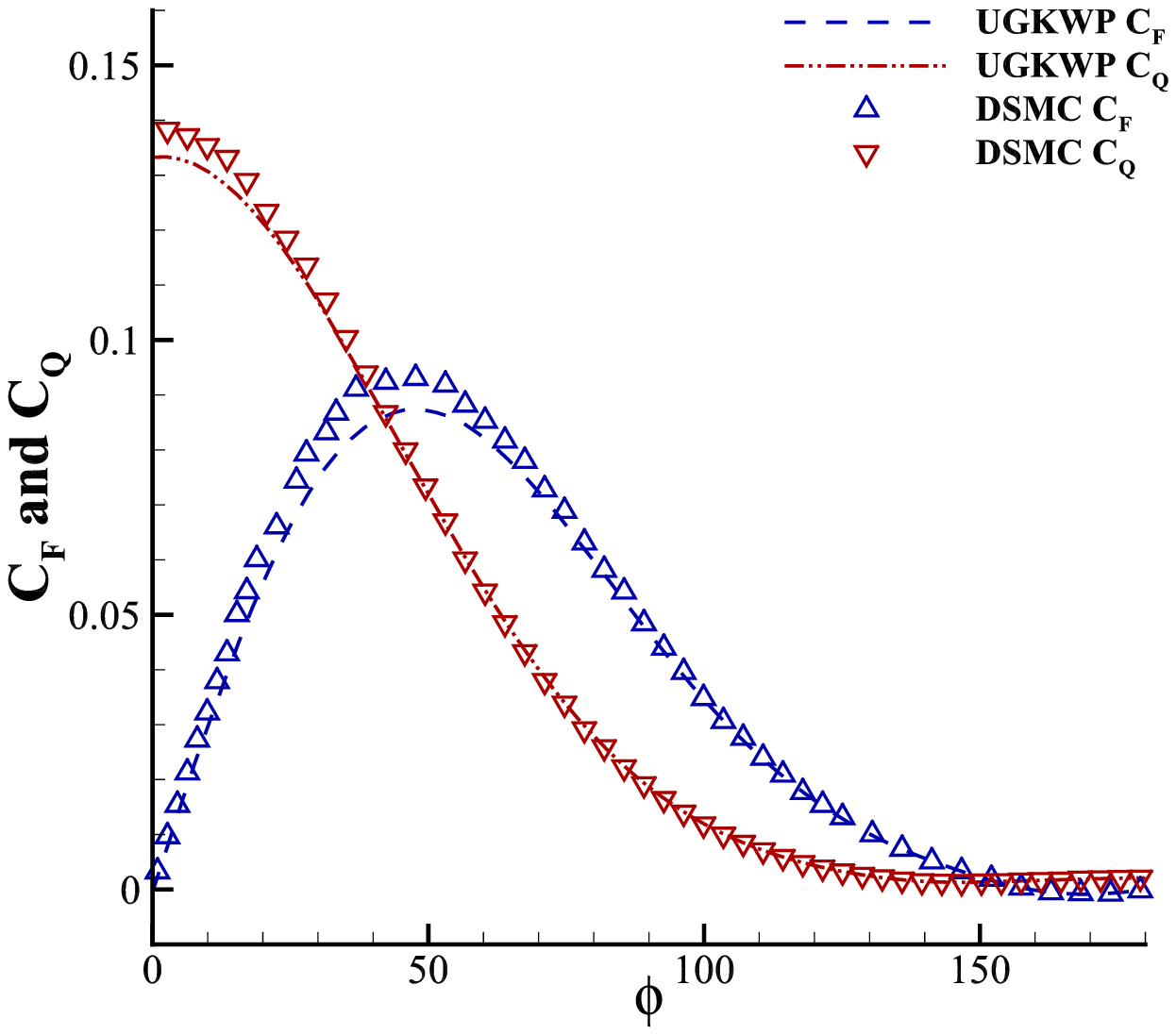}
    	}
	\caption{\label{ma9kn0.01} Hypersonic flow around a cylinder in an Ar-Ne mixture at ${\rm{Ma}}_{\infty}=9$, ${\rm{Kn}}_{\infty}=0.01$: (a) ${\rm{Ma}}$ contour and streamline, (b) velocity along the stagnation line, (c) temperature along the stagnation line, (d) density along the stagnation line, (e) pressure coefficient at the wall, (f) shear stress and heat flux coefficients at the wall.}
\end{figure}

The hypersonic flow around a cylinder is a typical multiscale case from continuum to rarefied regime. The high-$\rm{Ma}$ shock strongly compresses the flow at the windward wall, which makes the local scale tends more towards continuum, while the flow at leeward side is more rarefied due to the expansion. The characteristics of this case are consistent with the hypersonic flow around near-space vehicles, and in this section a wide range of ${\rm{Ma}}_{\infty}$ and ${\rm{Kn}}_{\infty}$ is taken for validation. Both of dimensionless and dimensional parameters are given in Tab.~\ref{cylinder-tt} and Tab.~\ref{cylinder-rhou}, where the gas properties can be referred to Tab.~\ref{arne} and ${\rm{Kn}}_{\infty}$ is calculated as Eq.~\eqref{eq:kn}. The results are shown from Fig.~\ref{ma3kn0.1} to Fig.~\ref{ma9kn0.01}. With ${\rm{Ma}}_{\infty}$ increasing, the shock is compressed closer to the wall. At higher ${\rm{Kn}}_{\infty}$, the shock is broader and the flow is smoother, while at lower ${\rm{Kn}}_{\infty}$, the shock becomes sharper and there is even a small vortex at the wake area as in Fig.~\ref{ma9kn0.01-ma}, which brings about a negative shear stress as in Fig.~\ref{ma9kn0.01-cfcq}. Along the stagnation line there are significant deviations of velocity and temperature between different species in rarefied cases. While in the near-continuum regime they gather closer, and in this case the most obvious difference lies in the post-shock location as in Fig.~\ref{ma9kn0.01-t}. The DSMC simulations are performed using DS2V code~\cite{dsmc}, and the results of proposed method match well with the DSMC, including the pressure, shear stress and heat flux coefficients at the wall.

\begin{table}[H]
\centering
\caption{\label{cylinder-tt} Identical parameters in all cases of hypersonic flow around a cylinder}
\begin{tabular}{|c|c|c|c|}
\hline
    $\chi_{{\rm{Ar}},\infty}$  &$0.25$       &Inflow temperature     &$273{\rm{K}}$ \\
	$\chi_{{\rm{Ne}},\infty}$  &$0.75$       &Wall temperature       &$500{\rm{K}}$ \\
    Reference length (radius)  &$1{\rm{m}}$  &                       &              \\
\hline
\end{tabular}
\end{table}

\begin{table}[H]
\centering
\caption{\label{cylinder-rhou} Differing parameters in each case of hypersonic flow around a cylinder}
\begin{tabular}{|c|c|c|c|c|c|}
\hline
    Case number                                  &$1$                  &$2$                  &$3$                  &$4$                  &$5$                  \\ \hline
    ${\rm{Ma}}_{\infty}$                         &$3$                  &$9$                  &$18$                 &$9$                  &$9$                  \\
	${\rm{Kn}}_{\infty}$                         &$0.1$                &$0.1$                &$0.1$                &$1$                  &$0.01$               \\
    ${\rho}_{\rm{Ar}}\left({\rm{kg/m^3}}\right)$ &$2.352\times10^{-7}$ &$2.352\times10^{-7}$ &$2.352\times10^{-7}$ &$2.352\times10^{-8}$ &$2.352\times10^{-6}$ \\
    ${\rho}_{\rm{Ne}}\left({\rm{kg/m^3}}\right)$ &$3.565\times10^{-7}$ &$3.565\times10^{-7}$ &$3.565\times10^{-7}$ &$3.565\times10^{-8}$ &$3.565\times10^{-6}$ \\
    ${U}\left({\rm{m/s}}\right)$                 &$1164.396$           &$3493.188$           &$6986.376$           &$3493.188$           &$3493.188$           \\
    Result                                       &Fig.~\ref{ma3kn0.1}  &Fig.~\ref{ma9kn0.1}  &Fig.~\ref{ma18kn0.1} &Fig.~\ref{ma9kn1}    &Fig.~\ref{ma9kn0.01} \\
\hline
\end{tabular}
\end{table}

\section{Conclusions}\label{sec:conclusion}
In this study, viscosity, heat conduction, and diffusion effects are comprehensively incorporated into the UGKWP method for multiscale gas mixtures. Utilizing the model of Groppi et al., the velocity of the target equilibrium distribution function is calculated as a convex combination of the individual species velocity and the mixture velocity. On one hand, this formulation reduces the stiffness of the momentum-exchange source term, eliminating the need for an implicit solver and facilitating straightforward extensions to multi-species mixtures. On the other hand, by introducing a second relaxation parameter, both the diffusion and viscosity coefficients are successfully recovered in the continuum flow regime. Unlike traditional approaches that treat the diffusion term solely as a source term via operator splitting, the entire target equilibrium distribution function is utilized here to derive the characteristic integral solution. This yields a multiscale flux that inherently accounts for diffusion effects. Additionally, the target equilibrium distribution function is evaluated at the same time level for both the flux and source term evaluations of waves and particles. This ensures strict consistency between particle and wave evolution while guaranteeing conservation. To correct the Prandtl number ($\rm{Pr}$), the Shakhov model formulation is incorporated. Consequently, the proposed method performs well in predicting both aerodynamic forces and heat fluxes in hypersonic test cases. Furthermore, the microscopic model for high-speed particles is modified by extending the formulation in Ref.\cite{xu-tau} to gas mixtures, moving beyond a single relaxation time. This extension leads to a significant improvement in capturing the temperature profile in the pre-shock region. Through binary-species test cases spanning multiple flow regimes, including shock structures, mass diffusion, Couette flows, and hypersonic flow around a cylinder, the results obtained from the proposed UGKWP method agree well with reference solutions. These benchmarks include solutions from the Boltzmann equation, the DSMC method, the UGKS based on the AAP model, and the DVM based on the McCormack model. The method accurately captures complex flow structures, particularly species-specific differences in mole fractions, velocities, and temperatures, as well as wall pressure, shear stress, and heat flux coefficients under hypersonic conditions. These findings demonstrate that the proposed method meets the rigorous requirements for simulating near-space hypersonic flows and possesses strong potential for extensions to multi-component systems and more complex physical couplings, such as thermochemical non-equilibrium and plasma transport.

\section*{Acknowledgements}
The current research is supported by National Key R$\&$D Program of China (Grant Nos. 2022YFA1004500), National Natural Science Foundation of China (92371107), and Hong Kong research grant council (16208324). Helpful discussions with Prof. Z. Guo and Dr. Z. Pu are gratefully acknowledged.

\section*{Reference}
\bibliography{wp-binary-ref}

\end{document}